\documentclass[12pt]{article}
\usepackage{a4}
\usepackage{epsfig}
\def\Journal#1#2#3#4{{#1} {\bf #2}, #3 (#4)}

\def\NPB{{\em Nucl. Phys.} B}
\def\PLB{{\em Phys. Lett.}  B}
\def\PRL{\em Phys. Rev. Lett.}
\def\PRD{{\em Phys. Rev.} D}
\def\PRP{{\em Phys. Reports}}
\def\ZPC{{\em Z. Phys.} C}
\def\COMP{{\em Comput. Phys. Commun.} }
\begin{document}
\newcommand{\newc}{\newcommand}
\newc{\R}{$R$}
\newc{\charginom}{M_{\tilde \chi}^{+}}
\newc{\mue}{\mu_{\tilde{e}_{iL}}}
\newc{\mud}{\mu_{\tilde{d}_{jL}}}
\newc{\barr}{\begin{eqnarray}}
\newc{\earr}{\end{eqnarray}}
\newc{\beq}{\begin{equation}}
\newc{\eeq}{\end{equation}}
\newc{\ra}{\rightarrow}
\newc{\lam}{\lambda}
\newc{\eps}{\epsilon}
\newc{\gev}{\,GeV}
\newc{\tev}{\,TeV}
\newc{\eq}[1]{(\ref{eq:#1})}
\newc{\eqs}[2]{(\ref{eq:#1},\ref{eq:#2})}
\newc{\etal}{{\it et al.}\ }
\newc{\eg}{{\it e.g.}\ }
\newc{\ie}{{\it i.e.}\ }
\newc{\nonum}{\nonumber}
\newc{\lab}[1]{\label{eq:#1}}
\newc{\dpr}[2]{({#1}\cdot{#2})}
\newc{\gsim}{\stackrel{>}{\sim}}
\newc{\lsim}{\stackrel{<}{\sim}}
\begin{titlepage}
\begin{flushright}
{ETHZ-IPP  PR-98-10} \\
{December 11, 1998}\\
\end{flushright}
\begin{center}
{\bf \LARGE Searching for the Higgs and other Exotic Objects} \\
\end{center}
\begin{center}
{\bf \large (A ``How to'' Guide from LEP to the LHC)} \\
\end{center}

\smallskip \smallskip \bigskip
\begin{center}
{\Large Michael Dittmar}
\end{center}
\bigskip
\begin{center}
Institute for Particle Physics (IPP), ETH Z\"{u}rich, \\
CH-8093 Z\"{u}rich, Switzerland
\end{center}

\bigskip
\begin{abstract}
\noindent 
Methods and ideas to search for new phenomena at existing and future collider 
facilities like LEPII and the LHC are analysed. Emphasis is put on the 
experimental aspects of discovery strategies for 
the Higgs Boson of the Standard Model. This is followed by a critical 
analysis of search methods for the extended Higgs sector and the direct 
detection of Supersymmetric Particles within the   
Minimal Supersymmetric Standard Model.
We also discuss methods and the potential mass 
reach of the LHC experiments to discover new Bosons and Fermions.
\end{abstract}
\vspace{2cm}

\begin{center}
{\it \large Lectures given at the \\
30~$^{eme}$ \'{E}cole d'\'{E}t\'{e} \\
de Physique des particules \\
Marseille 7-11 septembre 1998}
\end{center}
\end{titlepage}

\section{Introduction}
The search for new physics phenomena is often defined as the main motivation
for new experiments at higher center of mass energies.
This is especially true for the LHC project, with its two general purpose 
experiments ATLAS and CMS. The prime motivation of the LHC physics 
program is to discover the ``mechanism of electroweak symmetry breaking'' 
usually associated with a scalar particle, the Higgs boson. 
Theoretical ideas suggest that this hypothetical particle with a 
mass of less than roughly 1 TeV could explain the observed mass spectrum 
of bosons and fermions. The simplicity and mathematical elegance of
this model leads however to its own problems, the so called hierarchy problem 
or fine tuning problem of the Standard Model. 

These problems originate 
from theoretical ideas to extrapolate todays knowledge at mass scales of 
a few 100 GeV to energy scales which existed shortly after the Big Bang, 
e.g. to energies of about $10^{15}$ GeV and more.  
A purely theoretical approach to this extrapolation has lead theorist 
to SUPERSYMMETRY which could solve these conceptual problems by the 
introduction of supersymmetric partners to every known 
boson and fermion and at least an additional Higgs supermultiplett. 
Despite the largely unconstrained masses of these new partners,
the potential to discover such new objects has become a central question
for many design issues of future high energy particle physics experiments.  

The search for the Higgs boson(s) and the search for the supersymmetric 
particles can be considered 
as ``safe searches'', as they fit well into todays theoretical fashions.
In addition to safe searches one might be tempted 
to search for less fashionable exotic new phenomena. 
Such new phenomena, like new forces leading to CP violation or 
lepton number violation, might simply exist because they are 
not forbidden by any fundamental reason.
Such searches are often motivated by the possibility that the guidance 
from our theoretical methods has not yet reached a mature status.
These exotic searches require certainly a ``gambling mentality''
as they lead to ``all or nothing'' results.  

Having currently no clear experimental evidence for any exotic new phenomena
or anomalies, our discussion on search strategies for new phenomena at  
future high energy colliders should be quite general. However, despite 
the few ideas discussed in section 4, we follow todays fashion and discuss 
mainly the ideas and detector requirements 
discussed for future succesfull searches for the Higgs
and for Supersymmetry.
The presented ideas and methods should nevertheless give
a good guidance to other ``all or nothing'' searches. 
Consequently, todays simulation results 
of future LHC experimentation provide an important   
guidance on why one wants to participate in future LHC experiments 
and on how these experiments should look like.   

This ``how to guide'' on searches is structured as follows. 
First we discuss some general ,``how to'', experimental methods of searches
and compare these with a few examples. The proposed 
methods to search for the Standard Model Higgs boson at LEPII, at the LHC 
and at the TeV33 project are described in section 3. Search methods 
for additional heavy W$^{\prime}$ and Z$^{\prime}$ bosons are discussed 
in section 4. 
Finally in sections 5--8
we analyse SUSY discovery prospects in the extended Higgs sector
or with direct SUSY particle searches at the LHC. 

\section {How to discover new physics?}

Discovering new phenomena in high energy physics experiments 
means obviously to separate ``new''
from ``known'' phenomena. The used methods exploit the 
different kinematics of signals and backgrounds 
by looking directly for new mass peaks, or indirectly for measurable 
quantities like the $p_{t}$ spectra of leptons, photons and jets and their 
angular correlations. Other searches exploit 
the missing energy and momentum signature, which might either originate from 
unknown neutrino like objects, extra neutrino sources or simply 
from detector imperfections. Depending on the particular search project
different aspects of the detector design become important. 
The search for mass peaks requires in general excellent 
energy and momentum resolution for individual particles 
with less stringent requirements on the angular acceptance.
In contrast, searches based on the missing energy signature 
or the rate of events with special kinematic properties
like events with inclusive leptons and multi--jets with high mass 
require robust detectors with almost perfect angular coverage.  

Table 1 and Figure 1 combine required experimental observables 
with new physics possibilities. 
\begin{table}[htb]
\begin{center}
\begin{tabular}{|c|c|c|}
\hline
Type of measurements & indicates & required for \\
\hline
isolated high $p_{t}$ $e^{\pm}, \mu^{\pm}$ & $W^{(*)}$, $Z^{(*)}$ decays &  
Higgs search \\ 
& &top physics, ``all'' searches \\
\hline
isolated high $p_{t}$ $\gamma$'s & electro-magnetic process & Higgs search \\
\hline
$\tau$ and $b$-quark tagging & ``rare'' processes & 
special Higgs like searches \\
\hline
large missing    & $\nu$ like events & Higgs, Supersymmetry, \\
$p_{t}, E_{t}$   & $W, Z$ decays     & exotic ``exotica''    \\
\hline
jets                   & quarks and gluons & QCD, understanding of \\
                       &                   & backgrounds/efficiencies \\
\hline
\end{tabular}
\caption{New physics and some required detector capabilities.}
\label{tab:table1}
\vspace{0.2cm}
\end{center}
\end{table}

\begin{figure}[htb]
\begin{center}
\rotatebox{-90}{
\epsfig{figure=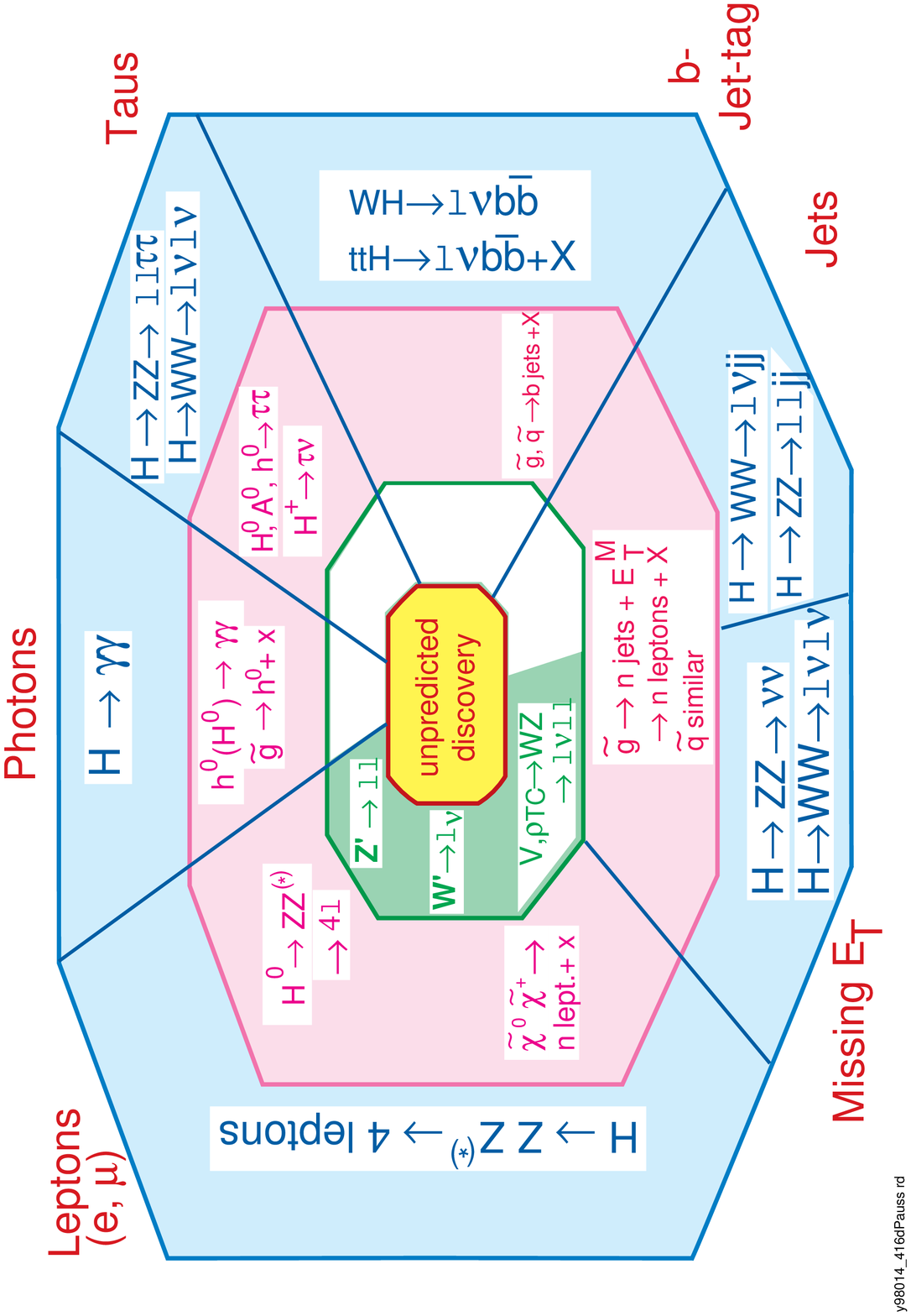,
width=8.cm,height=13.cm}}
\end{center}
\caption[fig1]{Possible new physics signatures and the 
corresponding detector needs}  
\end{figure}

Obviously, real experiments like the LEP or Tevatron 
experiments have to be a compromise between these different requirements.
Nevertheless, the existing experiments have proven to work 
according or better than specified in their technical proposals.
Especially astonishing results have been achieved 
with silicon  
micro vertex detectors, which allow to identify b--flavoured jets 
with high efficiencies and 
excellent purity. In addition, quite accurate 
calorimetric measurements allow to measure the missing
transverse energy in complicated events. Such indirect 
identification of energetic neutrino like objects is now 
routinely used by essentially all large collider experiments.   

The design objectives of the future ATLAS~\cite{atlastp} and 
CMS~\cite{cmstp} LHC experiments 
follow the above desired detector capabilities with
emphasis on high precision measurements with electrons, muons and 
photons and large angular coverage for jets. 
According to their technical proposals     
both collaborations 
expect to identify isolated electrons and muons, 
with $p_{t} > 10$ GeV and small backgrounds up to a pseudorapidity 
($\eta = -ln \tan (\Theta/2)$) of $|\eta| \leq 2.5$
and efficiencies of $\approx$ 90\%. Furthermore, both experiments 
expect to achieve b-jet tagging with up to 50\% efficiency and  
light flavour jet rejection factors of up to 100.   
These expectations are used for essentially all simulations of 
LHC measurements. For justifications of these figures 
we refer to the various ATLAS and CMS 
technical design reports and internal technical 
notes~\cite{atlascms}.  

\subsection {New Physics from mass peaks and from tails?}

Peaks in the invariant mass spectrum of assumed decay 
products are an unambiguous signature for new unknown particles. 
Narrow mass peaks do in principle not even require
any theoretical background estimates  
as the signal significance can be estimated directly from the data and
significance estimations for future experiments can be quite reliable. 
The reason becomes quite obvious from the following example with  
an expected Signal of 1000 events above a 
background of 10000. Such a deviation from known sources
could be claimed with a significance of about 10 standard deviations
($N(\sigma) = S/\sqrt{B} = 1000/\sqrt{10000}$). Assuming a relatively 
smooth flat background over many non signal bins, 
background extrapolations to the signal region 
can reach systematic accuracies of less than a percent.
Under such ideal conditions, even large background uncertainties are 
acceptable. The significance of the above example would still correspond to
about 5 standard deviations if the background would be 
increased by a factor of 4. Once a mass peak
has been observed, one needs detailed signal Monte Carlos to
determine cross sections and perhaps 
other quantities like spin and parity.
Furthermore, signal and background Monte Carlos are usually required 
to eliminate obvious backgrounds with some kinematic selection criteria.
Searchers should also   
remember that the advantages of optimised efficiencies, obtained with
complicated selection methods, 
are easily destroyed by uncontrolled systematic errors.
Other disadvantages of too much optimisation are     
model dependent phase space restrictions and the introduction 
of possible statistical fluctuations which 
increase with the number of studied cuts and mass bins.

In addition, it is not always an advantage to reduce signal and backgrounds 
to relatively small numbers when the significance has to be calculated 
from Poisson statistics! For example a simple $\sqrt{B}$ estimate 
for 9 expected background events requires an observation of 
at least 24 events, e.g. an excess of 15 events above 9 background events,
to claim a 5 $\sigma$ excess above background. However, for small event 
numbers one finds that the $\sigma = \sqrt{B}$ approximation is not 
good enough.  
A 5 $\sigma$ excess requirement is equivalent to 
a background fluctuation probability of  
less than $6\times 10^{-7}$. One finds, 
using Poisson statistics, that the required 5 $\sigma$ excess 
corresponds to an observation  
of more than 27 events! Despite this reduced significance 
(roughly $1\sigma$), systematic errors
start to become important. For small background numbers 
the sideband method is limited by statistics and direct and clean 
background estimates from data might increase/decrease backgrounds 
and might be larger than Monte Carlo background  
estimates. The method to determine backgrounds,  
either from data or from Monte Carlo might 
thus hide or enhance a real signal and 
artificial good or bad limits can be 
obtained\footnote{It is surprising that most searches 
appear to be ``lucky''; e.g. the number of observed 
events is smaller than the number of expected background events.}. 

The possibility to observe fluctuations 
due to many mass bins appears nicely 
in a CMS simulation~\cite{ecaltdr} of the  
two photons mass distribution for a SM Higgs with a 
mass of 130 GeV and backgrounds.
\begin{figure}[htb]
\begin{center}
\mbox{
\epsfig{file=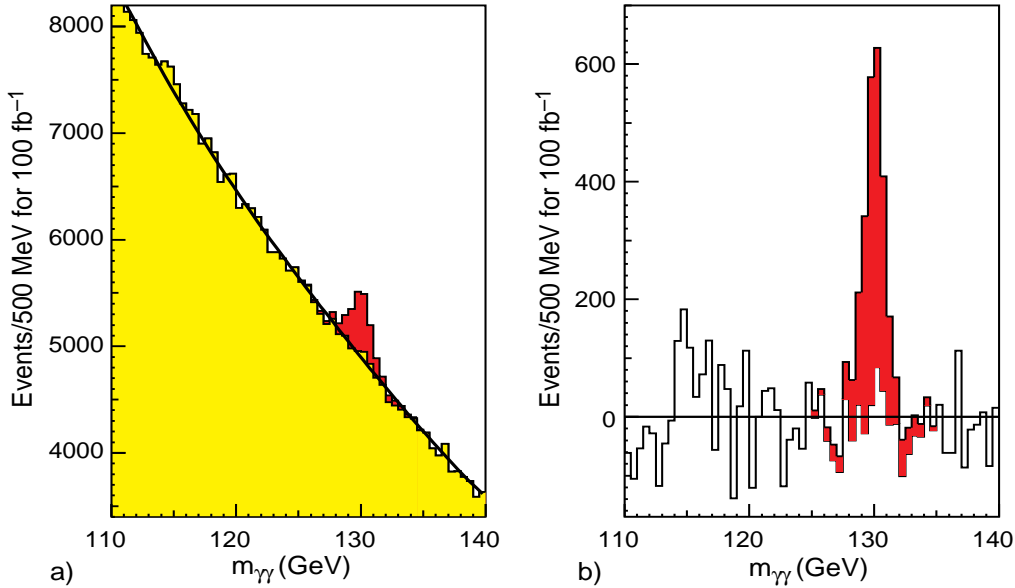,
height=8 cm,width=14cm}
}
\end{center}
\caption[fig2]
{CMS simulation~\cite{ecaltdr} for the decay  
$H \rightarrow \gamma \gamma$ before and after background 
subtraction.}  
\end{figure}
Figure 2 shows 
a clear narrow peak at 130 GeV. The observed signal,  
assuming a simple straight line to estimate the background, 
corresponds to about 10 standard deviations.
A more careful analysis of the mass distribution shows 
why at least five standard deviations are required to 
establish the existence of a new particle.  
Ignoring for example the simulated Higgs signal at 130 GeV,
one might try to look for an excess of events at an arbitrary 
mass. The largest excess of events appears at a mass of about 115 GeV. 
Taking the background from the sidebands one finds 
a statistical fluctuation with a significance of about 
three standard deviations. Thus, many possible mass bins combined with 
various event selection criteria are a remaining danger for mass peak hunters.

Despite the simplicity to discover new physics with mass peaks, 
most searches for new physics phenomena 
require an excess of events in special kinematic 
regions or tails of distributions. Some difficulties of such searches are 
indicated in Figure 3.
The figure shows a random simulation of missing transverse energy 
events from $pp \rightarrow Z X \rightarrow \nu \bar{\nu} X$  
and small statistics which is compared to  
a large statistic background simulation of the same process.   
Depending on the new physics signature, the small excess 
of tail events might coincide with a signal, expected for a certain
range of missing transverse momentum $p_{t}$. For a missing $p_{t}$ 
between 600-720 GeV one could quote an excess of almost 3 sigma, e.g.
6 events are seen while a background of only 2 events are predicted.
``Good arguments'' might increase 
the significance for new physics further.  
For example one might argue that the Monte Carlo overestimates the 
backgrounds, as the sideband region between 400--500 GeV 
shows about 50\% more events than found with the ``pseudo data''. 
Some additional unexpected features of the 6 events might
further be found to increase the significance further.
Of course, in case one wants to exclude new physics, one would 
argue that the number of 7.63 predicted events with missing $p_{t}$ 
above 500 GeV is in perfect agreement with the observed 8 events.   
The above example justifies the statement 
{\bf ``never search in tails''}. 
Unfortunately, most new physics scenarios, like SUPERSYMMETRY, would 
appear as rare events and in tails of distributions. 
\begin{figure}[htb]
\begin{center}
\mbox{
\epsfig{file=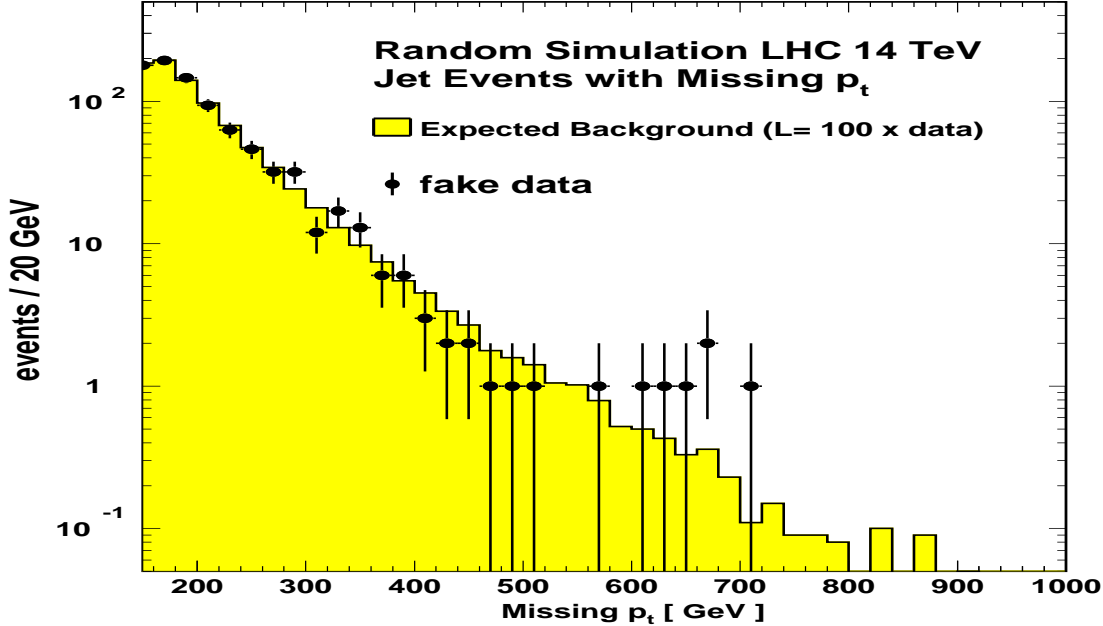,
 bbllx=100pt,bblly=150pt,bburx=500pt,bbury=680pt,
height=9 cm,width=11cm}
}
\end{center}
\caption[fig3]
{Simulation of a background fluctuation for events with 
missing transverse energy.}  
\end{figure}

Thus, ingenuity is required to separate new physics from tails of known 
processes. Such searches require not only to have enough statistical 
significance but {\bf a method to determine backgrounds}. 
The difficulty to 
establish a signal becomes clear from the following two examples.
{\bf Case a} is for a comfortable signal to background ratio of 1:1
while {\bf case b} is for a ratio of 1:10. The required minimal statistics 
is easy to estimate. A 5 sigma excess needs a statistics of
roughly 25 Signal events on top of a background 
of 25$\pm$5 for {\bf case a} while {\bf case b} 
needs about 250 signal events above a background of 2500$\pm$50.
The statistical excess however is not enough as 
the expected background has some systematic errors 
like uncertainties from the efficiency, the luminosity and the 
theoretical background model. Assuming that all these uncertainties 
are known with an accuracy of $\pm 5$\%, the 
significance of {\bf case a} is essentially unchanged while 
the significance of {\bf case b} is reduced to about 2 standard deviations.

It is worth noting that some studies claim to be {\bf ``conservative''} 
by multiplying backgrounds by arbitrary factors (method 1)  
or by using the error estimate from $\sqrt{S+B}$ (method 2).

Using {\bf case b} and method 1 one sees no 
dramatic change of the estimated sensitivity. One finds that only the 
minimal luminosity requirement has to be increased.
At the same time, the signal to background ratio became 1:20 and a $\pm$5\%
systematic error would result in almost meaningless results!

The estimated sensitivity, 
using method 2, does essentially not change for a bad signal to background 
ratio. However, method 2 reduces a clear signal, like 10 observed events with 
one expected background event, to a modest 3 standard deviation signal. 

We thus disagree with the 
claim that the above methods are conservative and reliable. 
In contrast, a correct approach 
to a possible significance figure should give the 
statistical sensitivity for new physics, estimated with 
$\sigma \approx \sqrt{B}$, and 
should describe how backgrounds can be estimated or at least  
how well they need to be known. Most sensitivity estimates 
do not provide answers to the latter requirements.
Attention should thus be paid to the estimated signal 
to background ratio, which allows to estimate the 
required systematic accuracies.

\subsection {Simulating the Future?}

A growing fraction of the preparation time for a modern collider 
experiment deals with the simulation of case studies and especially 
the search sensitivity for new physics. Such studies are 
not only required in order to motivate the required effort, time and money
but should also provide some guidance on ``what is possible'' and 
how a real detector should look like. 
Close to data taking,   
these studies can be considered as the last preparation step towards a
fast data analysis and the resulting discoveries.  

Furthermore, such early ``theoretical'' case studies, wanted or not, 
define very often {\bf new and original ideas and methods} which  
are rarely quoted when used later in real experiments. 

Among the many possible case studies, essentially all simulation studies 
concentrate on the SM Higgs search, ``question number 1'' and on 
SUSY particle searches. The investigated signatures provide thus not only 
``dream-land'' possibilities for a future collider but, as will 
become clear in the following,     
cover a wide range of detector requirements which help to shape the
final and real experiment. 

\subsubsection {The Higgs Search at LEP}

The possibility to search for the Higgs particle at LEP 
has been a central question for the LEP physics program.
With the now finished LEP I phase, it is interesting to compare 
the results of early simulation studies with the actual searches 
used at LEP.

The first studies, 
performed well before the discovery of the $W^\pm$ and $Z^0$,
are described in a DESY preprint from 1979~\cite{desy79}. The studies  
concentrated, among other possible channels on the 
signature $e^+e^- \rightarrow Z^{0(*)} H^0$.
Their conclusions on the neutral Higgs were: \newline

{\it ``The best production process seems to be 
$e^+e^- \rightarrow Z^{0(*)} H^0$ , which should give reasonable rates 
for $M_{H}$ up to $\sqrt{s}-M_Z$. Even if a clear signature 
from the $H^0$ is not available, the decays of the 
$Z^0 \rightarrow e^+e^-$ or $\mu^+\mu^-$ should provide a clear 
signature for this reaction. The mass and width of the Higgs could be 
measured quite precisely from the threshold behaviour.''}
\newline

The study concluded further that 
$Z^0$ decays at LEPI should allow to detect 
a Higgs with a mass of about 50 GeV using the 
$e^+e^- H^0$ or $\mu^+\mu^- H^0$ channel.

This early study was followed by more detailed ``LEP Yellow book''
studies in 1985/86~\cite{cern85} and in 1989~\cite{cern89} 
which essentially confirmed the 
1979 studies. The most promising signature was identified to be the
dilepton channel, $e^+e^- H^0$ or $\mu^+\mu^- H^0$, 
resulting in a mass sensitivity of about 
35 (55) GeV for $10^6$ ($10^7$) produced $Z^0$'s. 
The 6 times larger cross section of the 
neutrino channel, $\nu \nu H^0$, was also discussed.  
However, it was believed that this channel could 
at best confirm the results from 
the dilepton channel. Following the discussion 
in the 1989 yellow report it appears that 
the superiority of the neutrino channel,
at least for Higgs masses between 5-20 GeV, 
has first been realized in a study by Duchovni, Gross and 
Mikenberg~\cite{hnunu89}.

The actual performed searches at LEPI~\cite{hlep1} 
gave negative results. It was nevertheless somehow surprising that 
the most significant results where obtained from the
neutrino channel with individual experimental limits as 
high as 60-62 GeV.
In contrast, the ``golden'' dilepton channel gave roughly the expected 
sensitivity of up to about 50 GeV. The reasons for these essentially 
wrong sensitivity estimates are perhaps the unexpected performance of the LEP
experiments with respect to the complete angular coverage and the 
resulting missing energy detection and the absence of $t \bar{t}$ production
in $Z^0$ decays. Furthermore, one should keep in mind that the 
early studies were aiming for an unambiguous signal to measure 
the mass and width and not for the discovery signature. 
It was thus natural to concentrate on the simplest
Higgs signature provided by the dilepton channel.

\subsubsection{SM Higgs search at LEP II}
The first detailed studies for the Higgs search at LEPII
were performed for the 1986 LEPII workshop~\cite{lep286} 
in Aachen\footnote{These studies were much more detailed 
than the comparable 1985/86 and 1989 studies for the LEPI phase.}.
Besides the b-lifetime tagging, essentially 
all required techniques and channels were identified and studied in detail.
Without the tagging of b-jets the conclusion was that all channels
provide signals up to masses of about 80 GeV. The accuracy of the 
dilepton channel combined with high statistics was thought to give 
results up to masses of up to 90 GeV.

The 1996 LEP II studies~\cite{lep296} 
could benefit from the now well understood 
experiments, combined with excellent b-lifetime tagging results. 
Consequently, the obtained Higgs sensitivity results from the 1997 
data taking at  
$\sqrt{s} = 183 $ GeV agree well with the expectations, 
but do still not show any sign of excess as shown in Figure 4~\cite{vanc98h}.
\begin{figure}[htb]
 \begin{center}
\mbox{
\epsfig{file=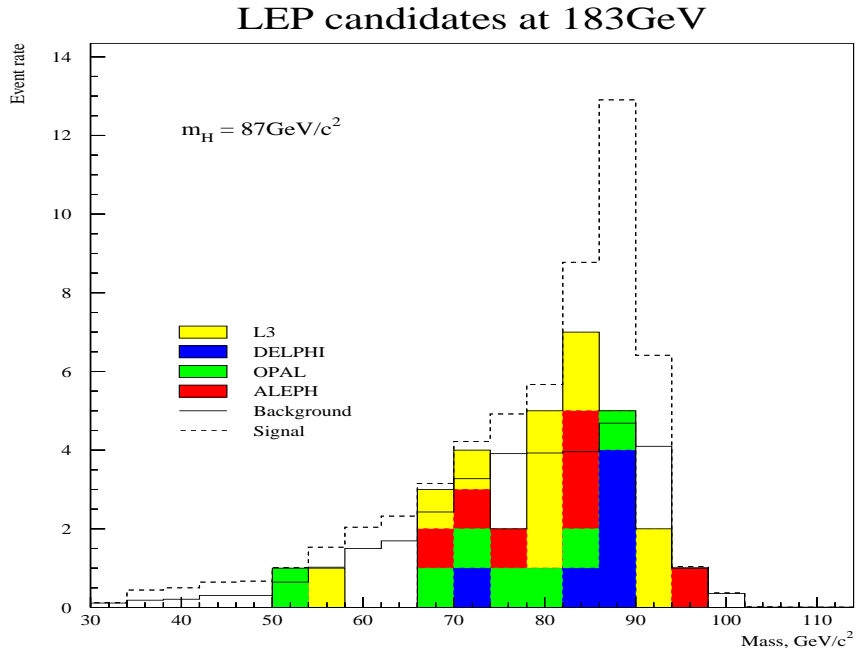,
height=9 cm,width=12cm}
}
 \end{center}
\caption[fig4]{Observed mass distribution of Higgs candidates 
from the four LEP experiments and $\sqrt{s} = 183 $ 
GeV (1997 data)~\cite{vanc98h}.}  
\end{figure}
 
In fact, one finds in general 
a reasonable agreement between the data and Monte Carlo expectations
without any sizeable excess at a fixed mass.
A more careful investigation shows however some remarkable 
effects. The observed mass distribution for Higgs candidates from the
four LEP experiments and an early fraction of the 1998 data is shown 
in Figure 5~\cite{vanc98h}. 
\begin{figure}[htb]
\mbox{
\epsfig{file=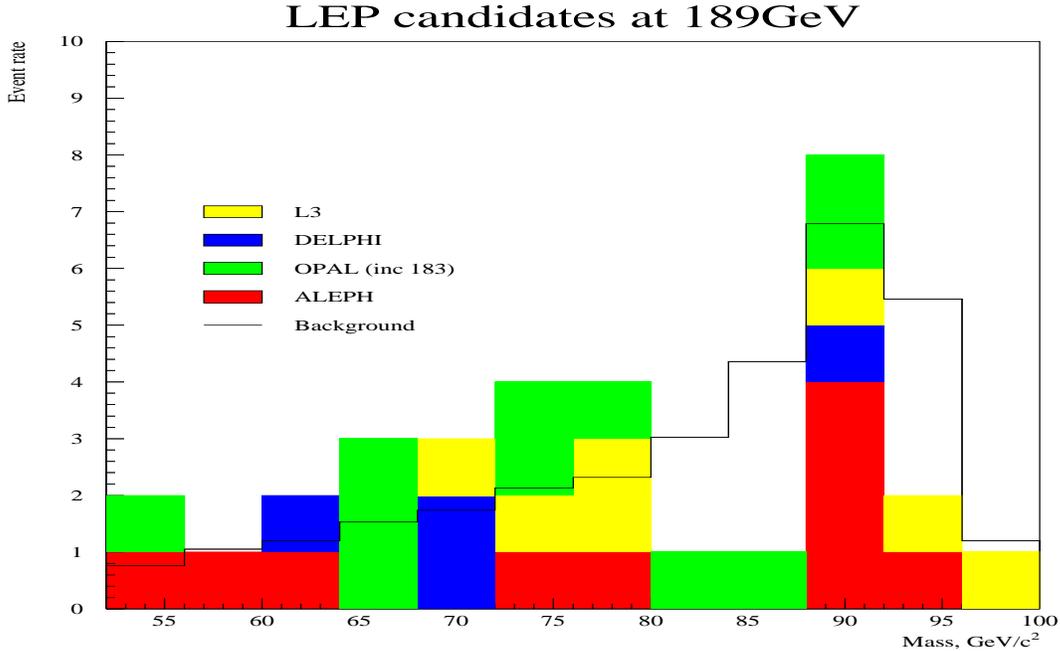,
height=9 cm,width=12cm}
}
\caption[fig5]{Observed mass distribution of Higgs candidates 
from the four LEP experiments and an early fraction of the $\sqrt{s} = 189 $ 
GeV data~\cite{vanc98h}.}  
\end{figure}

The total number of  
32 seen events agrees with an expectation of about 31 events.
This good agreement is somehow spoiled by the fact that 
about 11 candidates are expected for masses below 80 GeV while 
19 events are found from the data. This excess is compensated by a deficit 
for masses above 80 GeV with 20 expected and 13 seen events.
It appears to be tempting to use this mass distribution 
not only to exclude the SM Higgs production up 
to a mass of roughly 95 GeV, but also to put doubts on the 
Monte Carlo background expectations. The simple splitting might 
perhaps exclude the SM background estimation with
a confidence level of about 95\%.   
  
While preparing this article, not even rumours of excess Higgs candidates
are known. One might thus guess that this years 
combined LEP limit might be as high as 97 GeV, using a luminosity of about 
four times 180 pb$^{-1}$ at $\sqrt{s} = 189 $ GeV. 
The possible final Higgs sensitivity from the 1999/2000 
data taking at LEPII 
has been estimated to be 
about 105 GeV, using a luminosity of about 
$ 4 \times 200$ pb$^{-1}$ at $\sqrt{s} = 200$ GeV~\cite{hlep2f}.
 
\section {The future of the SM Higgs search}

As discussed above, one hopes that the 
4 LEP experiments should have a combined sensitivity to 
a SM Higgs with a mass close to 105 GeV.
Assuming that nothing will be found at LEPII, 
SM calculations can be used 
as a guidance for the Higgs search at future colliders.
We thus start this section with a discussion on  
calculations about the expected SM Higgs mass. We then describe 
Higgs search strategies at the LHC and discuss
the question of a potential Higgs window at the Tevatron.

\subsection {The Higgs Mass and Electroweak precision tests}

Starting with the assumption that the SM is a good approximation of 
nature, the Higgs boson and its mass remains to be discovered.
The mass of the Higgs can be constrained indirectly by 
the requirement that all measurements of electroweak observables, 
like the various asymmetry measurements at LEP and SLD, as well as 
the mass of the top quark and the $W^{\pm}$ are consistent. 
The accuracy of this procedure is however limited as there is only a soft 
logarithmic Higgs mass dependence. 
Nevertheless, a fit to all precision data 
constrains the SM Higgs mass to  
$ 76 \pm^{85}_{47} \pm 10 $ GeV and less than about 300 GeV.
One should however mention 
that the same procedure leads to a prediction of the top mass
of $ 161 \pm^{8}_{7}$ GeV, slightly below  
the measured value of $ 173.8 \pm 5$~\cite{mhiggssm1}. 
\begin{figure}[htb]
\begin{center}
\mbox{
\epsfig{file=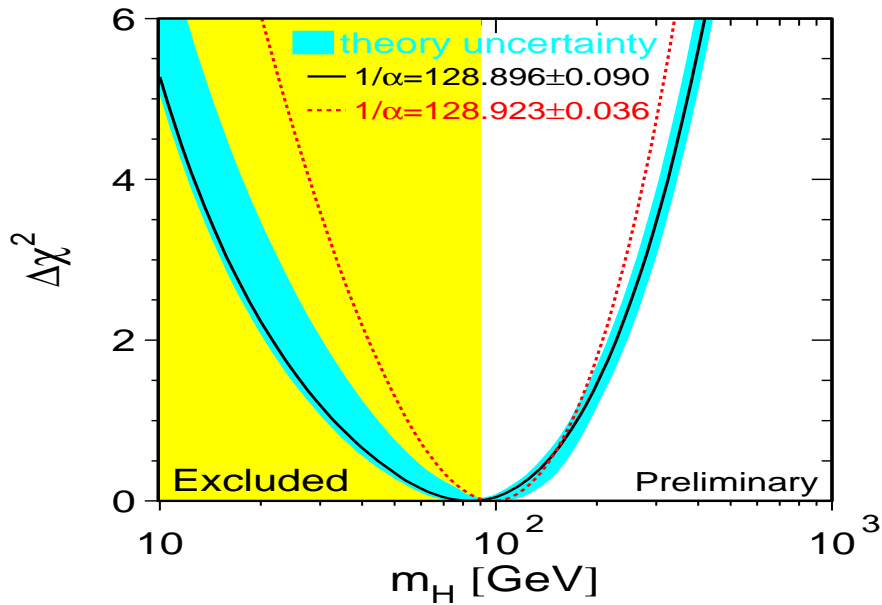,
height=8 cm,width=12cm}
}
\end{center}
\caption[fig6]{$\Delta \chi^{2}$ result of 
a SM fit to all electroweak observables assuming to have the 
Higgs mass as the only remaining free parameter~\cite{mhiggssm1}.}
\end{figure}

Furthermore, a recent analysis by J.H. Field~\cite{field}
shows that $\sin^{2} \Theta_{W}$ measured from lepton asymmetries 
differs by more than 3 sigma from the  
$\sin^{2} \Theta_{W}$ measurements with b--quarks.
It appears that the simplest 
explanation
are unknown systematic errors for asymmetry measurements with 
b--quarks. It could thus be justified to exclude 
the b--asymmetry measurements from the precision data. 
Such an approach~\cite{field} reduces 
the 95\% upper limit for the possible Higgs masses
to values lower than 200 GeV.
Alternatively, one could argue that the data 
indicate some new physics in the b--sector and do therefor not allow 
to draw strong conclusions for the SM Higgs sector.

Thus, ignoring the problems related to the b--sector, the 
1998 electroweak precision data result in upper limits for the 
SM Higgs mass somewhere between 200--300 GeV. 
This value, assuming that only the Higgs remains to be discovered, 
appears to be in nice agreement 
with consistency calculations of the SM~\cite{mhiggssm2},
which leads to a Higgs mass prediction 
of roughly $160 \pm 20$ GeV as shown in Figure 7.
\begin{figure}[htb]
\begin{center}
\rotatebox{90}{
\epsfig{figure=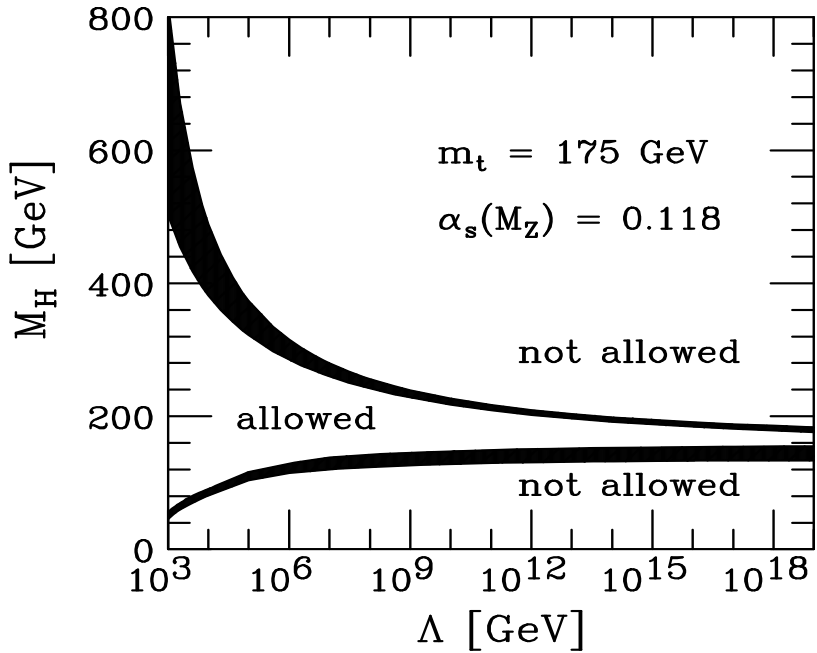,
 bbllx=100pt,bblly=150pt,bburx=500pt,bbury=680pt,
width=9.5cm,height=14.cm}}
\end{center}
\caption[fig7]
{The area between the two black curves shows the 
allowed Higgs mass range assuming the validity of the 
Standard Model up to a scale $\Lambda$~\cite{mhiggssm2}.}
\end{figure}

Such calculations lead to interesting implications. 
For example, a Higgs discovery with a mass of 100 GeV at LEPII or with a mass 
of 300 GeV at the LHC would immediately imply to have other new physics 
perhaps within the reach of the LHC experiments. 
Furthermore, once new physics is 
introduced, the assumptions used to constrain the Higgs mass from 
SM precision measurements are not valid. Thus even todays excluded 
Higgs, with a mass of 500 GeV might be in perfect agreement with 
todays precision data and a ``slightly'' 
enlarged model. Without new ideas in sight, the expected small improvements 
of the electroweak parameters will neither result in   
a precise Higgs mass prediction nor will allow to show unambiguously 
an inconsistency with the SM.

One can thus conclude that a Higgs or other new particles 
have to be discovered directly.   

\newpage

\subsection {SM Higgs Search at the LHC}
Figure 8 shows the results of  
Higgs cross section calculations~\cite{zoltan} at the LHC for 
various production processes as a function of the Higgs mass.
By far the largest contribution comes from the 
gluon--gluon fusion process~\cite{georgi}. 
\begin{figure}[htb]
\begin{center}
\includegraphics*[scale=0.6,angle=-90,bb=50 0 480 765 ]
{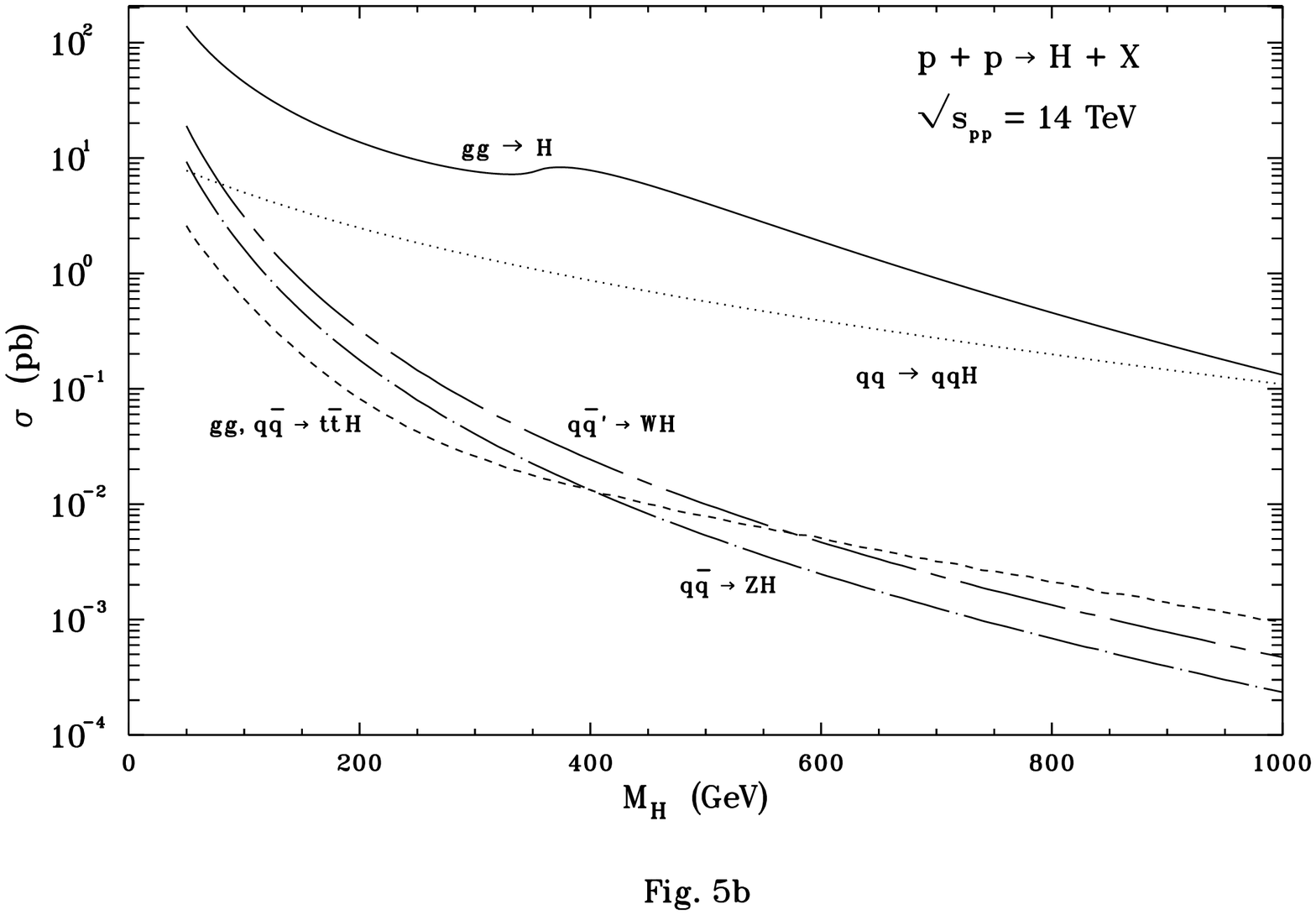}
\caption[fig8]
{Recent NLO cross section estimate for the SM 
Higgs~\cite{zoltan}.} 
\end{center}
\end{figure}

To study the different Higgs signatures, 
the total cross section has to be multiplied 
with the various branching ratios~\cite{spirabr}.
Figure 9 shows estimated 
$\sigma \times BR$ for promising Higgs search 
modes, $H \rightarrow \gamma \gamma$,
$H \rightarrow Z Z^{(*)} \rightarrow 4 \ell^{\pm}$,
and $H \rightarrow W W^{(*)} 
\rightarrow \ell^{+} \nu \ell^{-} \bar{\nu}$.
The first two signatures allow a precise and direct mass 
reconstruction, but require a high luminosity due to the
small detectable cross section. The third signature
was recently studied and found to be very sensitive
especially for the  
Higgs mass range between 155--180 GeV~\cite{hww1}.
It was found that the absence of a narrow mass peak 
can be compensated by the large event rate.

\begin{figure}[htb]
\begin{center}
\epsfig{file=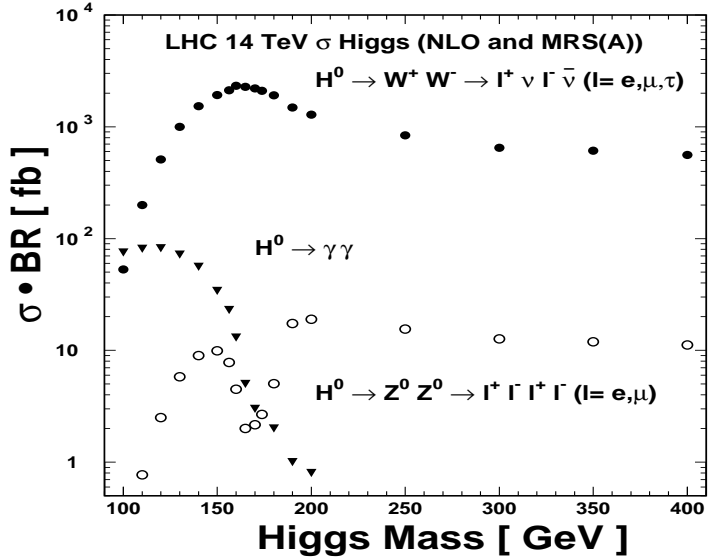,
width=10.cm,height=8.cm}
\end{center}
\caption[fig9]
{Expected $\sigma \times BR$ for different detectable 
SM Higgs decay modes.} 
\end{figure}

For Higgs masses above $\approx$ 500 GeV several 
additional and promising 
signatures involving hadronic $W$ and $Z$ decays 
as well as invisible $Z$ decays like
$H \rightarrow Z Z \rightarrow \ell^{+}\ell^{-} \nu \bar{\nu}$ 
have been discussed. The advantages of much larger 
branching ratios are compensated 
by serious backgrounds from events of the 
type $t\bar{t}$, $W X$ and $Z X$. 
These high mass Higgs signatures involve missing transverse energy 
and jet--jet masses and require thus hermetic detectors with
good jet energy reconstruction.

A few simulated promising Higgs signals for different 
masses and signatures are shown in Figures 2 and 10-14.
Recent estimates from ATLAS~\cite{atlastp} and 
CMS~\cite{cmstp} indicate that a luminosity of about
10 fb$^{-1}$, about 1 year of running with the 
initial ``low" luminosity of $10^{33}~sec^{-1}~cm^{-2}$
is required to discover a SM Higgs with masses
between 200--500 GeV with at least 5 standard deviation
in the four charged lepton channel. 

\begin{figure}[htb]
\begin{center}
\epsfig{file=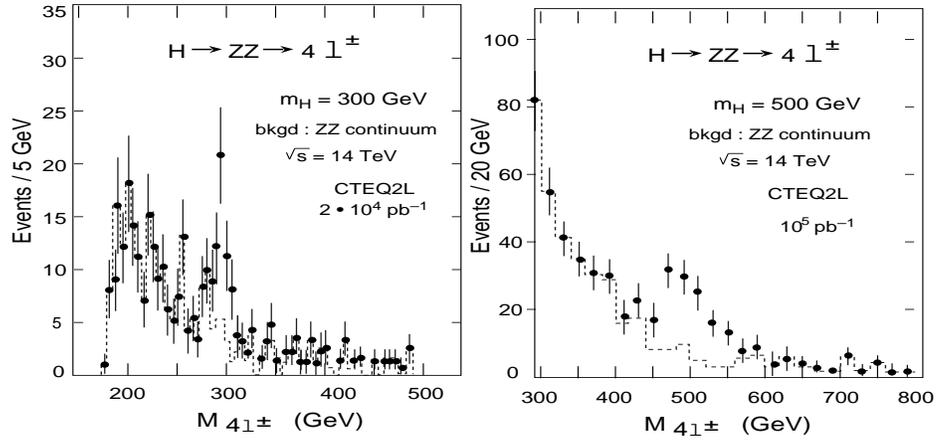,  
bbllx=150pt,bblly=380pt,bburx=450pt,bbury=550pt,
height=5. cm,width=12.cm}
\end{center}
\caption[fig10]
{CMS simulation results for $H \rightarrow Z Z \rightarrow
\ell^{+}\ell^{-} \ell^{+}\ell^{-}$ and $M_{H} = 300$ GeV
and $M_{H} = 500$ GeV.
}
\end{figure}
For example, the ATLAS study~\cite{atlasnote1} shows that
a Higgs ($M_{H}=300$ GeV and 
$H \rightarrow ZZ \rightarrow 4 \ell^{\pm}$) 
should be seen with 
35 signal events above a continuum 
background of $\approx 13 \pm 4$ events and 10 fb$^{-1}$.
The ATLAS study indicates also that the signal to background 
rate can be improved dramatically by requiring
that one reconstructed $Z$ has a $p_{t}$ of more than $M_{H} / 2$.
The corresponding event numbers for a 300 GeV Higgs  
would be 13 above a very small background of 0.6 events.

The most promising signature for a SM Higgs 
with masses between the expected LEPII limit, $\approx$ 100 GeV,
and 130 GeV is the decay $H \rightarrow \gamma \gamma$ 
with a branching ratio of only $\approx 2 \times 10^{-3}$.
As such a signal has to be found above a 
huge background of continuum $\gamma \gamma$ events, 
as shown in Figures 2 and 11,  
an excellent $\pi^{0}$ rejection and $\gamma \gamma$ mass resolution of 
$\approx$ 1\%, e.g. 1 GeV for $M_{H} \approx 100$ GeV, 
is required for a 5 standard deviation signal.
\begin{figure}[htb]
\begin{center}
\epsfig{file=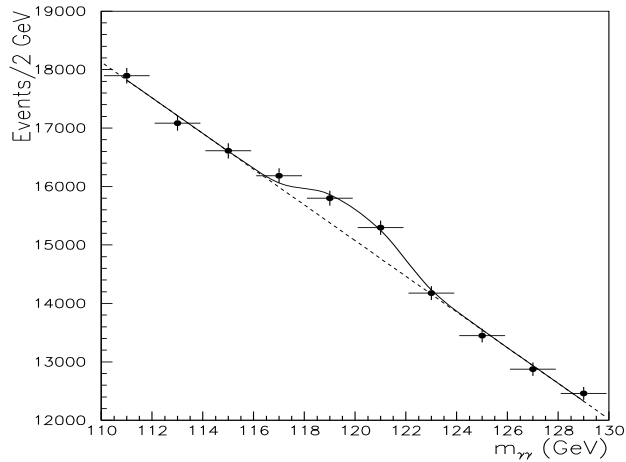,  
height=7. cm,width=9.cm}
\end{center}
\caption[fig11]
{ATLAS simulation for $H \rightarrow \gamma \gamma$
with $M_{H} = 120$ GeV and backgrounds~\cite{atlastp}.
}
\end{figure}
\clearpage
For Higgs masses between 130 GeV and 200 GeV
the $4 \ell^{\pm}$ signature suffers from very low 
branching ratios and a 5 standard deviation signal 
requires high luminosities of at least 30--100 fb$^{-1}$.
\begin{figure}[htb]
\begin{center}
\epsfig{file=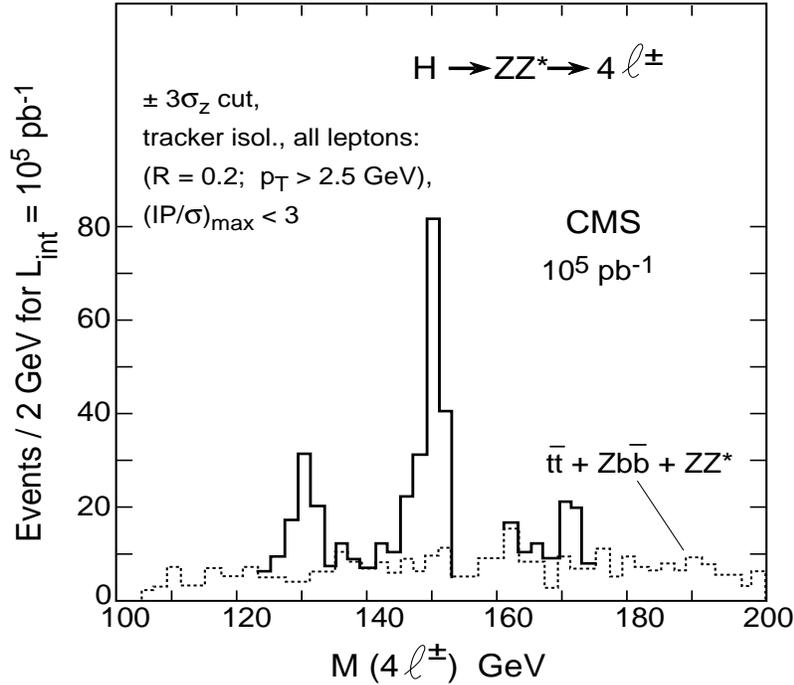,  
bbllx=70pt,bblly=240pt,bburx=500pt,bbury=630pt,
height=10.cm,width=12.cm}
\end{center}
\caption[fig12]
{CMS simulation for $H \rightarrow Z Z^{*} \rightarrow
\ell^{+}\ell^{-} \ell^{+}\ell^{-}$ and $M_{H}=130, 
150$ and $170$ GeV.
}
\end{figure}
 
A recent study has demonstrated that this Higgs mass region 
can be covered by the channel 
$H \rightarrow W W(^{*}) \rightarrow
\ell^{+} \nu \ell^{-} \bar{\nu}$~\cite{hww1},~\cite{hww2}. 
The performed analysis, described in section 3.3,
shows that this channel should allow to discover 
a SM Higgs with 5 standard deviation for a Higgs mass 
between 140--200 GeV and luminosities below 5 fb$^{-1}$.  

For Higgs masses above 500 GeV the natural width is already quite
large and the mass resolution becomes less important. 
Therefore, additional signatures with neutrinos and 
jets from hadronic $W$ and $Z$ 
indicate very promising and competitive Higgs discovery
channels as indicated in Figures 13 and 14. 

\begin{figure}[htb]
\begin{center}
\includegraphics*[scale=0.6,bb=0 180 600 650 ]
{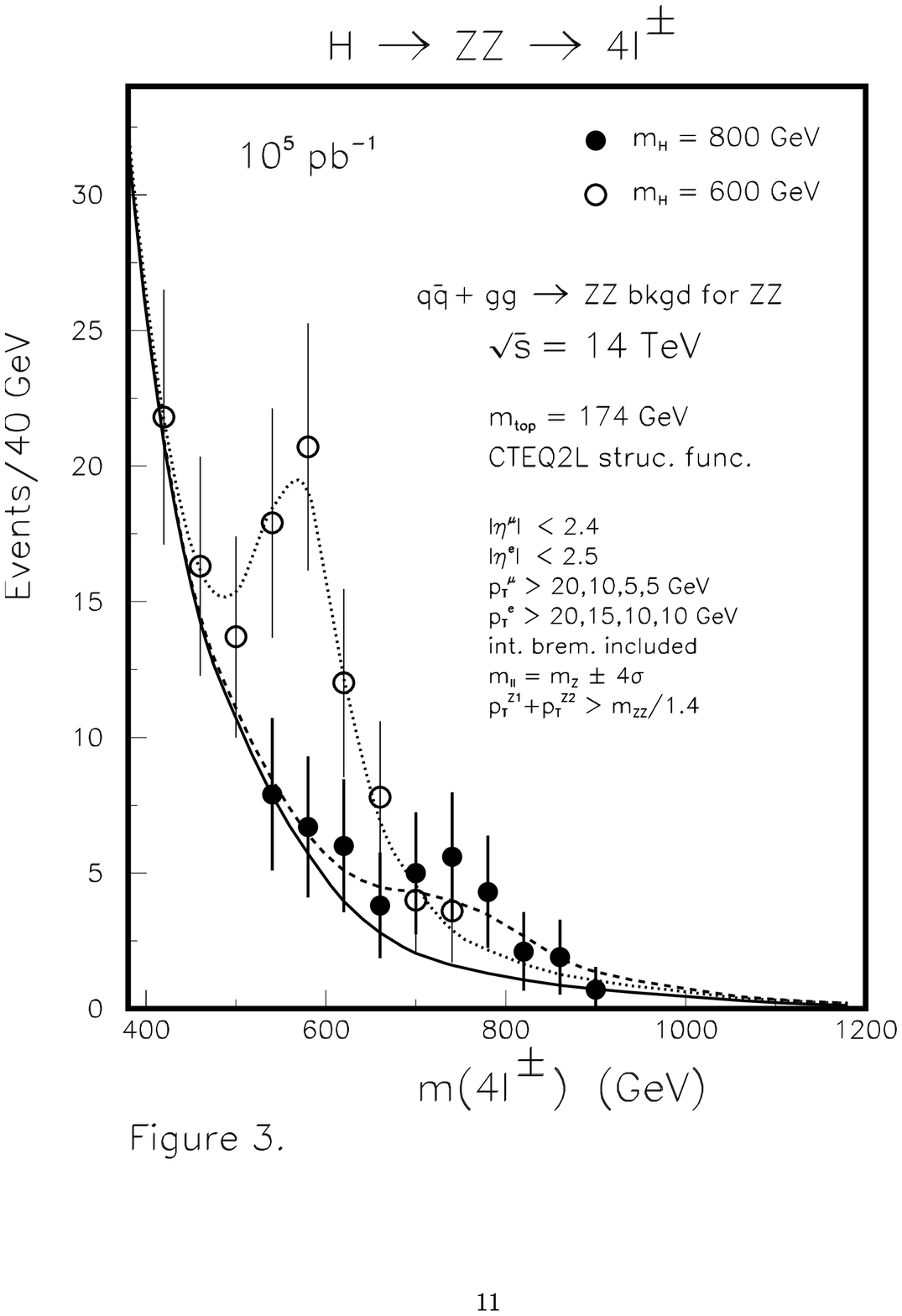}
\end{center}

\caption[fig13]
{CMS simulation results for $H \rightarrow Z Z \rightarrow
\ell^{+}\ell^{-} \ell^{+}\ell^{-}$.}
\end{figure}

\begin{figure}[htb]
\begin{center}
\epsfig{file=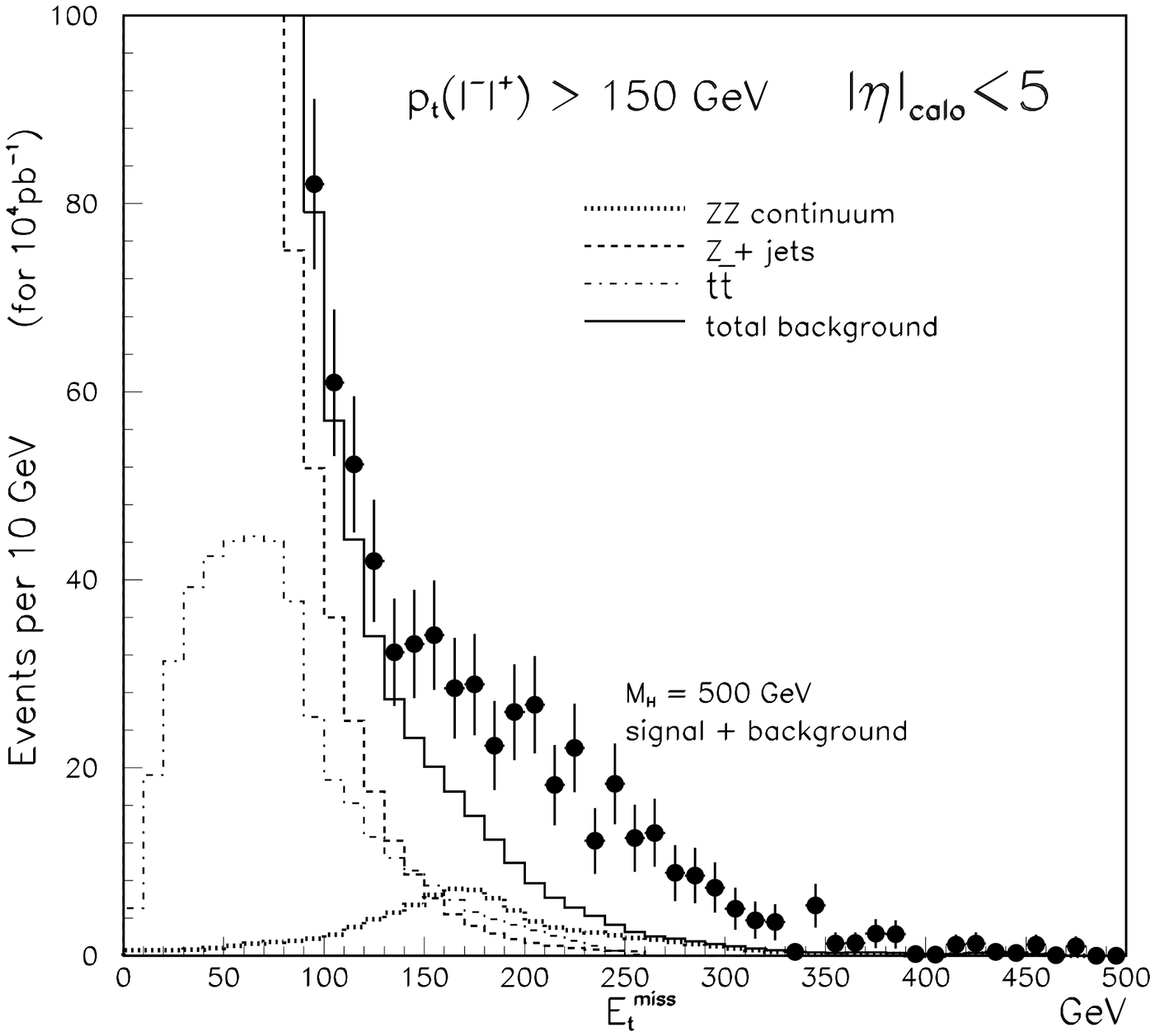,
height=8. cm,width=5.5cm}
\epsfig{file=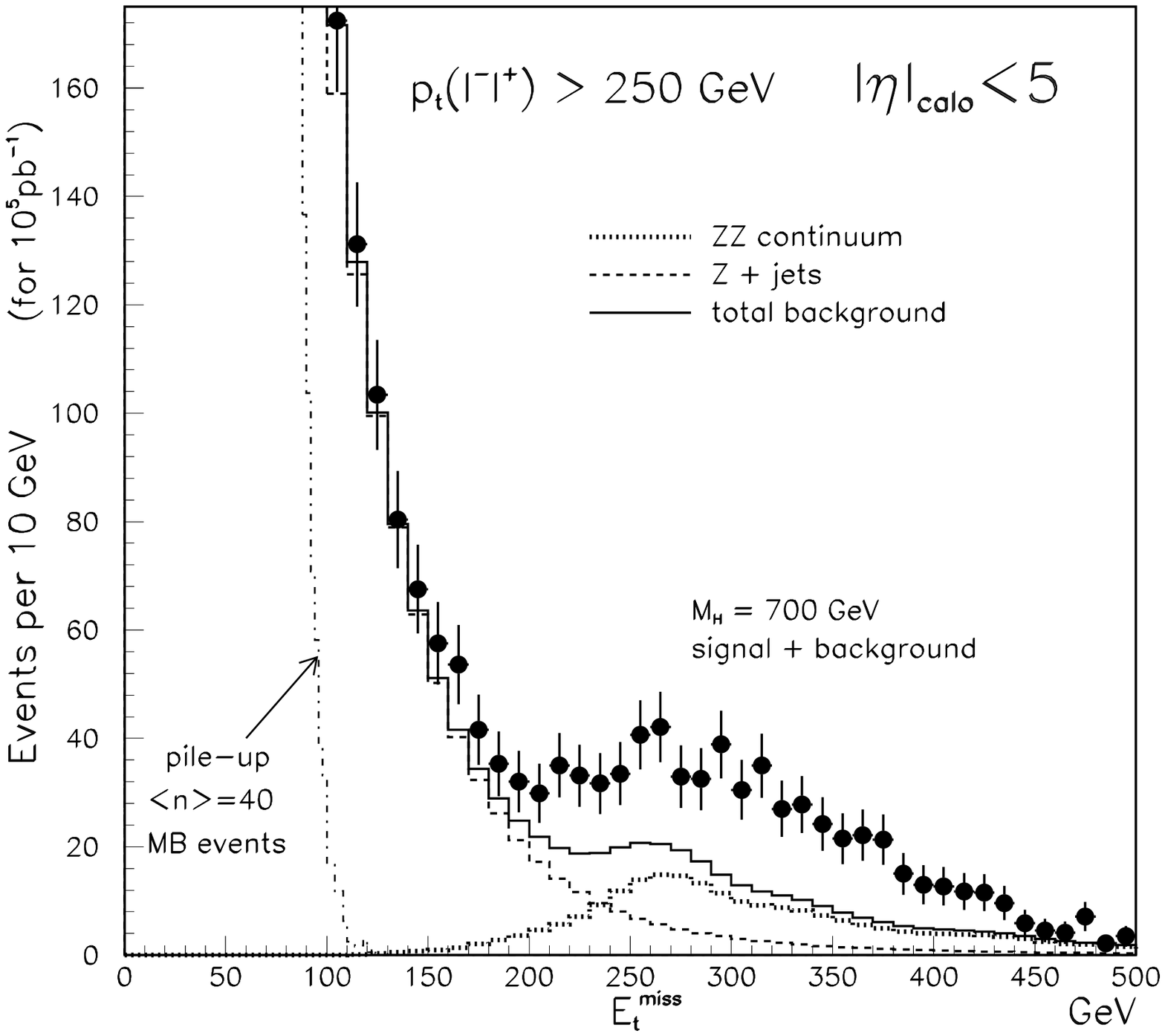,
height=8. cm,width=5.5cm}
\end{center}
\caption[fig14]
{ATLAS simulation results for $H \rightarrow Z Z \rightarrow
\ell^{+}\ell^{-} \nu \bar{\nu}$.}
\end{figure}

\clearpage
\newpage
\subsection{SM Higgs search with $H \rightarrow W^{+}W^{-}$}

A recent simulation analysis has demonstrated that the 
previously ignored signature 
$H \rightarrow W^{+}W^{-} \rightarrow 
\ell^{+} \nu \ell^{-} \bar{\nu}$ allows to close the 
``last" LHC Higgs detection gap. Furthermore, it has been 
shown that this signature should provide even the first sensitive 
Higgs results at the LHC.

The analysis exploits mainly two differences between a SM signal
and the non resonant background from $pp \rightarrow W^{+}W^{-} X$.
The two most important criteria from the analysis~\cite{hww1}
were:
\begin{enumerate}
\item
As shown in the left part of Figure 15, the 
signal events from gluon--gluon scattering are  
more central than the $W^{+}W^{-}$ 
background from $q\bar{q}$ scattering.
This difference is exploited by the requirement that 
the polar angle $\theta$ 
of the reconstructed dilepton momentum vector,
with respect to the beam direction, satisfies $|\cos \theta| < 0.8$.
As a result, both leptons are found essentially within the 
barrel region of the experiments with $|\eta| < 1.5$.
\item
The $W^{+}W^{-}$ spin correlations
and the V--A structure of the $W$ decays result in a 
distinctive signature for $W^{+}W^{-}$ pairs produced in
Higgs decays. As shown in Figure 15 (right side), for  
a Higgs mass close to $2 \times M_{W}$ the 
$W^{\pm}$ boost is small and the opening 
angle between the two charged leptons in the plane 
transverse to the beam direction is very small.
\end{enumerate}
\begin{figure}[htb]
\begin{center}\mbox{
\epsfig{file=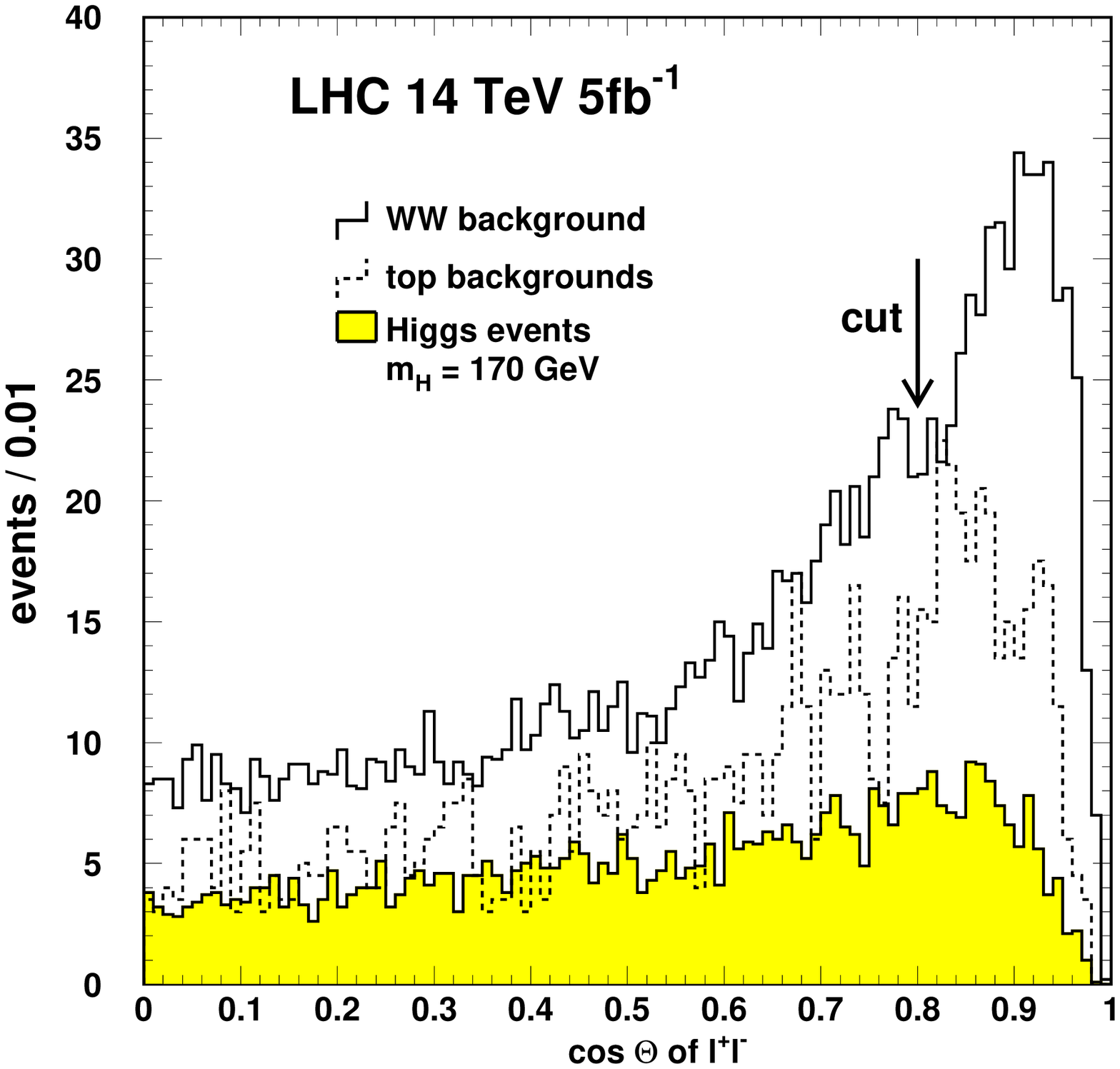,
height=7 cm,width=7cm}
\epsfig{file=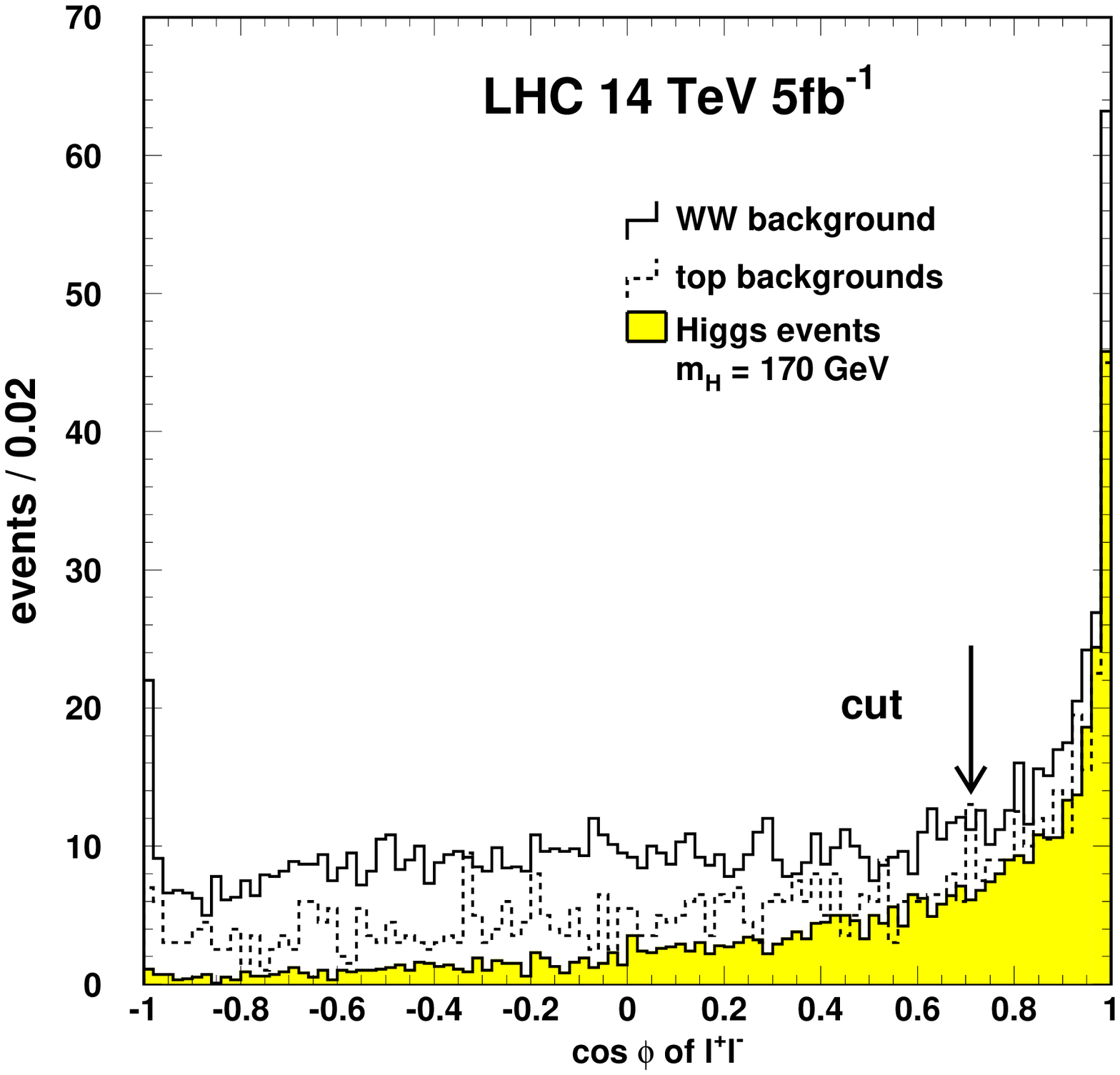,
height=7 cm,width=7cm}}
\end{center}
\caption[fig15]{Signal and background distributions for 
the $| \cos \theta |$ 
distribution of the dilepton  
system with respect to the beam direction (left)
and for $ \cos \phi$, where 
$\phi$ is the angle between the two leptons 
in the plane transverse
to the beam, after central dilepton events 
are selected (right).}
\end{figure}

The other proposed criteria enhance the signal to background ratio  
by using indirectly the 
slightly different lepton momentum spectra.
Following the proposed strategy~\cite{hww2}  
statistical significant Higgs signals can be obtained with 
a good signal to background ratio 
and a mass range between roughly 130--200 GeV as shown in Figure 
16. The studied distributions indicate also that backgrounds 
can be determined, with good accuracy, directly from the data. 
\begin{figure}[htb]
\begin{center}
\mbox{
\epsfig{file=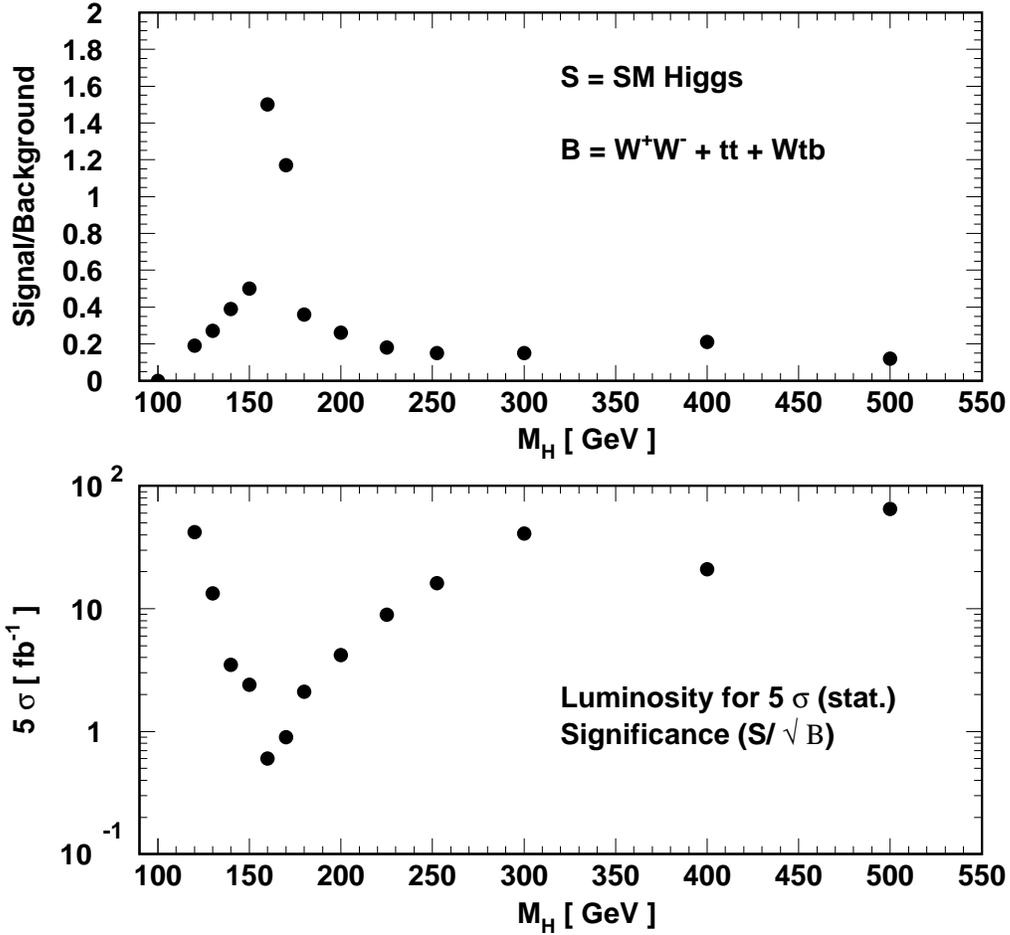,
height=14 cm,width=14cm}}
\end{center}
\caption[fig16]{SM Higgs signal over background
ratio (a) and (b) the required luminosity to obtain
a 5 standard deviation statistical significance signal
with $pp \rightarrow H \rightarrow W^{+}W^{-} \rightarrow
\ell^{+} \nu \ell^{-} \bar{\nu}$ for $M_{H}$ between 120 GeV
and 500 GeV.
}
\end{figure}

The resulting lepton $p_{t}$ spectra are shown in Figure 17 for 
a Higgs mass of 170 GeV. One finds that the lepton $p_{t}$ spectra are 
very sensitive to the Higgs mass as indicated 
in Figure 18.
\begin{figure}[htb]
\begin{center}
\epsfig{file=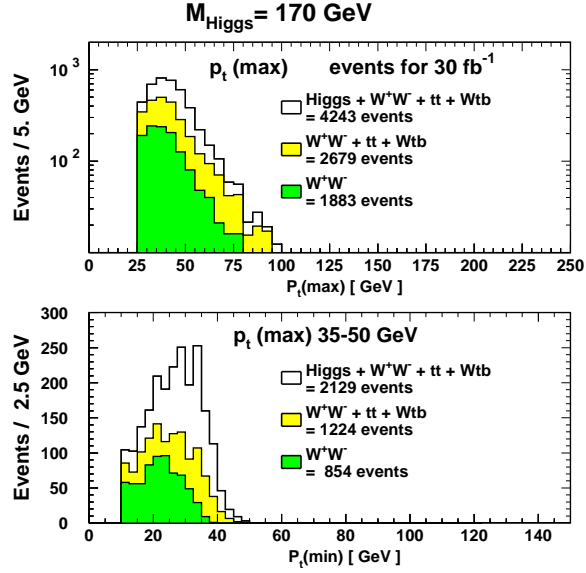,
width=8.cm,height=8.cm}
\end{center}
\caption[fig17]
{Expected lepton $p_{t}$ spectra for 
$H\rightarrow W^{+}W^{-} \rightarrow 
\ell^{+} \nu \ell^{-} \bar{\nu}$ and a mass of 170 GeV. The signal is 
superimposed to various SM backgrounds.}
\end{figure}
\begin{figure}[htb]
\begin{center}\mbox{
\epsfig{file=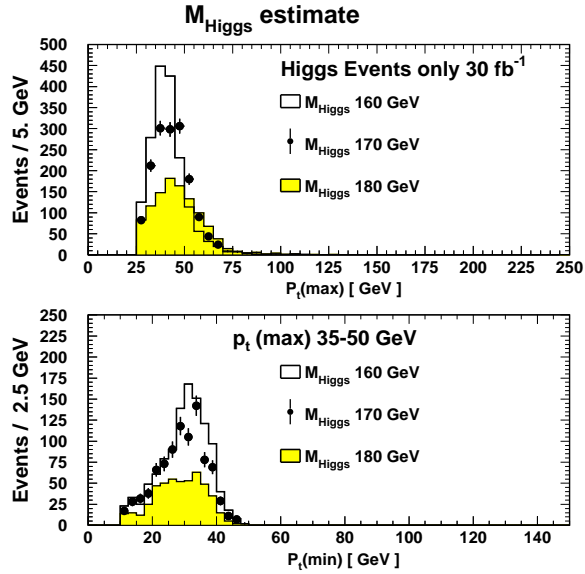,
width=8.cm,height=8.cm}}
\end{center}
\caption[fig18]
{Expected lepton $p_{t}$ spectra for 
$H\rightarrow W^{+}W^{-} \rightarrow \ell^{+} \nu \ell^{-} \bar{\nu}$ 
and three different Higgs masses.}
\end{figure}
\clearpage

\subsubsection{New ideas for LHC Higgs searches?}

The search for a Higgs with a mass below $\approx$130 GeV 
relies currently only on the ability to measure the rare decay
$H \rightarrow \gamma \gamma$ with a branching ratio of about
$1-2 \times  10^{-3}$ above a $\approx 10-20$ times larger $\gamma \gamma$
continuum background. This channel requires background free 
photon detection with high efficiency and excellent 
mass resolution of $\approx$ 1 GeV. 
It is thus interesting to combine this channel with 
possible alternatives. Recent theoretical studies, presented at 
the 1998 CERN theory workshop~\cite{cernth98}, discuss the possibility 
to improve the signal to background ratio drastically using jet tagging. 

For example, D. Zeppenfeld~\cite{zepp98} discussed a parton level study 
of the vector boson fusion Higgs production $q\bar{q} \rightarrow q\bar{q} H$
using double jet tagging and the decay $H\rightarrow \gamma \gamma$.
Additional background suppression is obtained 
from the different rapidity distributions of the underlying event. 
Their study indicates that a signal to background ratio of about 1:1 
is obtainable, paying however the price of very low signal rates 
of about 11 events for a luminosity of 10 fb$^{-1}$.
Some experimental studies within ATLAS and CMS have tried 
a similar approach to exploit the forward jet tagging. Unfortunately 
the obtained backgrounds from the ATLAS study~\cite{atlasggtag} 
are roughly a factor of four larger than the ones obtained 
with similar criteria performed by the CMS group~\cite{cmsggtag}. 

Another approach tries to exploit the $H\rightarrow \gamma \gamma$
signature using the different $p_{t}$ spectra
for a Higgs signal and backgrounds~\cite{hggpt}. 
Furthermore, very different 
angular distributions between jets and photons 
from signal and background where shown. This study indicates 
an interesting potential for a considerable 
improved signal to background ratio to at least 1:3 
for a $\pm$ 1 GeV signal window. This ratio 
should be compared to the inclusive study which results 
in a ratio of about 1:20 for the same mass window. The price to 
be paid seems to be a factor of 10 smaller efficiency.
Consequently, taking just a statistical error estimation, the numbers 
do not really imply an improvement. However it appears 
possible that the suggested kinematic differences between signal and 
backgrounds allow rather nice systematic studies of a potential
signal and should encourage further investigations.

Other ideas suggest to exploit the $H\rightarrow \gamma \gamma$
channel in the associated Higgs production channels 
$WH$ and $t \bar{t} H$~\cite{aachen90h}.
The performed studies indicate signal to background ratios of roughly 
6:1. The signal rate is however reduced 
to cross sections of  
only 0.1-0.2 fb, resulting in 1-2 accepted events for 10 fb$^{-1}$. 
Assuming that a high photon detection efficiency combined with low 
misidentification can be obtained, this 
channel might provide important additional significance. 

We conclude, that a possible discovery signal from 
the $H\rightarrow \gamma \gamma$ signature 
relies not only on an excellent electro-magnetic 
calorimeter, but perhaps also on unknown improved analysis strategies.
 
Having discussed the rare decay $H\rightarrow \gamma \gamma$
one is tempted to ask about the possibility to search for the dominant decays  
$H\rightarrow b \bar{b}$ and $H\rightarrow \tau \tau$.
Early estimates have studied the potential of the above decay modes 
using the inclusive production of $gg \rightarrow H$ and concluded 
that even the most optimistic signals would be 
much to small~\cite{aachen90h}.
Nevertheless, the potential of the associated Higgs production 
$WH$ and $t \bar{t} H$ and $H\rightarrow b \bar{b}$ has been studied in 
quite some detail~\cite{whbb}. 
Assuming an almost perfect b-jet identification
with efficiencies between 30-50\% and background rejection factors of about
100 only small signal to background ratios of about 1/50 have been 
obtained. The claimed statistical significance for 30 fb$^{-1}$
and a 120 GeV mass Higgs reaches at most 2-3 standard deviations.
A judgement of the proposed signal relies on 
the possible systematic uncertainties which 
unfortunately, are not discussed. 
 
The search for the  
associated Higgs production $WH$ and $t \bar{t} H$ and the decay  
$H\rightarrow \tau \tau$ suffers from the difficulty to reconstruct 
a mass peak. Nevertheless, a detailed study 
might still show that some signal indications can be obtained. 
Thus, after Higgs indications have been seen with other channels,
the $H\rightarrow \tau \tau$ might give additional information.

Recently, the topology of events $gg \rightarrow H\rightarrow b \bar{b}$ 
has been compared to the one from $gg (q\bar{q}) \rightarrow b\bar{b}$ 
continuum production. The study 
shows that the hadrons produced between the observable jets
and between the jets and the beam direction 
should be quite different for signal 
and background~\cite{hbbflow}.
Keeping the expected excellent b-tagging capabilities of ATLAS and CMS 
and the large Higgs rate, $\sigma \approx  50$ pb in mind,   
one might eventually reconsider the 
inclusive $H\rightarrow b \bar{b}$ signature.
A possible strategy might combine the 
different hadron production between the jets with 
the ideas discussed for the selection of 
Higgs events with large $p_{t}$ as presented 
above for the $H\rightarrow \gamma \gamma$ channel~\cite{hggpt}.

\subsection {A Higgs window at the Tevatron Run III?}

While patient physicist are waiting and preparing 
for a Higgs discovery at the LHC, others are trying to find the Higgs 
at LEPII or to investigate Higgs possibilities with Run III at the 
Tevatron collider. 
The upgraded Tevatron machine and 
experiments expect to analyse proton-antiproton collisions 
at a center of mass energy of 2 TeV and a yearly luminosity 
of about 1 fb$^{-1}$/year, Run II, starting in the year 2000.
Machine physicists are looking into possibilities to 
increase this luminosity further.  It seems possible to 
increase the luminosity by another factor of 10, the so called 
RUN III or TeV33 phase which should 
lead to roughly 10 fb$^{-1}$/year. Such a luminosity 
matches the requirements from theoretical parton level studies
which show some SM Higgs sensitivity in the channel 
$WH \rightarrow \ell \nu b\bar{b}$~\cite{tevhiggs1}. 
As the search for the Higgs appears 
to be the main issue of future collider physics,
it is important to study this TeV33 
possibility in some detail. In the following we will discuss the 
results of the TeV2000 study group~\cite{tev2000h} and compare them with todays
experimental facts. 

The basics for the discussion about the 
Higgs sensitivity at the Tevatron Run III rely 
essentially on the demonstrated b--jet tagging capabilities of 
the modern silicon micro vertex detectors with single 
hit resolutions of 10-20 $\mu$ with high efficiencies
and the experimental results on b physics from the CDF experiment. 
A further very encouraging CDF result~\cite{cdfzbb} shows 
a $\approx$ 3 sigma signal for the decay $Z^{0} \rightarrow b \bar{b}$.
Despite the low signal efficiency, roughly 100 events are extracted from 
a total estimated production rate of about $10^{5}$,   
the obtained signal demonstrates the possibility to select 
an object decaying to $b \bar{b}$ jets.
  
\subsubsection{The Higgs sensitivity claim}

A detailed experimental analysis~\cite{tev2000h}, following the 
original parton level study~\cite{tevhiggs1}
has been performed for the TeV2000 study. 
The investigated signature consists of events with 
one ``isolated'' electron or 
muon and a two-jet system with a mass window of roughly $\pm$ 20 GeV
around the assumed Higgs mass. Both jets are tagged as 
b--flavoured jets assuming a tagging efficiency of 50\% per jet 
while the efficiency for other jets is assumed to be roughly 0.5\%.
Combining all the criteria 
a signal efficiency of 10\%  is obtained for 
$WH \rightarrow \ell \nu b\bar{b}$ and Higgs masses between 100-120 GeV.
The expected SM Higgs signal for 10 fb$^{-1}$ drops from 52 events above a 
background of 249 events for a mass of 100 GeV to a signal of 27 events 
above a background of 130 events. The corresponding expected invariant 
mass distributions are shown in Figure 19.
\begin{figure}[htb]
\begin{center}
\epsfig{file=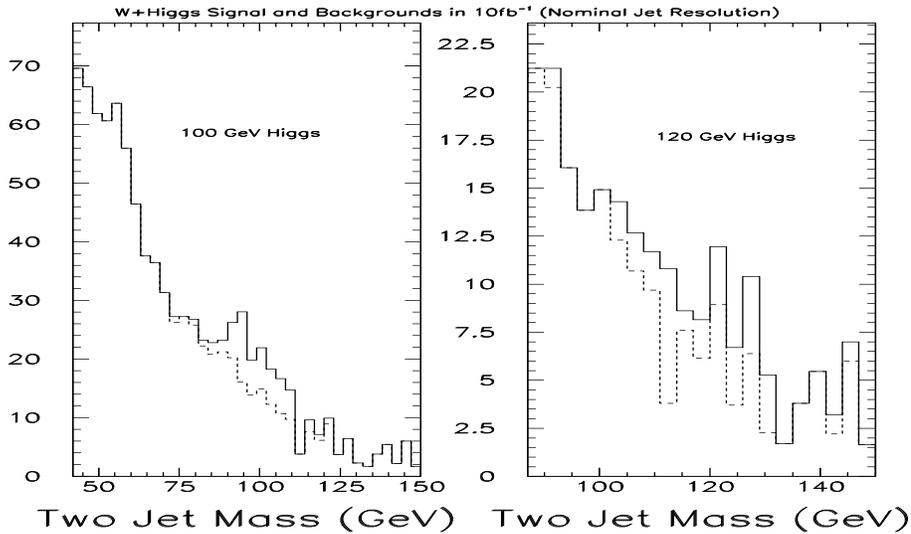,  
bbllx=150pt,bblly=140pt,bburx=500pt,bbury=580pt,
height=8.cm,width=12.cm}
\end{center}
\caption[fig19]
{Expected Higgs signal plus background mass distribution for 10$fb^{-1}$ 
at the upgraded Tevatron~\cite{tev2000h}.}
\end{figure}

As a result one finds a statistical 
significance (assuming $\pm \sqrt{N_{back}}$) of  
3.3 sigma for a mass of 100 GeV decreasing to 2.4 sigma for a 
mass of 120 GeV. No comments are made on possible systematic uncertainties. 
These somehow encouraging results might be slightly improved, if combined 
with other W decay modes and with $ZH \rightarrow \ell \ell b\bar{b}$.
Taking these estimates, one might indeed conclude that there is 
some Higgs sensitivity for the Run III program.   

To investigate the claim a little more we have compared the 
Run III assumptions with a real analysis from the existing CDF experiment
published in 1997~\cite{cdfhbb}.
This analysis was based on a total luminosity of $109 \pm 7$ pb$^{-1}$
collected during the RUN I. The used event selection criteria
are quite similar to the ones from the TeV2000 study group.
The main difference are the lepton acceptance window of $|\eta| < 1$
compared to an assumed window of $|\eta| < 2.5$. 
Figure 20 shows the observed jet--jet mass distribution 
for the selected candidates.    
\begin{figure}[htb]
\begin{center}
\epsfig{file=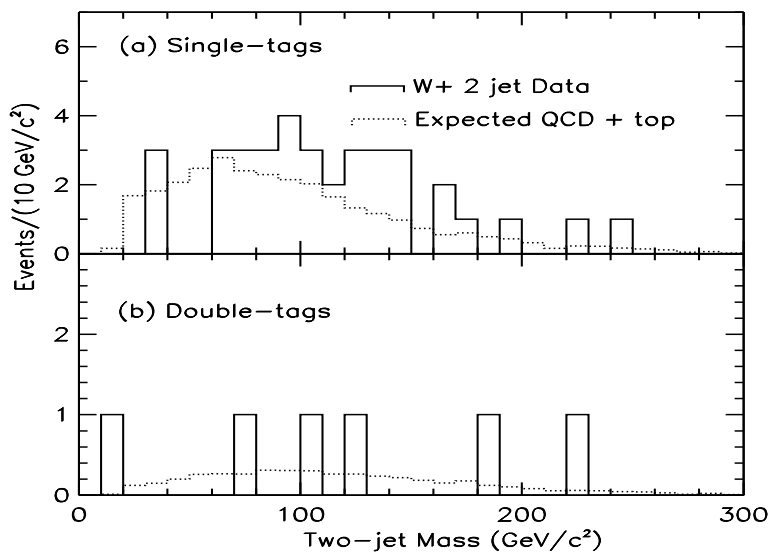,  
bbllx=100pt,bblly=160pt,bburx=500pt,bbury=600pt,
height=8.cm,width=10.cm}
\end{center}
\caption[fig20]
{Observed and expected 
two-jet mass distribution from CDF in W$\rightarrow \ell \nu$ 
+ 2 jet events from single (a) and double (b) tagged b-jets 
events and 109 pb$^{-1}$~\cite{cdfhbb}.} 
\end{figure}

Using the achieved signal efficiency of 0.4$\pm$0.11\% 
a cross section limit of about 10 pb is obtained.  
This is about a factor of 100 larger than the expected
SM Higgs cross section, as shown in Figure 21.
\begin{figure}[htb]
\begin{center}
\includegraphics*[scale=0.55,bb=70 180 600 600 ]
{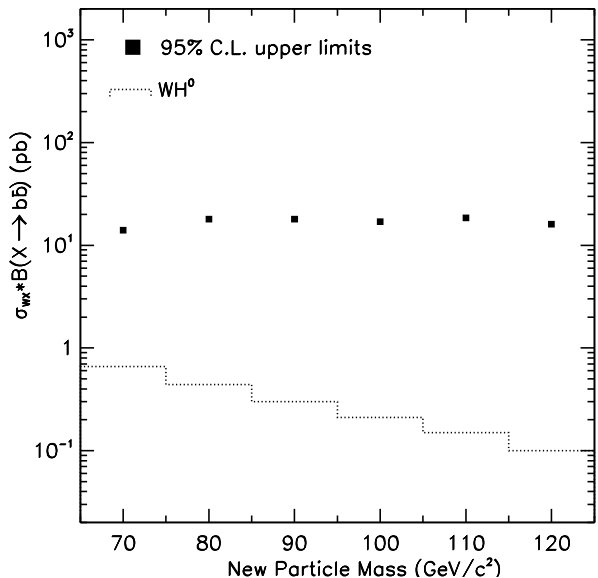}
\end{center}
\caption[fig21]
{The 1997 CDF experimental 95\% C. L. upper cross section limit, 
using a luminosity of 109 pb$^{-1}$,
for the associated production of a resonance X in 
WX events (black squares). The dotted line shows 
the theoretical SM WH cross section~\cite{cdfhbb}.} 
\end{figure}
 
Furthermore, the observed candidates agree with the 
background expectation of roughly 
one event for a $\pm$ 20 GeV mass window. One thus finds that  
todays background, obtained for an efficiency of 0.4\%, should 
give an expected background of about 100 events for a luminosity of 
10 fb$^{-1}$. Such a background rate agrees with the 
theoretical estimates, which assume however a factor
20 higher signal efficiency. The expected detector improvements for RunII,
as given in~\cite{cdfhbb},    
are estimated to increase a signal efficiency to about 1\%.

We are thus tempted to conclude that some factors, summarised in table 2, 
are still missing before one could claim that a SM Higgs window
exists at the Tevatron with 30 fb$^{-1}$. 
This is even more true for theoretical optimists which hope  
to have some sensitivity for supersymmetric Higgs particles with 
lower detectable cross sections~\cite{thhopes}. 
\begin{table}[htb]
\begin{center}
\begin{tabular}{|c|c|c|c|}
\hline
           & existing & expected & required  \\
\hline
efficiency & 0.4 $\pm$ 0.11 \% & 1\% & $\approx$ 10\% \\
\hline
$\Delta M$ (b-b jet) & 15 GeV & 15 GeV (?) & 11 GeV \\
\hline
background  & $\approx 100$  & ??? & $\approx$ 100 \\
events    & $\epsilon =0.4$ \% &   & $\epsilon =10$ \% \\
\hline
systematic error & $\pm$ 25\% & ??? & $ <<$ 10\% \\
\hline
\end{tabular}
\caption{Comparison of the existing, expected and required 
detector capabilities for the Tevatron Higgs potential.}
\label{tab:table2}
\end{center}
\end{table}
\clearpage

\section {Aspects of Searches for non Mainstream Exotica}

The combination of high energies and luminosity allows 
to dream even of a discovery of unfashionable and unpredicted 
new phenomena. The more conventional searches hope 
for fourth family quarks and leptons or additional bosons 
like a $Z^{\prime}$ or $W^{\pm \prime}$ and compositeness. 
Other searches look for 
new objects with fancy names like axigluons, mirror fermions, 
color octet technirho, massive stable sextets and octets etc.
The motivation to search for such objects is mostly a {\bf \it ``why not''}
or {\bf \it ``what is not forbidden might be allowed''}. 
\begin{figure}[htb]
\begin{center}
\epsfig{file=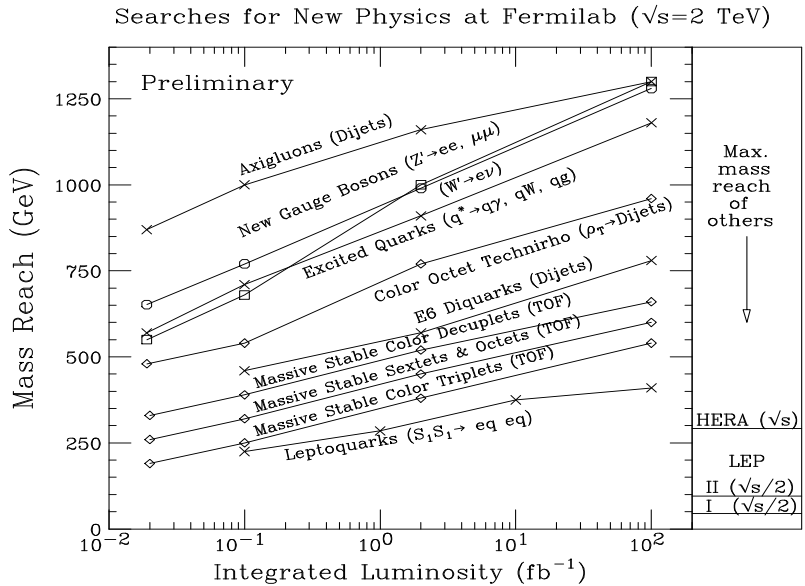,  
bbllx=70pt,bblly=240pt,bburx=500pt,bbury=630pt,
height=8.cm,width=10.cm}
\end{center}
\caption[fig22]
{Existing and expected CDF sensitivity for massive 
exotica~\cite{cdfupgrade}. The maximum mass reach for LEP 
and HERA experiments is also indicated.}
\end{figure}

Many such negative searches have been performed and published by essentially
all experiments at high energy collider experiments. 
The main simplified search method (should) proceeds along the 
following steps:
\begin{enumerate}
\item
Find a particular attractive and unexplored window 
to search for new physics.
\item
Identify a ``clean'' signature for the exotic object which 
separates the new from the old.
\item
Compare data and Monte Carlo using common sense. 
\item 
Publish the result and start again. 
\end{enumerate}

Having almost an infinite list of possibilities, we prefer  
not to describe any details of such exotic searches
but refer to the latest particle data book~\cite{pdg98} 
and the references therein.
In most cases it appears to be relatively easy to extrapolate 
the existing null results to future experiments at higher energies
and luminosities. In general one expects a factor of $\approx$ 2 in mass
reach from the upgraded Tevatron (RUN II) as shown in Figure 22.
Assuming no additional magic backgrounds at the LHC, ATLAS and CMS
should increase the sensitivity well into the TeV mass range, 
about a factor of $\approx$ 7 larger than the Tevatron RUN II.

Following our believe that  
the search for additional vector bosons at higher masses 
is a particular interesting area we will describe such possibilities 
in some detail.

\subsubsection{LHC signals for $W^{\prime}$ and $Z^{\prime}$} 

The sensitivity of ATLAS~\cite{wprime} for
heavy $W^{\prime}$ bosons, $W^{\prime} \rightarrow e^{\pm} \nu $ 
has been studied using
events with isolated high $p_{t}$ electron with 
large missing transverse momentum.   
A $W^{\prime}$ would show up
like a ``peak" in the transverse mass spectrum
above the steeply falling  
$W^{*}$ continuum background, as shown in Figure 23. 
The analysis shows 
good sensitivity for $W^{\prime}$ bosons with masses up to 6 TeV
and an integrated luminosity of 100 fb$^{-1}$.
\begin{figure}[htb]
\begin{center}
\epsfig{file=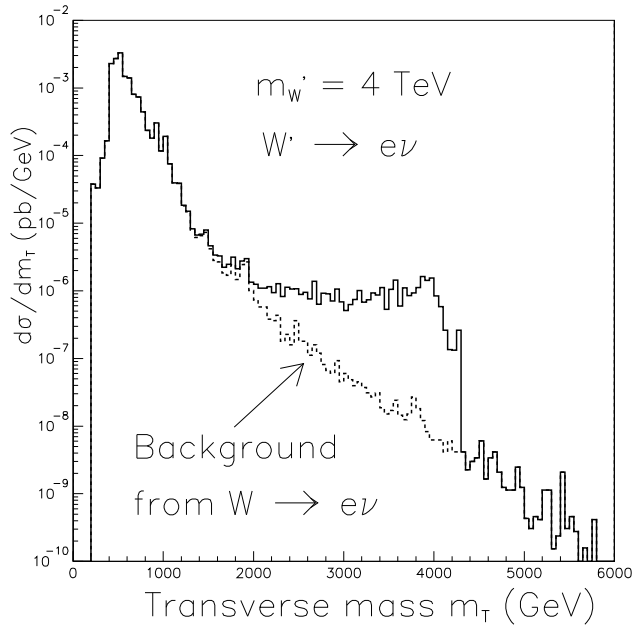,
height=9 cm,width=9cm}
\end{center}
\caption[fig23]{ATLAS simulation of the 
transverse mass distribution
for the Standard Model $W$ production and an
exotic $W^{\prime}$ scenario.
}
\end{figure}

Heavy additional $Z^{\prime}$ bosons with 
TeV masses might for example show up as a mass peak 
in the dilepton invariant mass spectrum.
Such a $Z^{\prime}$ might be discovered,
depending slightly on its couplings to fermions,
up to masses of about 4 TeV. 
Assuming that such a $Z^{\prime}$ would couple 
to quark pairs and lepton pairs like the SM $Z^{0}$, its
mass and width could be measured at the LHC.
Furthermore, the different $x$ distribution of valence 
quarks and sea antiquarks allow to analyse the
forward backward charge asymmetry and 
thus study the couplings to quarks and leptons and 
the interference with the $Z^{0}$ and $\gamma$
in quite some detail.
The results of a simulation~\cite{lhcafb}, including realistic 
experimental criteria, for 
the $M_{\ell^{+}\ell^{-}}$ mass distribution 
and the corresponding forward backward lepton 
charge asymmetry are shown in Figure 24 for 
the SM and two different $Z^{\prime}$ models.
\begin{figure}[htb]
\begin{center}
\epsfig{file=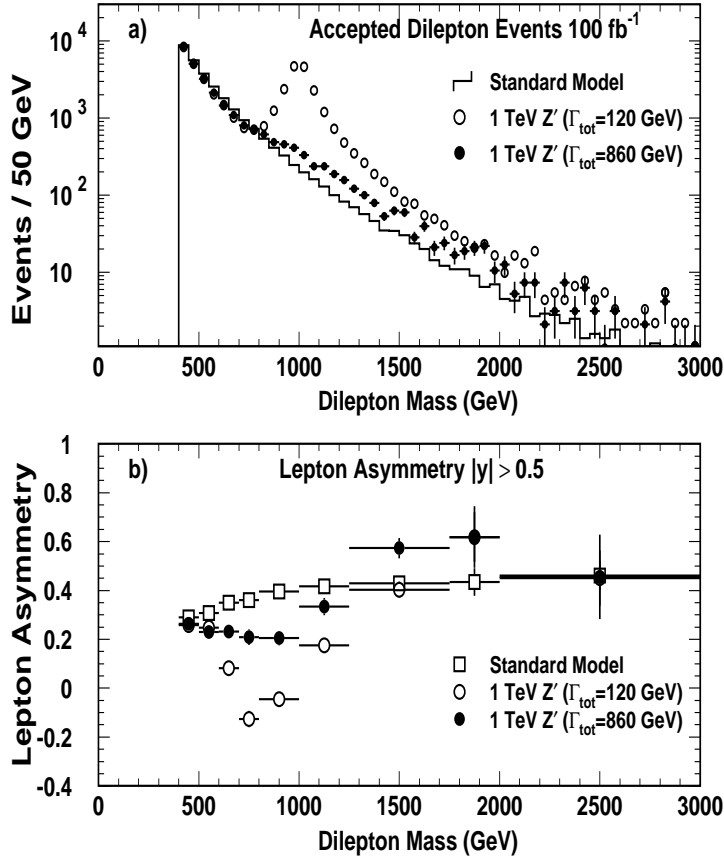,
height=13 cm,width=10cm}
\end{center}
\caption[fig24]{a) Expected dilepton mass distributions (a) 
and asymmetries (b) 
for the Standard Model and for two exotic $Z^{\prime}$ scenarios.
}
\end{figure}

\clearpage
\newpage

\section {Mainstream MSSM SUSY Searches}
Among the many possible extensions of the Standard Model the 
Minimal Supersymmetric Standard Model (MSSM) is usually considered 
to be the most serious theoretical frame. The attractive features 
of this approach are:
\begin{itemize}
\item It is quite close to the existing Standard Model. 
\item It explains the so called hierarchy problem of the Standard Model.
\item It allows to calculate.   
\item Predicts many new particles and thus ``Nobel Prizes'' for the 
masses.
\end{itemize}
These attractive features of the MSSM are nicely described 
in a Physics Report from 1984 by H. P. Nilles~\cite{nilles}. 
We repeat here some of his 
arguments given in the introduction: \newline

{\it ``Since its discovery some ten years ago, supersymmetry has 
fascinated many physicists. This has happened despite the absence of 
even the slightest phenomenological indication that it might be relevant 
for nature. .... Let us suppose that the standard model is valid up 
to a grand unification scale or even the Planck scale $10^{19}$ GeV.
The weak interaction scale of 100 GeV is very tiny compared to these 
two scales. If these scales were input parameters of the theory 
the (mass)$^2$ of the scalar particles in the Higgs sector have to 
be chosen with an accuracy of $10^{-34}$ compared to the Planck Mass. 
Theories where such adjustments of incredible accuracy have to 
be made are sometimes called unnatural.... 
Supersymmetry might render the standard model natural... 
To render the standard model supersymmetric a price has to be paid. For 
every boson (fermion) in the standard model, a supersymmetric partner 
fermion (boson) has to be introduced and to construct phenomenological
acceptable models an additional Higgs supermultiplett is needed.''}

Figures 25 and 26~\cite{hollik98} compare the consistency of the various 
electroweak measurements with the SM and the MSSM. 
\begin{figure}[htb]
\begin{center}
\epsfig{file=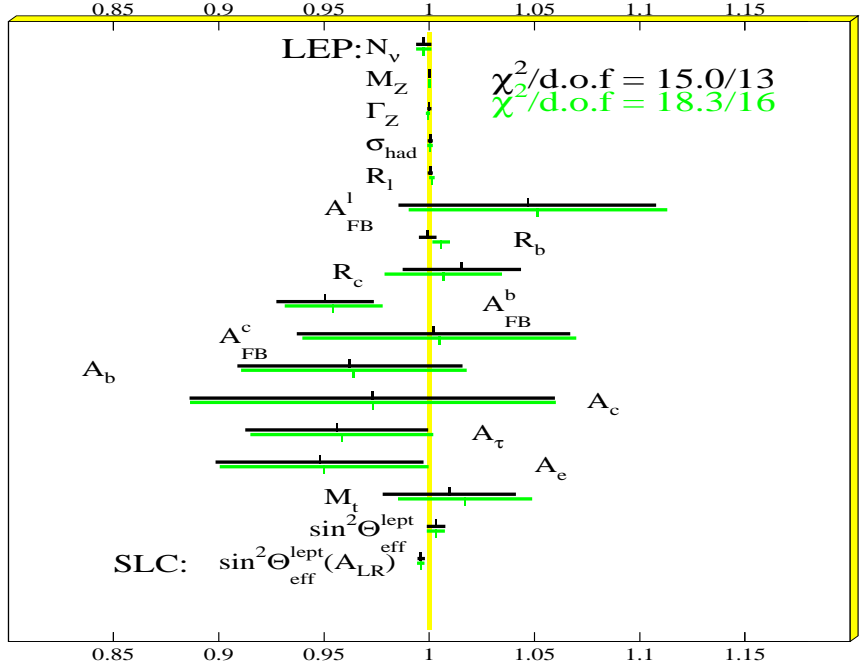,
height=12 cm,width=14cm}
\end{center}
\caption[fig25]{Comparison of $Z^{0}$ precision 
measurements with the Standard Model and the MSSM
with $\tan \beta = 1.6$ and very heavy SUSY particles~\cite{hollik98}.
}
\end{figure}
\begin{figure}[htb]
\begin{center}
\includegraphics*[scale=0.6,bb=0 150 600 650 ]
{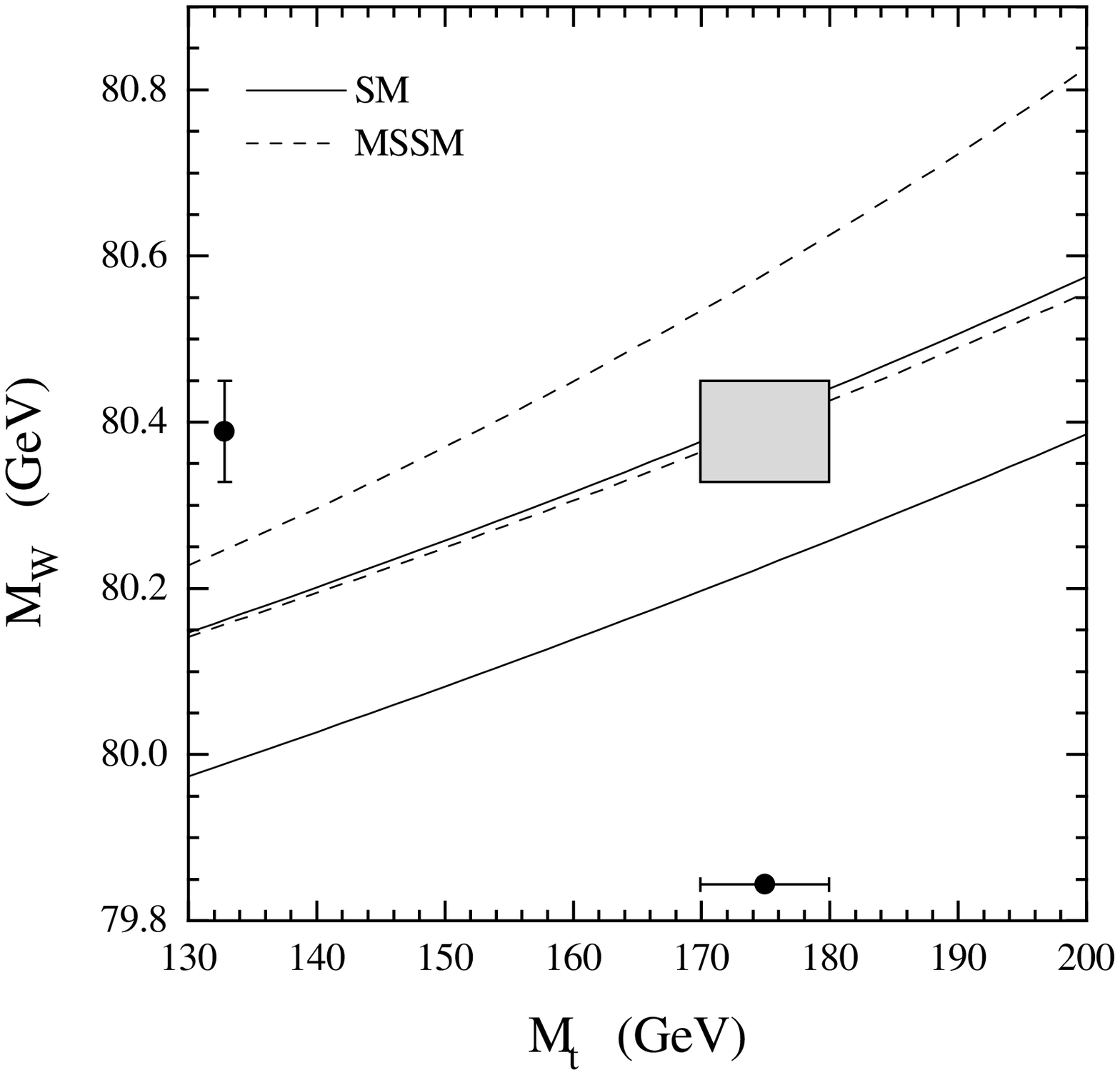}
\end{center}
\caption[fig26]{Expected relation between 
$M_{W}$ and $M_{top}$ in the 
Standard Model and  
the MSSM, the bounds are from the non-observation of Higgs or SUSY 
particles at LEPII~\cite{hollik98}.
The 1998 experimental area, with 
$M_{W}=80.39 \pm 0.06$ GeV and $M_{top}=174 \pm 5$ GeV
is also indicated.
}
\end{figure}

The largest difference, shown in Figure 26, appears in the relation 
between the $W^{\pm}$ mass and the top mass. Unfortunately todays data,  
$M_{W}=80.39 \pm 0.06$ GeV and $M_{top}=174 \pm 5$ GeV,   
favour an area which is perfectly consistent with both models.
One might thus conclude that it is not possible to decide between 
the SM and the MSSM without finding direct evidence for SUSY particles.

As mentioned above, SUSY predicts a doubling of the fundamental 
fermions and bosons and requires at least 5 Higgs bosons.
Beside the lightest, possibly invisible SUSY particle, one knows from  
the absence of such new particles that their masses have to be heavier
than $\approx$ 100 GeV. SUSY searches can be divided into 
a) the MSSM Higgs sector and b) the direct SUSY particle search.  

\section {Searching for the MSSM Higgs sector}

The MSSM Higgs sector is highly constraint. 
With the known mass of the top quark, 
all Higgs masses are strongly related. For a fixed 
mixing angle between the stop quark and the Higgs one 
usually expresses all other Higgs masses as a function of 
$\tan \beta$ and $M_{A}$. The relations become particular easy 
for masses of $M_{A}$ larger than $\approx$ 200 GeV when the 
masses of $M_{A}$, $M_{H^0}$ and $M_{H^\pm}$ are essentially 
degenerate and the mass of the lightest scalar Higgs, $h^{0}$, depends only on 
$\tan \beta$ and the mixing angle, resulting in an upper mass limit
of about 120-130 GeV for at least one Higgs boson~\cite{lowmassh}.
Thus, the search for a fundamental scalar 
particle with a mass below 130 GeV is often considered to be the
most important test of the MSSM~\cite{thhopes}. 
However, recent theoretical calculations show that 
this upper mass limit can be increased to masses of up to 200 GeV if 
additional Higgs doublets are introduced 
into the model~\cite{newlowmassh}. 
One might argue that at least the ``minimal'' of the MSSM model 
remains testable 
\footnote{Instead of discussing the meaning of the word {\it minimal}
we remind the reader about 
the (three) quark model which was destroyed 
and accepted with the observation of the charm (the fourth) quark.}.

For studies of the MSSM Higgs sector one 
usually assumes that the Higgs bosons 
can decay only to SM particles, e.g. that  
all SUSY particles are heavy.
Furthermore, once the Higgs masses are fixed, 
couplings and kinematically possible decay modes 
are constrained mainly from $\tan \beta$ and the mixing angle.
Detailed branching ratio calculations can be 
found in reference~\cite{hdecays}.
Qualitatively one finds that 
the lightest Higgs, $h^{0}$, looks in all respects like the SM Higgs 
if the mass of the $A$ is large, $M_{A} > 400-500$ GeV. 
Consequently, the lightest MSSM Higgs $h^{0}$
should be discovered at LEPII if $\tan \beta$ is smaller than about 4, 
e.g. $m_{h}$ is smaller than $\approx$ 100 GeV or 
at the LHC with the channel $h \rightarrow \gamma \gamma$.
if $\tan \beta$ is larger than about 4.

For a smaller mass of $M_{A}$ the $h^{0}$ cross section 
and possible decay modes depend strongly on $\tan \beta$.
Again, the lightest MSSM Higgs $h^{0}$
should be discovered at LEPII if $m_{h}$ is smaller 
than $\approx$ 100 GeV and if 
$\tan \beta$ is smaller than about 4.
For $m_{h}$ larger than 100 GeV and larger 
values of $\tan \beta$ the expected LHC rate 
for the signature $h \rightarrow \gamma \gamma$
appears to be strongly suppressed.  

The possible signatures for the other Higgs bosons 
depend strongly on their masses, the kinematically allowed decay 
products and the choice of $\tan \beta$. 

For small values of $\tan \beta$
and a mass smaller than twice the top quark mass, $m_{H^{0}} < 350$ GeV, 
and smaller than twice the mass of $h^{0}$,  
the $H^{0}$ appears to behave like the SM Higgs.
Once kinematically 
allowed, the branching ratio $H^{0} \rightarrow h^{0}h^{0}$ might 
become large. As the  
couplings to the third quark and lepton family are enhanced 
proportional to $(\tan \beta)^{2}$ one expects that 
roughly 90\% of the $H^{0}$ and $A^{0}$ decay to $b\bar{b}$ jets  
and 10\% to $\tau^{+}\tau^{-}$ if $\tan \beta$ is large.

The couplings of the charged Higgs $H^{\pm}$
are dominated by the third fermion family. Up to a $H^{\pm}$ mass of roughly 
$m_{t}+m_{b}$ the dominant decay mode is 
$H^{\pm} \rightarrow \tau \nu$. For larger masses only the 
decay mode $H^{\pm} \rightarrow t~b $ appears to be relevant. 
Having large couplings to the $tb$ system, direct 
searches can be performed in $t$ decays 
$t \rightarrow b H^{+}$ with $H^{+} \rightarrow \tau^{+} \nu$. 
Furthermore, b--decays, like  
$b \rightarrow s \gamma $, 
provide strong indirect constraints on the charged Higgs as discussed 
in section 8. 

\subsection{MSSM Higgs search at LEPII}

Experiments at LEPII and $\sqrt{s} \approx 200$ GeV
will have an excellent sensitivity to the SM Higgs with 
masses of about 100 GeV. One finds that this sensitivity 
translates to a Higgs sensitivity of the MSSM  
for values of $\tan \beta$ of about roughly 4 (3) with no (maximal) mixing 
using the process $e^{+}e^{-} \rightarrow Z^* \rightarrow Z h^{0}$.
For larger $\tan \beta$ values the couplings of the h to the weak bosons 
are reduced proportional to $\cos \beta$ 
and the predicted mass value of $m_{h^{0}}$ increases.  
However, the $h^{0},A^{0}$ Higgs pair production 
$\rightarrow Z^* \rightarrow h^{0} A^{0}$, 
if kinematically allowed, appears to be detectable. This process results in   
a distinct signature of events with 
four b-jets. The search for such 4 b-jet events during the 
future LEPII running will thus give sensitivity to
masses of $M_{h}, M_{A} < 90$ GeV and all $\tan \beta$ values. 

Searches for Higgs bosons with masses beyond the kinematic LEPII limit 
have to wait, either for the LHC, as will be discussed below, 
or for a future high luminosity high energy linear $e^{+}e^{-}$ collider. 

\subsection {MSSM Higgs search at the LHC}

Current LHC studies show that the sensitivity to the MSSM Higgs sector is
somehow restricted. One finds that one either needs Higgs particles
with essentially SM like couplings, e.g. $M_{A} >500$ GeV
or one needs large $\tan \beta$ values. 
This sensitivity is usually shown in a complicated two--dimensional  
multi--line contour plot\footnote{Also called spaghetti plot.}
like the one in Figure 27.  
\begin{figure}[htb]
\begin{center}
\includegraphics*[scale=0.6,bb=0 180 600 590 ]
{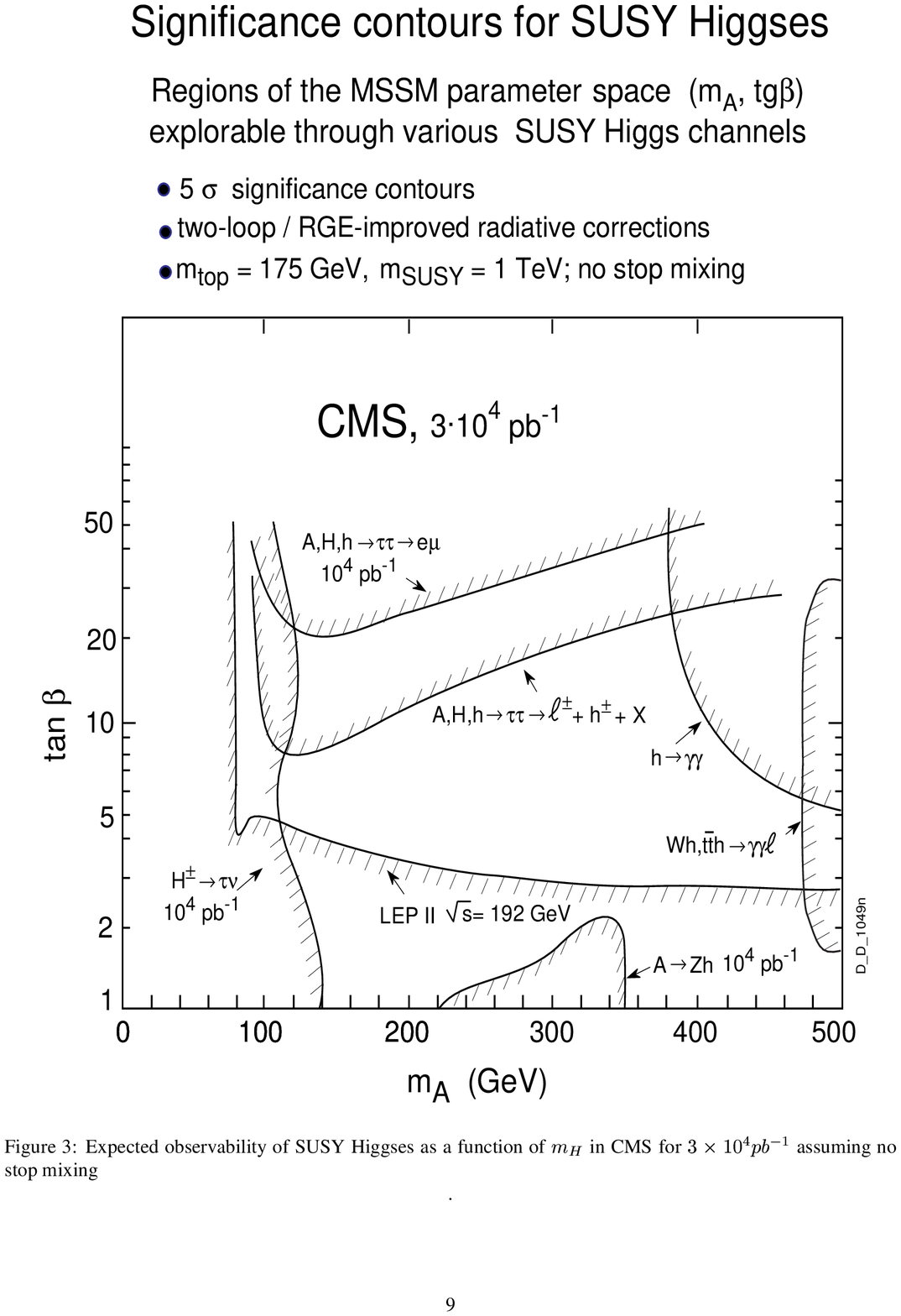}
\caption[fig27]
{CMS 5 sigma significance contour plot for the different MSSM Higgs 
sector in the $M_{A}$ - $\tan\beta$ plane~\cite{cmsmssmh}.
Each curve indicates the sensitivity for different Higgs search modes.} 
\end{center}
\end{figure}

\subsubsection{The lightest neutral Higgs $h^{0}$}

For the lightest Higgs, with a mass below 120-130 GeV, the only established
signature appears to be the decay $h^{0} \rightarrow \gamma \gamma$. 
For masses of $M_{A}$, larger than 400 GeV one finds essentially the 
SM rates and its sensitivity. For smaller masses of $M_{A}$, 
the branching ratio $h \rightarrow \gamma \gamma$ becomes 
too small to observe 5 standard deviations signals.
The combination of the $h^{0} \rightarrow \gamma \gamma$ search 
with other $h^{0}$ decay modes, like 
$h^{0} \rightarrow b \bar{b}$, $h^{0} \rightarrow ZZ^{*}$ and 
$h^{0} \rightarrow WW^{*}$ should help to enlarge the 5 sigma domain.   

\subsubsection{The heavy neutral Higgse $H^{0}, A^{0}$ and low $\tan \beta$}

For values of $\tan \beta$ smaller than $\approx$4 
one expects that the lightest Higgs will soon be discovered at LEPII. 
For such a scenario one
finds that the $H^{0}$ might be visible for some masses and decays. 
For example a $H^{0}$ with a mass close to 170 GeV  
appears to be detectable with the channel 
$H^{0} \rightarrow WW^{*} \rightarrow \ell \nu \ell \nu$. 
Other studies indicate possible $H^{0}$ signals with
$H^{0} \rightarrow hh \rightarrow \gamma \gamma b \bar{b}$ and 
$A \rightarrow Z h \rightarrow \ell^{+} \ell^{-} b \bar{b}$ and $H^{0}$ 
masses between 200-350 GeV.  
We will not go into further details here as the relevance of such 
``utopic'' studies for low values of $\tan \beta$ depends 
so strongly on the near future LEPII results.
  
\subsubsection{The heavy neutral Higgse $H^{0}, A^{0}$ and large $\tan \beta$}

For large values of $\tan \beta$ the Higgs production 
cross sections, especially the ones for $b\bar{b}H^{0}$ and $b\bar{b}A^{0}$ 
are much larger than the ones for the SM Higgs with similar masses.  
The only relevant Higgs decays are 
$H^{0}, A^{0} \rightarrow \tau \tau$ and 
$H^{0}, A^{0} \rightarrow b \bar{b}$. While it is generally assumed that 
the Higgs decays to $b \bar{b}$ jets can not be seen at the LHC, 
the decays to $\tau^{+} \tau^{-}$ are believed to give a detectable  
signature.    
 
The analysis of the $\tau^{+} \tau^{-}$ final states proceeds
along the following ideas. The decay products of $\tau^{\pm}$,
can be separated from quark and gluon jets by the low mass, the
low charged multiplicity and the missing transverse energy. 
Studies indicate that hadronic 
$\tau$ decays with a $p_{t}$ of the observable hadrons above 20 GeV 
can be separated with good efficiency and a quark/gluon jet rejection factor 
of more than 100. Leptonic $\tau$ decays  
to electrons, muons allow even stronger jet rejection factors and 
provide in addition a straight forward trigger signal.

The proposed analysis proceeds along the following lines:
\begin{itemize}
\item
The event should contain two opposite charged isolated $\tau$
candidates with high $p_{t}$. At least one of the two $\tau$'s should 
be an electron or a muon.
\item
The usual $\tau$ signature is an isolated charged hadron with a 
minimum $p_{t}$ of 5 GeV or more, combined 
eventually with some associated $\pi^{0}$ activity in the calorimeter.   
The reconstructed $\tau$ energy and momentum vector is obtained from 
the sum of the associated track and calorimeter measurements
plus the assigned missing neutrino energy.    
\item 
The two $\tau$ candidates should not be back to back in the plane 
transverse to the beam direction.  
This requirement results in a considerable efficiency reduction, 
but is required to allow a $\tau\tau$ mass reconstruction and  
to reduce the background from leptonic $Z^{0}$ decays 
and higher mass Drell-Yan lepton pair events.
\item
The event should contain at least one jet with a 
large transverse energy of at least $\geq$ 40-50 GeV, to 
balance the required $p_{t}$ of the $\tau\tau$ system.
\item 
The invariant mass of the $\tau\tau$ system is reconstructed under the 
assumption that the reconstructed $\tau$ direction agrees 
with the true $\tau$ direction. The measured missing transverse 
energy, using all other measured particles in the event, 
is assumed to originate from the two $\tau$ decays.
The missing transverse event energy is thus split 
and added to each reconstructed $\tau$ decay products.
The mass is determined according to 
$m_{\tau\tau}=\sqrt{2 \times E_{\tau}^{1}\times E_{\tau}^{2}(1-\cos\theta)}$.
Simulations indicate mass resolutions of about 10-15\%, about 
$\pm$ 20-30 GeV for masses of 
about 200 GeV, and about $\pm$ 50 GeV for a mass of 300 GeV.  
\item
For values of $\tan \beta$ above 10, the cross section for the 
process $gg \rightarrow b \bar{b} A(H)$ becomes large enough to 
improve the signal to background ratio by requiring the presence of
additional b-jets.
\end{itemize}
Depending slightly on the Higgs mass, the efficiency and the  
signal to background ratio are quite small.
For example, a recent CMS study~\cite{cmsmssmh1} 
of the MSSM ($\tan \beta = 8 (15)$) 
Higgs search with $\tau \tau$ final states and  
a luminosity of 30 fb$^{-1}$,
expects a signal of roughly 500 (300) events for masses of 
140 (300) GeV above a background of 7500 (3000) events. 
About 50\% of the estimated background comes from 
high mass Drell-Yan $\tau \tau$ pair production.
Simulation studies indicate that the signal 
to background ratio can be strongly improved 
if one jet is identified as a b-jet as 
indicated in Figure 28~\cite{cmsmssmh2}. 
Unfortunately most of the associated b-jets are expected to have a very 
low $p_{t}$. Therefore, the good signal to background 
ratio can be achieved only with a small 
signal efficiency. As a result, statistical significance with 5 sigma 
requires large signal cross sections expected for 
$\tan \beta$ values larger than 15--20.  
\begin{figure}[htb]
\begin{center}
\includegraphics*[scale=0.6,bb=0 380 600 800 ]
{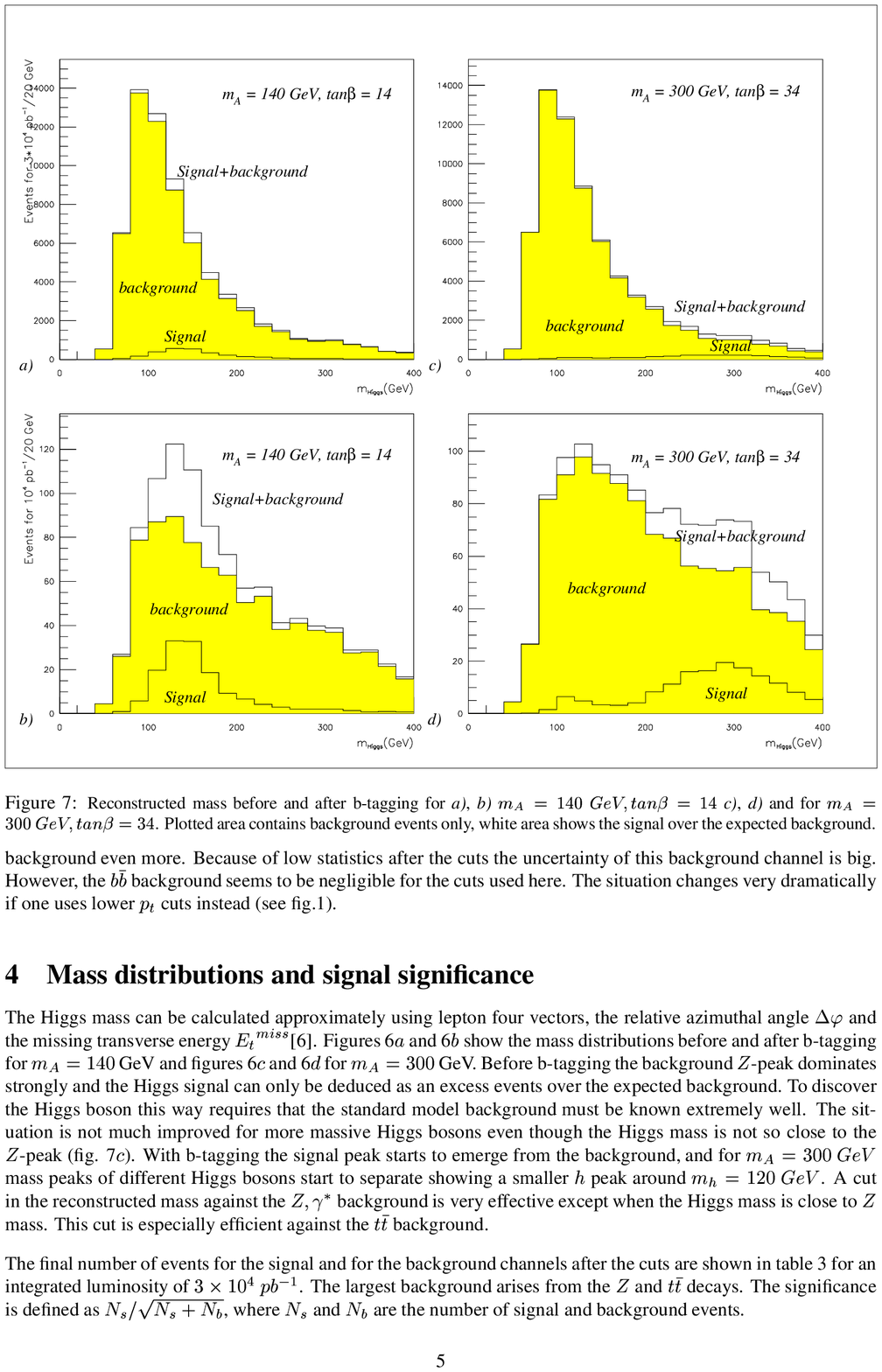}
\caption[fig28]
{CMS simulation of the MSSM Higgs search 
for $H, A \rightarrow \tau \tau$ with and without b-tagging~\cite{cmsmssmh}
for $M_{A} = 140 GeV$ and $\tan \beta = 14$ and 
for $M_{A} = 300 GeV$ and $\tan \beta = 34$}
\end{center}
\end{figure}

Optimists prefer therefore to assume
negligible systematics and 
search for an excess of $\tau \tau$ pairs
without b-tagging. Following this procedure one finds
statisticaly significant 5 sigma signals 
for $\tan \beta$ values larger than 8--10 as shown 
in Figure 27. 
The significance figures from    
similar studies with the ATLAS simulation appear to be almost 
identical. Unfortunately, a direct comparison of the obtained results 
is difficult as different backgrounds are considered
for both studies. We thus hope that the search for the MSSM Higgs with 
$A^{0} (H^{0}) \rightarrow \tau \tau$ final states
will not be spoiled by unforeseen backgrounds and that the 
more realistic Next-to-Leading-Order Monte Carlos will not result 
in systematic background uncertainties larger than 2-3\%.

Assuming large $\tan \beta$ values, the rare 
decay $A, H \rightarrow \mu \mu$\footnote{The branching ratio
is expected to be about a factor of 300 smaller than 
the one for the decay to $\tau\tau$ as it scales with $(m_{\mu}/m_{\tau})^{2}$.}  
might show up as a resonance peak
above a large background. Assuming excellent 
mass resolutions in the $\mu \mu$ channel of about 0.01-0.02$\times$ 
m(Higgs) [GeV], the performed studies  
indicate a Higgs discovery possibility in this channel 
for a luminosity of 30 fb$^{-1}$
and $\tan \beta$ larger than $\approx$ 20.  

\subsubsection{The charged Higgs $H^{\pm}$}

Depending only slightly on $\tan \beta$, the relevant 
MSSM charged Higgs decay modes are $H^{+} \rightarrow \tau^{+} \nu$
for masses below the $t \bar{b}$ threshold, and 
$H^{+} \rightarrow t \bar{b}$ above.  
Having large couplings to the $t b$ system, $t\bar{t} X$ events 
are a large source of charged Higgs events.
Inclusive $t\bar{t}$ events might thus provide 
a good experimental signature for $H^{\pm}$ with a mass below 
$m_{top} - 10$ GeV. One has to search for $t\bar{t}$
events with isolated $\tau$ candidates which originate from the decay chain
$t\bar{t} X \rightarrow b W^{\pm} b H^{\pm}$ and 
$H^{\pm} \rightarrow \tau \nu$. 

The studied signature requires events with:
\begin{itemize}
\item
an isolated high $p_{t}$ electron or muon from a $W$ decay,
\item  
the decay products from an isolated energetic $\tau$,
\item 
two b--flavoured jets and perhaps some missing $p_{t}$.
\end{itemize}
The performed ATLAS/CMS simulations~\cite{atlascmshplus}, 
using a luminosity of 10$fb^{-1}$,
indicate that signals of a few 100 events with a signal 
to background ratio of about 1/7 can be obtained.
Assuming that the backgrounds are well known, 
a sensitivity for $H^{\pm}$ masses
up to about 130-140 GeV is obtained for all values of $\tan \beta$.  

Another interesting process might be the 
production of a heavy $H^{\pm}$ in association with a top quark, 
$g b \rightarrow t H^{-} \rightarrow ttb \rightarrow WW bbb $. 
A parton level 
analysis of this channel~\cite{hplus} indicates the possibility 
to obtain $H^{\pm}$ mass peaks 
with reasonable signal to background ratios. 
The proposed analysis selects events 
with one leptonic and one hadronic $W$ decay and  
three identified b-jets. The used efficiencies for lepton tagging and
b-jet identification are close to the ones assumed in 
simulations from ATLAS and CMS for other LHC processes. 

The performed analysis obtains  
the mass peaks, shown in Figure 29, 
from the mass distribution of the reconstructed 
$tb$ jet system, where the top 
is reconstructed from the decay $t \rightarrow W b$.
\begin{figure}[htb]
\begin{center}
\includegraphics*[scale=0.555,bb=0 80 600 800 ]
{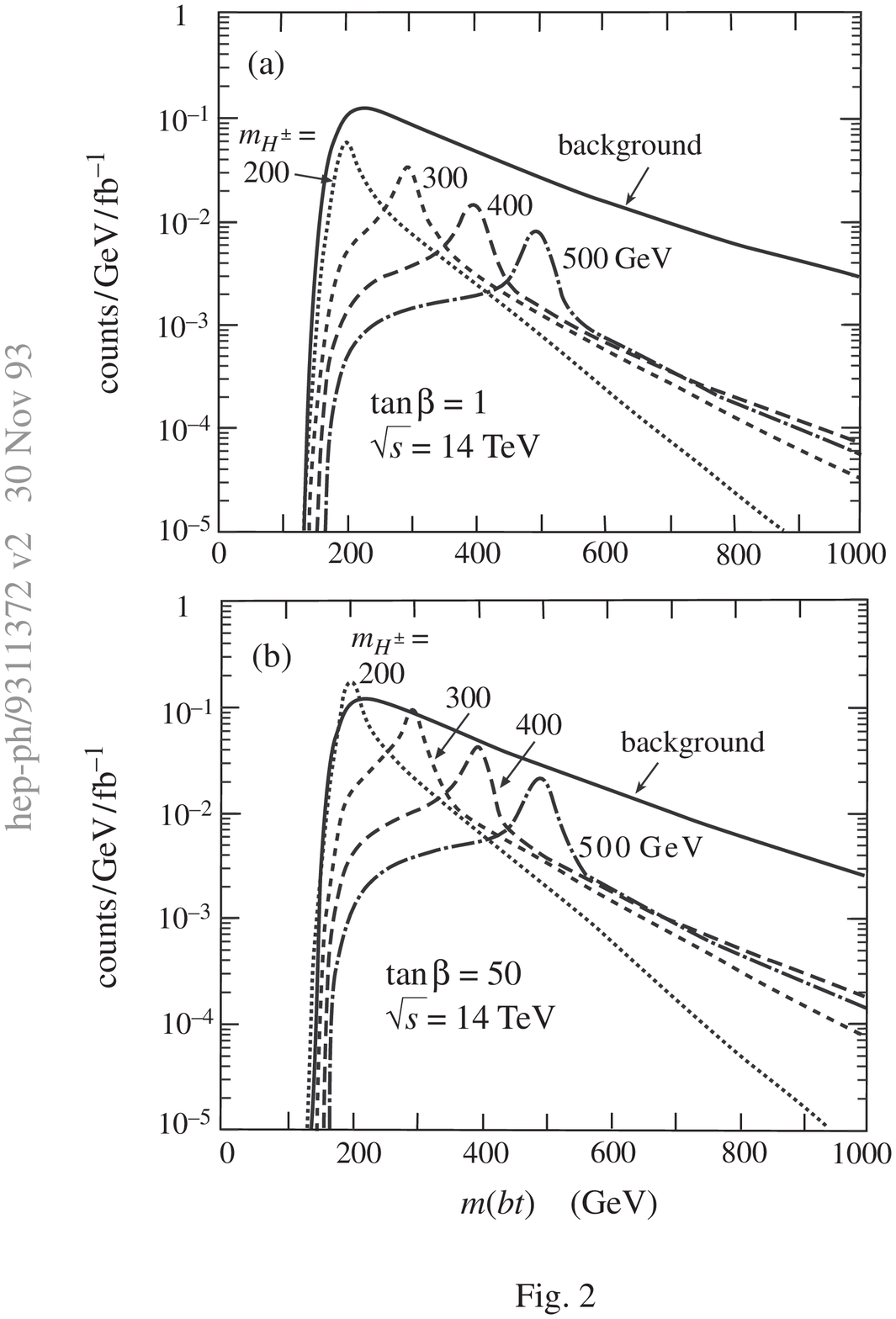}
\caption[fig29]
{Comparison of the reconstructed mass $H^{\pm}$ 
signals and backgrounds for $H^{\pm}$ masses of 
200, 300, 400 and 500 GeV. The upper and lower plots are for 
$\tan \beta = 1$ and $\tan \beta = 50$ respectively~\cite{hplus}.}
\end{center}
\end{figure}
Furthermore, the resolution and the combinatorial background 
is reduced using the known mass of the top quark.
The study indicates accepted signal cross sections of up to 1 fb 
above background cross section between 1-2 fb and a 
60 GeV mass bins for an interesting MSSM parameter range.
It would certainly be interesting
to see if this parton level result can be confirmed 
in a more detailed detector level study. 
  
\subsubsection{Are MSSM Higgse a proof of Supersymmetry?}

The discovery of at least one of the MSSM Higgs particles is often 
believed to be the proof of SUPERSYMMETRY. Figure 27 and 30 indicate 
the estimated sensitivity of the CMS and ATLAS experiments
to various Higgs decay channels and different luminosities.
These two dimensional multi line 
5 sigma (statistical) significance plots,
especially in the logarithmic version, indicate
sensitivity over almost the entire MSSM parameter space. 
\begin{figure}[htb]
\begin{center}
\epsfig{file=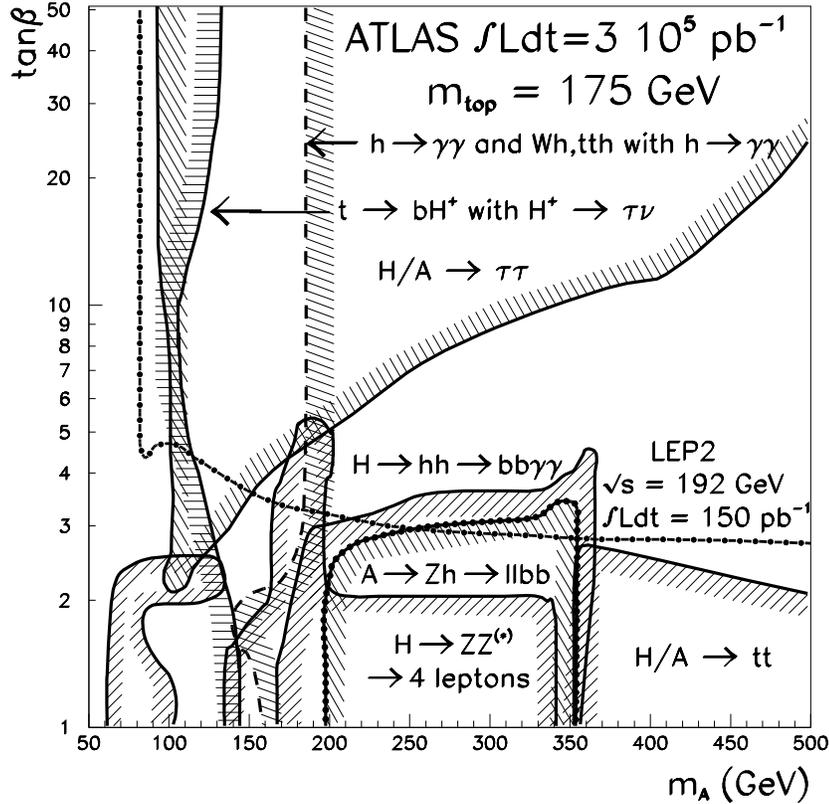,
height=12 cm,width=12cm}
\end{center}
\caption[fig30]
{Estimated ultimate ATLAS 5 sigma discovery sensitivity for 
the MSSM with a luminosity of 300 fb$^{-1}$~\cite{atlasmssmh}. 
The sensitivity of the different search signatures are shown 
in the $M_{A} - \tan \beta $ plane.}
\end{figure}
\clearpage
However, 
it is worth to remind the reader that the assumed sensitivity 
includes only statistical errors. As a consequence, the obtained 
curves, especially when extrapolated to larger integrated luminosities
and combined for ATLAS and CMS are doubtful. This is especially the 
case, as discussed above, for channels like 
$WH \rightarrow \ell \nu b \bar{b}$ and 
for $H^{0},A^{0} \rightarrow \tau\tau$ where the proposed signatures 
suffers certainly from the very bad signal to background ratio.
   
Despite these ``small'' problems, the performed studies indicate 
a hole for $\tan \beta$ values between roughly 4--10 and a mass of $M_{A}$
between 100--300(400) GeV. Furthermore, the different Higgs bosons
indicate only little overlap. For example, the remaining discovery 
potential of LEPII experiments for  
the lightest Higgs require $\tan \beta$ values between $\approx$
2--4. Todays LHC studies for this parameter range 
show a rather limited possibility to discover
any additional MSSM Higgs bosons at the LHC.

In contrast, one might assume that all Higgs bosons, beside $h^{0}$
are very heavy. Theoretical calculations show that
such a scenario results in a light Higgs with the couplings of the 
SM Higgs. This means that such a light Higgs can not be distinguished 
from the SM Higgs and does not prove or disprove SUPERSYMMETRY!

The remaining MSSM scenario for LHC experiments is 
a large $\tan \beta$ value combined with 
a $A^{0}$ with a mass below 300--400 GeV.   
For this scenario, the sensitivity plots indicate 
that one should look for the signature 
$H^{0},A^{0} \rightarrow \tau\tau$, which suffers unfortunately from 
the estimated bad signal to background ratio. 
It appears doubtful that 
this channel will convince anybody of SUPERSYMMETRY.

We thus conclude this section with the remark 
that neither the full MSSM Higgs parameter space can be covered at the LHC 
nor that the discovery of a single MSSM Higgs boson will allow to prove 
SUPERSYMMETRY. Furthermore, ``minor'' additions to the MSSM Higgs sector, 
like the existence of (one) additional Higgs doublet(s),
increases the expected upper Higgs mass limit to values 
of up to 200 GeV~\cite{newlowmassh}.   
Thus, even the discovery of a SM like Higgs with a mass of 160 GeV
will not allow to distinguish between the Standard Model, 
SUPERSYMMETRY and other new physics.

Consequently, the only way to prove SUPERSYMMETRY is the direct
unambiguous detection of at least one SUSY particle. 

\newpage
\section {Direct Searches for SUSY Particles}

The discussion in the previous chapter leads to the result 
that SUPERSYMMETRY can not be discovered or excluded from the 
Higgs sector. The experimental high energy physics community is 
thus forced to either discover at least one supersymmetric particle,  
or to show without any doubt that supersymmetric particles 
are much heavier than all known SM particles with masses of at least 
a few TeV. 

Not even a small indication for 
SUSY like particles has been found at LEPII, at the TEVATRON or at HERA.
Thus, the statement that detectable SUSY particles have to be heavier than 
$\approx$ 100 GeV appears to be quite safe.
The experimental obligation to search for SUSY requires thus 
(a) to reach higher center of mass energies 
combined with large luminosities and (b) to search 
for various SUSY signatures in a multi dimensional parameter space. 
Following this guideline we describe in the following the 
known most promising SUSY signatures discussed for the LHC and the 
upgraded TEVATRON. 
 
Starting from the MSSM, the so called minimal model, theoretical 
counting results in more than hundred free parameters.
So many free parameters do not offer a good
guidance for experimentalists, which prefer to use additional  
assumptions to constrain the parameter space.
The simplest approach is the so called MSUGRA 
(minimal supergravity model) model with 
only five parameters ($m_{0}, m_{1/2}, \tan \beta, A^{0}$ and $\mu$).

This SUSY model is used for most sensitivity estimates of future 
colliders and the obtained results for the LHC will be discussed below.
The main reason for this model choice is the existence of 
very advanced Monte Carlo programs~\cite{isajet}, ~\cite{spythia},  
required for detailed simulation studies.
This pragmatic choice of one approach to investigate 
the potential of a future experiment appears to be more than sufficient, 
as essentially all required detector features can be tested. 
 
However, such a pragmatic approach should not be considered 
as a too strong guidance principle if one wants to discover 
SUPERSYMMETRY with real experiments. Two recent examples show that   
the absence of any MSUGRA indications, enlarges the acceptance for more 
radical SUSY models. 

The first example is the famous lonely CDF event, 
which has large missing transverse energy,
2 high $p_{t}$ isolated photons and 2 isolated high $p_{t}$ 
electron candidates~\cite{cdfevent}.
The presence of high $p_{t}$ photons does not match MSUGRA
expectations but might fit 
into so called gauge mediated symmetry breaking models, GMSB~\cite{gmsbth}. 
This event has certainly motivated many additional, 
so far negative searches. 

The second example is related to the 1997 HERA excitement. 
The observed excess of a handful of events appeared 
to be consistent with either a 
lepton--quark resonance with a mass of roughly 200 GeV or with 
a signature predicted from R--parity violation SUSY 
models~\cite{rparity}. While this excess was not confirmed 
with larger statistics, the R--parity violation models became 
certainly much more attractive.

These modified searches indicate 
the discovery potential of searches which are 
not guided by todays fashion.
Having reminded the reader of potential shortcomings 
between a SUSY Nature and the studied SUGRA model, we now turn 
to future LHC (and Tevatron) search strategies for 
SUSY particles within SUGRA. 

\subsection{MSUGRA predictions}

Essentially all signatures related to the MSSM and 
in particular to MSUGRA searches are
based on the consequences of R--parity conservation. R--parity is a 
multiplicative quantum number like ordinary parity. The R--parity of 
the known SM
particles is 1, while the one for the SUSY partners is -1. 
As a consequence, SUSY particles have to be produced in pairs.
Unstable SUSY particles decay, either directly or via 
some cascades, to SM particles and the lightest supersymmetric
particle, LSP, required by cosmological arguments to be 
neutral. Such a massive LSP's, should have been abundantly produced 
after the Big Bang and is currently considered 
to be ``the cold dark matter'' candidate. 
This LSP, usually assumed to be the lightest neutralino 
$\tilde{\chi}^{0}_{1}$ has neutrino like interaction cross sections
and can not be observed in collider experiments. 
Events with a large amount of 
missing energy and momentum are thus the prime SUSY  
signature in collider experiments.

Possible examples are the pair production of sleptons with 
their subsequent decays, $pp \rightarrow \tilde{\ell}^{+}\tilde{\ell}^{-}$
and $ \tilde{\ell} \rightarrow \ell \tilde{\chi}^{0}_{1}$
which would appear as events with a pair of isolated electrons or muons 
with high $p_{t}$ and large missing transverse energy.  

Within the MSUGRA model, the masses of 
SUSY particles are strongly related to the so called universal fermion 
and scalar masses $m_{1/2}$ and $m_{0}$. The masses of the spin 1/2 
SUSY particles are directly related to $m_{1/2}$. 
One expects approximately the following mass hierarchy:
\begin{itemize}
\item
$\tilde{\chi}^{0}_{1} \approx 1/2 m_{1/2} $
\item
$\tilde{\chi}^{0}_{2}\approx \tilde{\chi}^{\pm}_{1} \approx m_{1/2} $
\item
$\tilde{g}$ (the gluino) $\approx 3 m_{1/2} $
\end{itemize}
The masses of the spin 0 
SUSY particles are related to $m_{0}$ and $m_{1/2}$ and 
allow, for some mass splitting between the ``left'' and ``right'' handed
scalar partners of the degenerated left and right handed fermions. 
One finds the following simplified mass relations:
\begin{itemize}
\item
$m(\tilde{q})$(with q=u,d,s,c and b) $ \approx \sqrt{m_{0}^{2} + 6 m_{1/2}^{2}}$
\item
$m(\tilde{\nu}) \approx m(\tilde{\ell^{\pm}})$ (left)
$ \approx \sqrt{m_{0}^{2} + 0.52 m_{1/2}^{2}}$
\item
$m(\tilde{\ell^{\pm}})$ (right) $\approx \sqrt{m_{0}^{2} + 0.15 m_{1/2}^{2}}$
\end{itemize}
The masses of the left and right handed stop quarks ($\tilde{t}_{\ell, r}$) 
might show, depending on other SUGRA parameters,  
a large splitting. As a result, the right handed stop quark might be 
the lightest of all squarks.   
 
Following the above mass relations and using the known SUSY couplings, 
possible SUSY decays and the related signatures can be defined.
Already with the simplest MSUGRA frame one finds   
a variety of decay chains.

For example the $\tilde{\chi}^{0}_{2}$ could decay 
to $\tilde{\chi}^{0}_{2}\rightarrow \tilde{\chi}^{0}_{1}+ X$
with $X$ being:
\begin{itemize}
\item
$X= \gamma^{*} Z^{*} \rightarrow \ell^{+}\ell^{-}$
\item
$X= h^{0} \rightarrow b \bar{b}$
\item
$X= Z  \rightarrow f \bar{f}$
\end{itemize}

Other possible $\tilde{\chi}^{0}_{2}$ decay chains are 
$\tilde{\chi}^{0}_{2}\rightarrow \tilde{\chi}^{\pm(*)}_{1}+ \ell^{\pm} \nu$ 
and 
$\tilde{\chi}^{\pm(*)}_{1} \rightarrow \tilde{\chi}^{0}_{1} \ell^{\pm} \nu$
or 
$\tilde{\chi}^{0}_{2}\rightarrow \tilde{\ell}^{\pm} \ell^{\mp}$.

Allowing for higher and higher masses, even more decay channels
might open up. It is thus not possible to define all search strategies
a priori. Furthermore, possible 
unconstrained mixing angles between neutralinos, lead to 
model dependent search strategy for squarks and gluinos as 
will be discussed below. 

Todays negative SUSY searches~\cite{treille} 
provide the following approximate lower mass 
limits:
\begin{itemize}
\item
$m(\tilde{\chi}^{\pm}_{1}) > 90$ GeV (LEPII) 
\item $m (\tilde{g}) $(gluino) $>$ 160-220 GeV depending slightly
on the assumed squark masses (TEVATRON).
\end{itemize}

One might argue, that the negative results of the chargino search at LEPII
imply that future gluino searches at the upgraded TEVATRON 
should not start for masses below $\approx$ 270 GeV.
However, the continuing TEVATRON searches indicate that many 
searchers do not follow too strictly specific mass relations of a
MSUGRA model. Current experimental results are usually shown  
as a function of the searched for SUSY masses.

In contrast, sensitivity estimates for future collider experiments are 
usually given in the $m_{0}$--$m_{1/2}$ parameter space. 
Despite the model dependence, such estimates allow to 
compare the possible significance of the different 
studied signatures. 
Having various proposed methods, the resulting 
sensitivity figures appear to be quite confusing and require 
some time for appreciation.   
A typical example is shown in Figure 31a-d~\cite{baer96}, 
where the different 
curves indicate the LHC sensitivity for different signatures 
and different SUSY particles. It is usually assumed that the maximum 
information about SUSY can be extracted in regions, covered 
by many signatures. The meaning of the various curves and their potential
significance should become clear from the following sections.  
\begin{figure}[htb]
\begin{center}
\includegraphics*[scale=0.65,bb=40 200 600 670 ]
{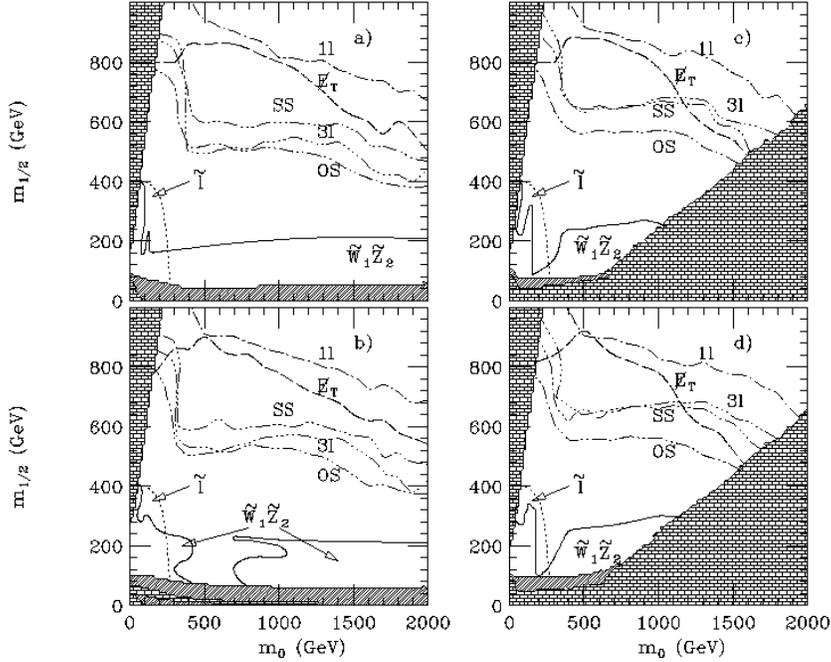}
\caption[fig31]
{Expected sensitivity for various SUSY particles and signatures 
in the $m_{0} - m_{1/2}$ plane using an integrated luminosity of 
10 fb$^{-1}$ at the LHC~\cite{baer96}. Figures a and b are for $\tan \beta$=2
with negative and positve $\mu$; the corresponding results for 
$\tan \beta$=10 and negative and positive $\mu$ are shown in c and d.
The different curves indicate 
the sensitivity for SUSY events with n leptons ($\ell$) 
and for events with lepton pairs with same charge (SS) and 
opposite charge (OS).}  
\end{center}
\end{figure}

\subsection{Anatomy of a Slepton Signature at the LHC}

Hadron colliders are certainly not a good source of sleptons. 
Nevertheless, we start our analysis of the various SUSY search strategies 
with an anatomy of the simplest possible SUSY signal.
Our discussion starts with the cross section and the 
expected decay modes. This is followed by a qualitative 
description of a possible discovery signature at the LHC
which is then compared to a detailed simulation of a search for sleptons
at the LHC. 

The pair production of sleptons at the LHC can easily be related to the 
production of Drell-Yan dilepton pairs with high mass.
The expected total slepton pair production cross section 
as a function of the slepton mass is shown in Figure 32~\cite{baer94}.
 
\begin{figure}[htb]
\begin{center}
\includegraphics*[scale=0.6,bb=40 350 600 800 ]
{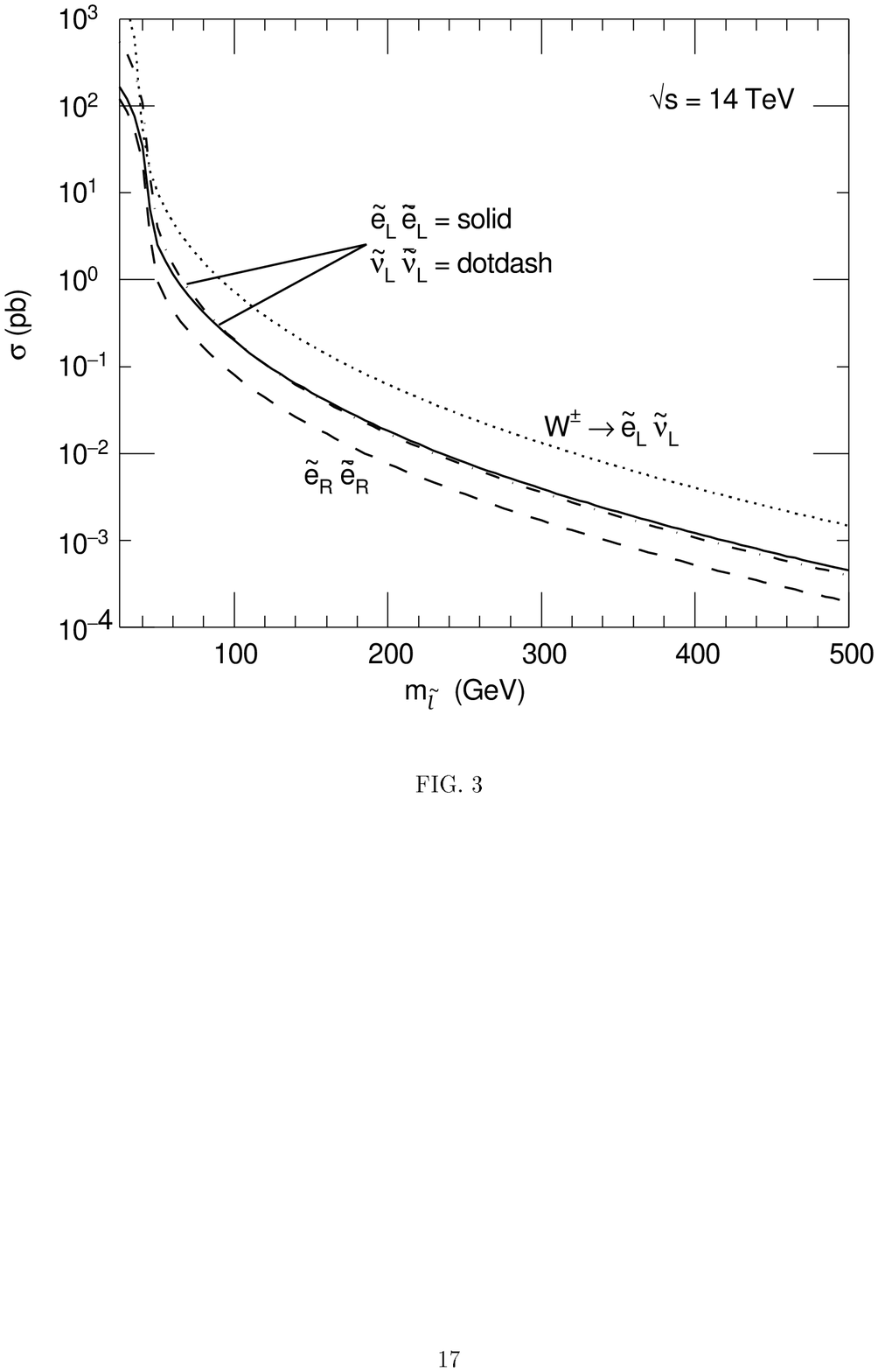}
\caption[fig32]
{Slepton mass dependence of the total pair production cross section 
at the LHC for various slepton combinations~\cite{baer94}.}
\end{center}
\end{figure}

Figure 33 shows the expected mass (with m$>$200 GeV) 
distribution of the virtual $\gamma^{*}, Z^{*}$ 
system leading to a lepton or slepton pair. 
The cross section for scalar charged sleptons has a simple relation 
to the production of the corresponding right and left handed lepton 
pair production 
$\sigma(\tilde{\ell} \tilde{\ell})=1/4 \beta^{3} \sigma(\ell\ell)$.
\begin{figure}[htb]
\begin{center}
\includegraphics*[scale=0.5,bb=40 150 600 700 ]
{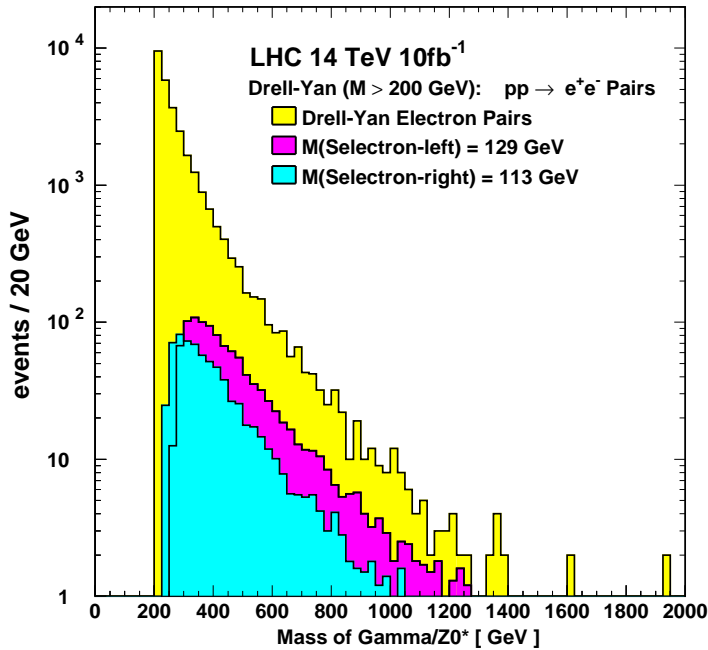}
\caption[fig33]
{Mass distribution of Drell--Yan electron pairs with a mass above 200 GeV 
and for left or right handed selectron pairs at the 
LHC with selectron masses of 129 GeV and 113 GeV respectively
as obtained with PYTHIA~\cite{pythia} and SPYTHIA~\cite{spythia}.}    
\end{center}
\end{figure}
For our example the masses of the left handed and right handed slepton 
where fixed to 129 GeV and 113 GeV respectively. The 
expected mass spectra show the $\beta^{3}$ cross section suppression 
close to threshold.  The larger rate for Drell-Yan pairs produced  
from the left handed virtual $\gamma^{*}, Z^{*}$ system results, despite the
larger mass, into a bigger cross section for left handed sleptons.
This simple relation between slepton mass and cross section
allows precise cross section predictions for slepton pairs,
once the corresponding 
mass spectrum of Drell-Yan lepton pairs has been measured.

As a next step one has to consider the possible slepton decay modes. 
While the right handed slepton can decay only to 
the lightest neutralino and the corresponding lepton 
$\tilde{\ell}^{\pm} \rightarrow \tilde{\chi}^{0}_{1} \ell^{\pm}$, 
several somehow model dependent decay modes, 
are possible for left handed sleptons:
\begin{center}
\begin{itemize}
\item
$\tilde{\ell}^{\pm} \rightarrow \tilde{\chi}^{0}_{1} \ell^{\pm}$
~~~or~~~ 
$\tilde{\ell}^{\pm} \rightarrow \tilde{\chi}^{0}_{2} \ell^{\pm}$
~~~or~~~ 
$\tilde{\ell}^{\pm} \rightarrow \tilde{\chi}^{\pm}_{1} \nu$
\item
$\tilde{\nu} \rightarrow \tilde{\chi}^{0}_{1} \nu~~$
~~~~or~~~ 
$\tilde{\nu} \rightarrow \tilde{\chi}^{\pm}_{1} \ell^{\mp}$
\end{itemize}
\end{center}

The best signature for  
slepton pair production appears to be the two--body 
decay $\tilde{\ell}^{\pm} \rightarrow \tilde{\chi}^{0}_{1} \ell^{\pm}$.
The resulting signature are events with a pair of two isolated 
same flavour leptons with opposite charged and some missing transverse 
momentum. To distinguish such signal events from various 
possible SM backgrounds several kinematic selection
criteria have to be applied. To identify good selection criteria
at the LHC it is useful to start with simplified     
kinematics in the center--of--mass frame.  
The observable leptons, originating from the decays of massive sleptons, 
should show: (1) a characteristic momentum spectrum; (2) should not 
balance their 
momenta and (3) should not be 
back to back. Furthermore, the measurable mass of the event should 
be much smaller than the original center--of--mass energy and the missing mass
should be much larger than zero. 

A possible selection requires thus the possibility to measure 
isolated leptons with good accuracy and to determine indirectly 
the missing energy and momentum from all detectable particles.
An accurate missing energy determination requires 
an almost 4$\pi$ acceptance for all visible particles.
Unfortunately, a realistic experiment has to live with several 
detection gaps especially the ones around the beam pipe. 
Consequently, missing momentum measurements along the beam direction 
are of limited use. In addition, the event kinematics 
at a Hadron Collider are very different from the 
center--of--mass frame. As a result, signal and background events have a
large and unknown momentum component along the beam direction. 
However, variables which exploit the missing transverse
energy and momentum remain very useful.

Furthermore, in contrast to a $e^{+}e^{-}$ collider with a fixed 
dilepton mass, hadron collider searches must consider 
the effects that sleptons pairs are not produced at a fixed $\sqrt{s}$
and show a wide longitudinal momentum range. 
As a result, good selection criteria exploit the differences between signal
and background in the plane transverse to the beam.
Such variables are (1) the transverse momenta of each lepton, 
(2) the opening angle between the two leptons in the plane transverse to the 
beam and (3) the missing transverse momentum. 
The specific choice of cuts depends strongly on the 
studied mass region and the relevant backgrounds.
 
The largest ``irreducible'' background for slepton pair production 
are events with leptonic $W^{\pm}$ decays from  
$W$--pair production $pp \rightarrow WW X$ with 
($\sigma \times BR(WW \rightarrow e^{+} \nu e^{-} \bar{\nu}$) of about 0.8 pb.
Another potentially very large background comes from 
$t \bar{t}$ production with a 
$\sigma \times BR (t\bar{t} \rightarrow 
WW b \bar{b} \rightarrow e^{+} \nu e^{-} \bar{\nu} X)$ 
with a cross section of about 7 pb. This background can be strongly reduced 
by applying a jet veto.
Other potential backgrounds are miss-measured Drell--Yan lepton pairs and 
electrons and muons from leptonic $\tau$ decays produced in the reaction   
$pp \rightarrow \tau \tau$. 
Additional backgrounds might come from  
events of the type $W^{\pm} X$ and $Z^{0} X$  
with one leptonic boson decay and one high $p_{t}$ hadron 
misidentified as an electron or a muon.   
In addition, other unknown sources of new physics, like 
$pp \rightarrow \tilde{\chi}^{+} \tilde{\chi}^{-}$ might also result
in events with two isolated leptons and missing transverse momentum.
These large background cross sections should be compared 
with the much smaller signal cross sections between about 0.2 and 0.02 pb
for the pair production of selectrons with masses 
between 100--200 GeV respectively as shown in Figure 32. 
 
The qualitative ideas discussed above, can now be compared with 
a quantitative simulation of a slepton search with the CMS experiment.
\begin{figure}[htb]
\begin{center}
\includegraphics*[scale=0.65,bb=180 120 415 650 ]
{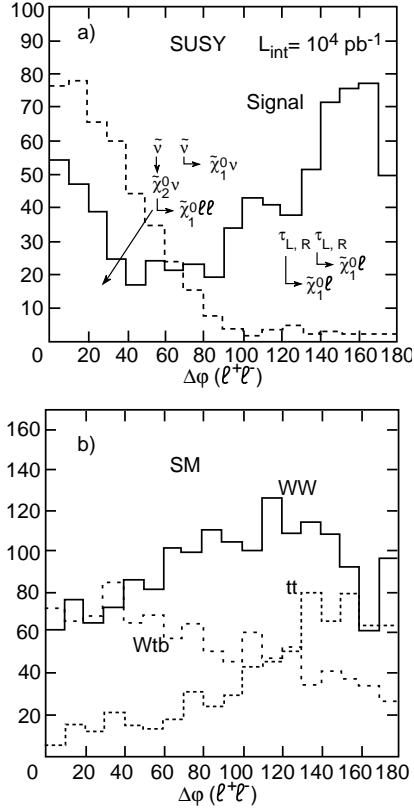}
\caption[fig34]
{Relative azimuthal angle $\phi$ between the two leptons 
for (a) sleptons and other SUSY signals and (b) for the main SM 
backgrounds~\cite{denegri96059}. The proposed cut is $\phi>130^{o}$}
\end{center}
\end{figure}

The analysis selects first events which contain 
a pair of opposite charged electrons or muons and no additional jets.
It is assumed that isolated electrons or muons with 
a minimum transverse momentum of 20 GeV and $|\eta| < 2.5$
can be identified with high efficiency ($ \epsilon > 90$ \% )
and small backgrounds.
One assumes also that jets with a transverse energy above 30 GeV
and $|\eta| < 4.5$ can be identified and vetoed. 
In addition, events from $pp \rightarrow Z \rightarrow \ell^{+}\ell^{-}$ 
are rejected by demanding that 
the invariant mass of the lepton pair should be inconsistent 
with a $Z^{0}$. Additional mass dependent 
selection criteria, specified in table 3, 
are required to improve the potential signal significance.

\begin{table}[htb]
\begin{center}
\begin{tabular}{|c|c|c|c|c|c|}
\hline
$m (\tilde{\ell}^{\pm})$ & $p_{t}^{lepton} $ & $E_{t}(miss) $ &
$\Delta \phi_{\ell^{+}\ell^{-}}$ & S (100 fb$^{-1}$) & 
B (100 fb$^{-1}$) \\
\hline
100 GeV &  $>$ 20 GeV & $>$ 50 GeV & $>$ 130$^{o}$ & $\approx$ 3200 & 
$\approx$ 10000 \\
\hline
200 GeV &  $>$ 50 GeV & $>$ 100 GeV & $<$ 130$^{o}$ & $\approx$ 230 & 
$\approx$ 170 \\
\hline
300 GeV &  $>$ 60 GeV & $>$ 150 GeV & $<$ 130$^{o}$ & $\approx$  67 & 
$\approx$ 45 \\
\hline
400 GeV &  $>$ 60 GeV & $>$ 150 GeV & $<$ 140$^{o}$ & $\approx$  24 & 
$\approx$ 53 \\
\hline
\end{tabular}
\caption{CMS simulation of the charged slepton search
at LHC~\cite{denegri96059}. The proposed selection criteria 
and signal (S) and background (B) rates are given for a luminosity 
of 100 fb$^{-1}$ and a few slepton masses.}
\label{tab:table3}
\end{center}
\end{table}

The lepton pairs from 
$W^{+}W^{-}$ and $t \bar{t}$ events appear to be the dominant backgrounds.
For a slepton mass of about 100 GeV one finds a statistical 
significant signal of $\approx$ 300 events 
above a background of about 1000 events and a luminosity of 10 fb$^{-1}$.
The expected signal and background distributions before the 
$\Delta \phi$ cut are shown in Figure 34a and b. 

The analysis shows that sleptons with masses between 200-300 GeV
can be selected with signal to background ratios of about 1:1. 
The low signal cross section requires however a large luminosity
of at least 30 fb$^{-1}$. 
For larger masses the slepton cross section becomes very small  
and seems to limit the mass reach to about 400 GeV 
with expected signal rates of 24 events and a total expected background    
of about 50 events for a luminosity of about 100 fb$^{-1}$.
In summary, pair production of charged sleptons at the LHC 
appears to be detectable from an excess of events above 
dominant backgrounds from leptonic decays of $W^{+}W^{-}$ 
and $t\bar{t}$ events. The expected mass reach starts from  
about 100 GeV, roughly the final LEPII reach, and is limited 
to masses of about 400 GeV. 
Particular problems are the 
small signal to background ratio for masses below 200 GeV an the
small signal rate for masses above 300 GeV. 

Other slepton signals, like the one from the reaction 
$pp \rightarrow W^{*} \rightarrow \tilde{\ell} \tilde{\nu}$ 
have been studied and were found to be hopeless~\cite{baer94}.

The investigated signature of
a single high $p_{t}$ lepton with large missing 
$E_{t}$ was found to be at least two orders of magnitude 
smaller than the event rate from single $W$'s as shown in Figure 35. 

They concluded further that a possible trilepton signal, from cascade decays 
of the sneutrino 
$\tilde{\nu} \rightarrow \tilde{\chi}^{0}_{2} \nu \rightarrow \ell \ell  
\tilde{\chi}^{0}_{1}$ is much smaller than a possible signal from 
the simultaneous produced trilepton events of the type
$\tilde{\chi}^{0}_{2}\tilde{\chi}^{\pm}_{1}$ as described in the next section. 

\begin{figure}[htb]
\begin{center}
\includegraphics*[scale=0.5,bb=40 350 600 800 ]
{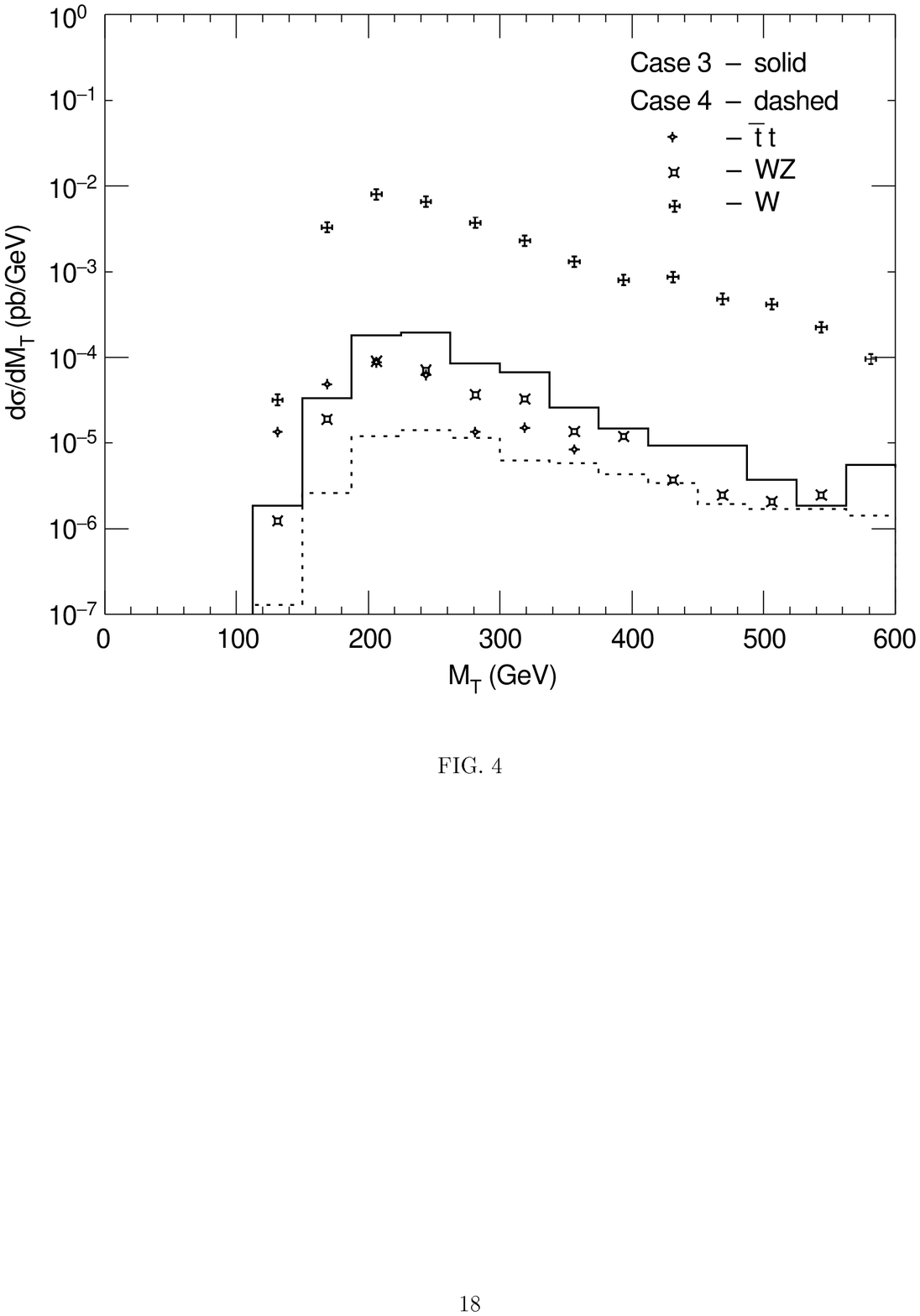}
\caption[fig35]
{Reconstructed transverse mass of single high $p_{t}$ lepton events
for signal events of the type 
$pp \rightarrow W^{*} \rightarrow \tilde{\ell}^{\pm} \tilde{\nu}
\rightarrow \ell^{\pm} \nu \tilde{\chi}_{1}^{0}\tilde{\chi}_{1}^{0}$ 
and various backgrounds. The studied slepton masses were
100 GeV (case 3) and 200 GeV (case 4)~\cite{baer94}.}
\end{center}
\end{figure}
\clearpage

\subsection{Chargino-neutralino searches; the Trilepton signature}

In analogy to the reaction 
$q_{i}\bar{q}_{j} \rightarrow Z^{0} W^{\pm}$, one might expect the production of 
$q_{i}\bar{q}_{j} \rightarrow \tilde{\chi}^{0}_{2}\tilde{\chi}^{\pm}_{1}$ 
events.
Such events can be detected from an analysis of events with 
three isolated high $p_{t}$ leptons and large missing transverse energy. 
The potential of this trilepton signature at hadron colliders 
like the LHC has been described in several phenomenological 
studies~\cite{trileptons}. It was found, that trilepton events 
with jets should be rejected to
distinguish signal events from SM and SUSY backgrounds.

After the removal of jet events, the only remaining relevant 
background comes from leptonic decays of $WZ$ events.
Potential backgrounds from dilepton events like 
$W^{+}W^{-} \rightarrow \ell^{+} \nu \ell^{-} \bar{\nu}$ and 
hadrons misidentified as electrons or muons 
are usually assumed to be negligible.
Depending on the analysed SUSY mass range, the background from 
leptonic decays of $WZ$ events, in contrast to a 
potential signal, will show a $Z^{0}$ mass peak in the dilepton 
spectrum. 

This signature is also used at the Tevatron. 
Estimates for RUN II (with a few fb$^{-1}$) hope for a 
$\tilde{\chi}^{0}_{2}\tilde{\chi}^{\pm}_{1}$ mass sensitivity  
of up to 130 GeV, which might be improved further to about 210 GeV
with RUN III (with 20--30 fb$^{-1}$)~\cite{tev2000susy}. 
These estimates assume 
that a background cross section of less than 0.5 fb.
This number can be compared to recent searches for trilepton events, 
optimised for masses of $\approx$ 80 GeV, from 
CDF~\cite{cdftrileptons}. 
Table 4, shows the current CDF background estimates for various applied cuts 
resulting in a final background cross section about 10 fb.
\begin{table}[htb]
\begin{center}
\begin{tabular}{|c|c|c|c|}
\hline
Cut           & observed  & SM Background & MSSM MC \\
Cut           & Events    & Expectation & $M(\tilde{\chi}_{1}^{\pm})=
M(\tilde{\chi}_{2}^{0})$ =70 GeV \\
\hline
Dilepton data                & 3270488 &  &   \\
\hline
Trilepton data               &     59  &  &  \\
\hline
Lepton Isolation             &     23  &  &   \\
\hline
$\Delta R_{\ell \ell} > 0.4$ &      9  &  &   \\
\hline
$\Delta \phi_{\ell\ell}<170^{o}$ &  8 & 9.6$\pm$1.5 & 6.2 $\pm$0.6  \\
\hline
$J/\Psi, \Upsilon, Z$ removal &     6 & 6.6$\pm$1.1 & 5.5 $\pm$0.5  \\
\hline
missing $E_{t}(miss) > 15 $  &      0 & 1.0$\pm$0.2 & 4.5 $\pm$0.4  \\
\hline
\end{tabular}
\caption{Results from a recent trilepton 
analysis from CDF with a dataset of $\approx$ 
100 pb$^{-1}$~\cite{cdftrileptons}.
The number of observed events shows good agreement with various SM 
background sources.}
\label{tab:table4}
\end{center}
\end{table}

A recent CMS simulation~\cite{cms97007} 
of the trilepton signal at the LHC proceeds as 
follows:

\begin{itemize}
\item
Events should contain three isolated leptons, all with $p_{t} > 15$ GeV
and $|\eta| < 2.5$ and no jets. 
\item
The missing transverse energy should exceed 15 GeV.
\item
The possible same flavour dilepton mass combinations should be 
inconsistent with a $Z^{0}$ decay.  
\end{itemize}

Depending on the studied mass range, additional or harder 
selection criteria are applied. Figure 36 shows the expected 
missing transverse energy distribution for trilepton signal events, 
with different choices of $m_{0}$ and $m_{1/2}$, and for 
background events.
Table 5 gives a few numbers for signal and backgrounds from the CMS
study and different SUSY masses. 
\begin{table}[htb]
\begin{center}
\begin{tabular}{|c|c|c|c|}
\hline
$M_{1/2} \approx M(\tilde{\chi}_{1}^{\pm})$ & $\sigma \times$ BR (trileptons)  
& Signal & SM Background  \\
          &              & (100 fb$^{-1}$) & (100 fb$^{-1}$) \\
\hline
100         & 0.8~~-1.3 pb & 4000--8000    &  900   \\
\hline
150         & 0.04-0.08 pb &    300-600    & 1000   \\
\hline
200         & 0.01-0.02 pb &  80-120       &  700   \\
\hline
300-400     & 0.01-0.02 pb &      50       &  100   \\
\hline
\end{tabular}
\end{center}
\caption{Expected signal and background numbers from a CMS trilepton 
study with different choices of $m_{0}$ and $m_{1/2}$ 
with $\tan \beta = 2$ and negative $\mu$~\cite{cms97007}.}
\label{tab:table5}
\vspace{0.4cm}
\end{table}
\begin{figure}[htb]
\begin{center}
\includegraphics*[scale=0.6,bb=40 100 600 750 ]
{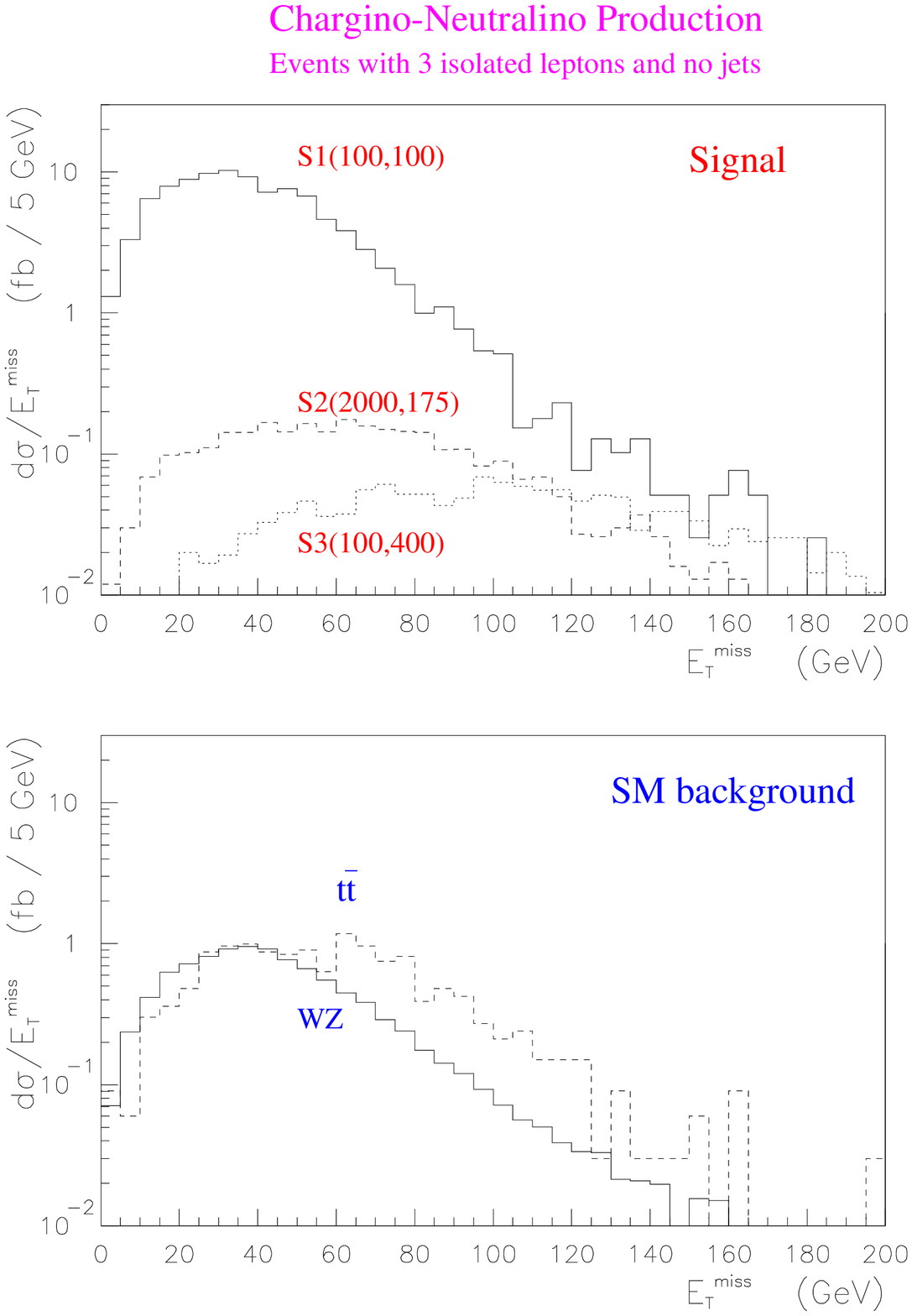}
\caption[fig36]
{Missing transverse energy distribution for 
trilepton events without jets from 
$pp \rightarrow \tilde{\chi}^{0}_{2}\tilde{\chi}^{\pm}_{1}$ signal events  
and background~\cite{cms97007}. The numbers in brackets 
give the used $m_{0}$ and $m_{1/2}$ values.}
\end{center}
\end{figure}

In all cases one finds signal efficiencies of $\approx$ 5\%. 
The best results are obtained for masses close to 100 GeV 
with expected signal rates of $\approx 40$ above a background of 10 events
per 1 fb$^{-1}$ of luminosity. The signal rate drops quickly
for higher masses and much higher luminosities are required 
to establish potential signals up to masses of at most 300-400 GeV. 
Furthermore, for some $m_{0}, m_{1/2}$ mass regions, the estimated 
leptonic branching ratios are very small and result in
signal to background ratios smaller than 0.2.
We conclude that the LHC experiments 
can measure excellent trileptons signals 
in mass and parameter regions where the discovery has most probably been made 
at the upgraded Tevatron.
Such high statistics signals will allow 
some detailed SUSY studies as described in section 7.5.
For chargino/neutralino masses above $\approx$ 200 GeV significant 
signals require at least 30 fb$^{-1}$ and a very good understanding of 
possible backgrounds. However, as will become clear from 
the next section, cascade decays of squarks and gluinos 
should provide a much better sensitivity for 
charginos and neutralinos with higher masses.

\clearpage

\subsection{Squark and Gluino searches; the showcase for a hadron collider}

The discussion in the previous sections covered the potential 
to study non-hadronic interacting SUSY particles with 
relatively small cross section. We now turn the discussion to the 
search for squarks and gluinos with large couplings to quarks and 
gluons.  
The cross section for strongly interacting particles at hadron colliders
like the LHC are quite large.
For example the pair production cross section of 
squarks and gluinos with a mass of $\approx$ 1 TeV has been estimated 
to be as large as 1 pb resulting in 
10$^{4}$ produced SUSY events for one ``low'' luminosity 
LHC year. Such high rates,
combined with the possibility to observe many different 
decay modes, is considered often 
as a ``raison d`\^{e}tre'' for the LHC.

Depending on the SUSY model parameters, a large variety of 
massive squark and gluino decay channels and signatures might exist.
A complete search analysis for squarks and gluons 
at the LHC should consider the various 
signatures resulting from the following decay channels. 
\begin{itemize}
\item $\tilde{g} \rightarrow \tilde{q} q$ and perhaps 
$\tilde{g} \rightarrow \tilde{t} t$ 
\item $\tilde{q} \rightarrow \tilde{\chi}^{0}_{1} q$ ~~~or~~~
$\tilde{q} \rightarrow \tilde{\chi}^{0}_{2} q$  ~~~or~~~
$\tilde{q} \rightarrow \tilde{\chi}^{\pm}_{1} q$
\item $\tilde{\chi}^{0}_{2} \rightarrow \tilde{\chi}^{0}_{1} \ell^{+}\ell^{-}$ 
 ~~~or~~~ $\tilde{\chi}^{0}_{2} \rightarrow \tilde{\chi}^{0}_{1} Z^{0} $  
~~~or~~~ $\tilde{\chi}^{0}_{2} \rightarrow \tilde{\chi}^{0}_{1} h^{0}$
\item $\tilde{\chi}^{\pm}_{1} \rightarrow \tilde{\chi}^{0}_{1} \ell^{\pm}\nu$
~~~or~~~$\tilde{\chi}^{\pm}_{1} \rightarrow \tilde{\chi}^{0}_{1} W^{\pm}$.
\end{itemize}  
The various decay channels 
can be separated into at least three distinct event signatures.
\begin{itemize}
\item Multi--jets plus missing transverse energy. These events 
should be spherical in the plane transverse to the beam.
\item Multi--jets plus missing transverse energy plus n(=1,2,3,4) 
isolated high $p_{t}$ leptons. These leptons originate 
from cascade decays of charginos and neutralinos.
\item
Multi--jets plus missing transverse energy plus same charge leptons pairs. 
Such events can be produced in events of the type 
$\tilde{g}\tilde{g} \rightarrow \tilde{u} \bar{u} \tilde{d} \bar{d}$
with subsequent decays of the squarks to 
$\tilde{u} \rightarrow \tilde{\chi}^{+}_{1} d$ 
and $\tilde{d} \rightarrow \tilde{\chi}^{+}_{1} u$ followed by
leptonic chargino decays  
$\tilde{\chi}^{+}_{1} \rightarrow \tilde{\chi}^{0}_{1} \ell^{+} \nu$.  
\end{itemize}
It is easy to imagine that the observation and detailed analysis
of the different types of squark and gluino signatures might allow 
to measure some of the many MSSM parameters.

The above signatures have already been investigated with the data from 
the Tevatron RUN I. The negative searches gave 
mass limits for squarks and gluinos 
as high as $\approx 200$ GeV. The estimated 
5--sigma sensitivity for RUN II and RUN III 
reaches values as high as 350--400 GeV.
More details about the considered signal and backgrounds 
can be found from the TeV2000 studies~\cite{tev2000susy} and 
the ongoing Tevatron workshop.

A simplified search strategy for squarks and gluinos 
at the LHC would study 
jet events with large visible transverse mass and some missing 
transverse energy. Such events can then be classified according 
to the number of isolated high $p_{t}$ leptons.
Once an excess above 
SM backgrounds is observed for any possible combination of the 
transverse energy spectra, one would try to explain the 
observed types of exotic events and their cross section(s) for
different SUSY $\tilde{g}, \tilde{q}$ masses and decay modes and models.
An interesting approach to such a multi--parameter analysis uses some 
simplified selection variables. For example one could use
the number of observed jets and leptons and their transverse energy, their
mass and the missing transverse energy to separate signal and 
backgrounds. Such an approach has been used to perform a ``complete''
systematic study of $\tilde{g}$ and $\tilde{q}$ decays~\cite{baer95}. 
The proposed variable $E_{t}^{c}$ is the value of the smallest of 
$E_{t}$(miss), $E_{t}$(jet1), $E_{t}$(jet2). The events are further 
separated into the number of isolated leptons. Events with lepton pairs 
are divided into same sign (charge) pairs (SS) and opposite 
charged pairs (OS). 
Signal and background distributions for various squark and gluinos 
masses, obtained with such an approach are shown in Figure 37.  
\begin{figure}[htb]
\begin{center}
\includegraphics*[scale=0.6,bb=40 200 600 700 ]
{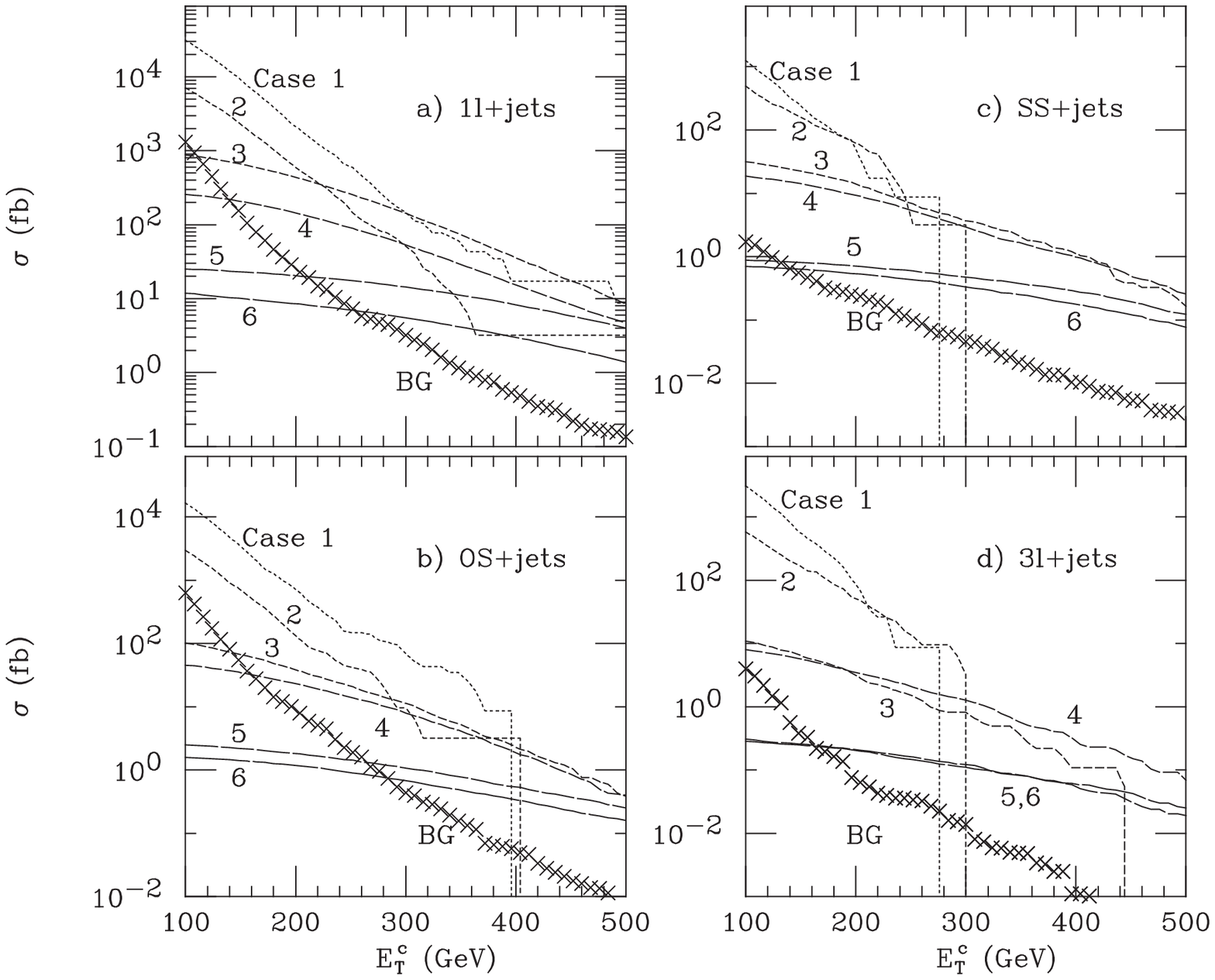}
\caption[fig37]
{Expected $E_{t}^{c}$ distributions for SUSY signal 
and background processes at the LHC and realistic experimental cuts
for $\tan \beta = 2$ and $\mu < 0$~\cite{baer95}.
The different cases are for: 
(1) $m_{\tilde{g}}$= 290 GeV and $m_{\tilde{q}}$= 270 GeV;  
(2) $m_{\tilde{g}}$= 310 GeV and $m_{\tilde{q}}$= 460 GeV;  
(3) $m_{\tilde{g}}$= 770 GeV and $m_{\tilde{q}}$= 720 GeV;
(4) $m_{\tilde{g}}$= 830 GeV and $m_{\tilde{q}}$= 1350 GeV;  
(5) $m_{\tilde{g}}$= 1400 GeV and $m_{\tilde{q}}$= 1300 GeV;
(6) $m_{\tilde{g}}$= 1300 GeV and $m_{\tilde{q}}$= 2200 GeV.}
\end{center}
\end{figure}

According to this classification the number of expected 
signal events can be compared with the various SM background processes.
The largest and most difficult backgrounds originate mainly from 
$W+$jet(s), $Z+$jet(s) and $t\bar{t}$ events. 
Using this approach, very encouraging
signal to background ratios, combined with quite large signal cross sections 
are obtainable for a large range of squark and gluino masses.
The simulation results of such studies indicate, as shown in Figure 38,
that the LHC experiments are sensitive to 
squark and gluinos masses up to masses of about 2 TeV and 100 fb$^{-1}$. 
\begin{figure}[htb]
\begin{center}
\includegraphics*[scale=0.5,bb=40 50 600 580 ]
{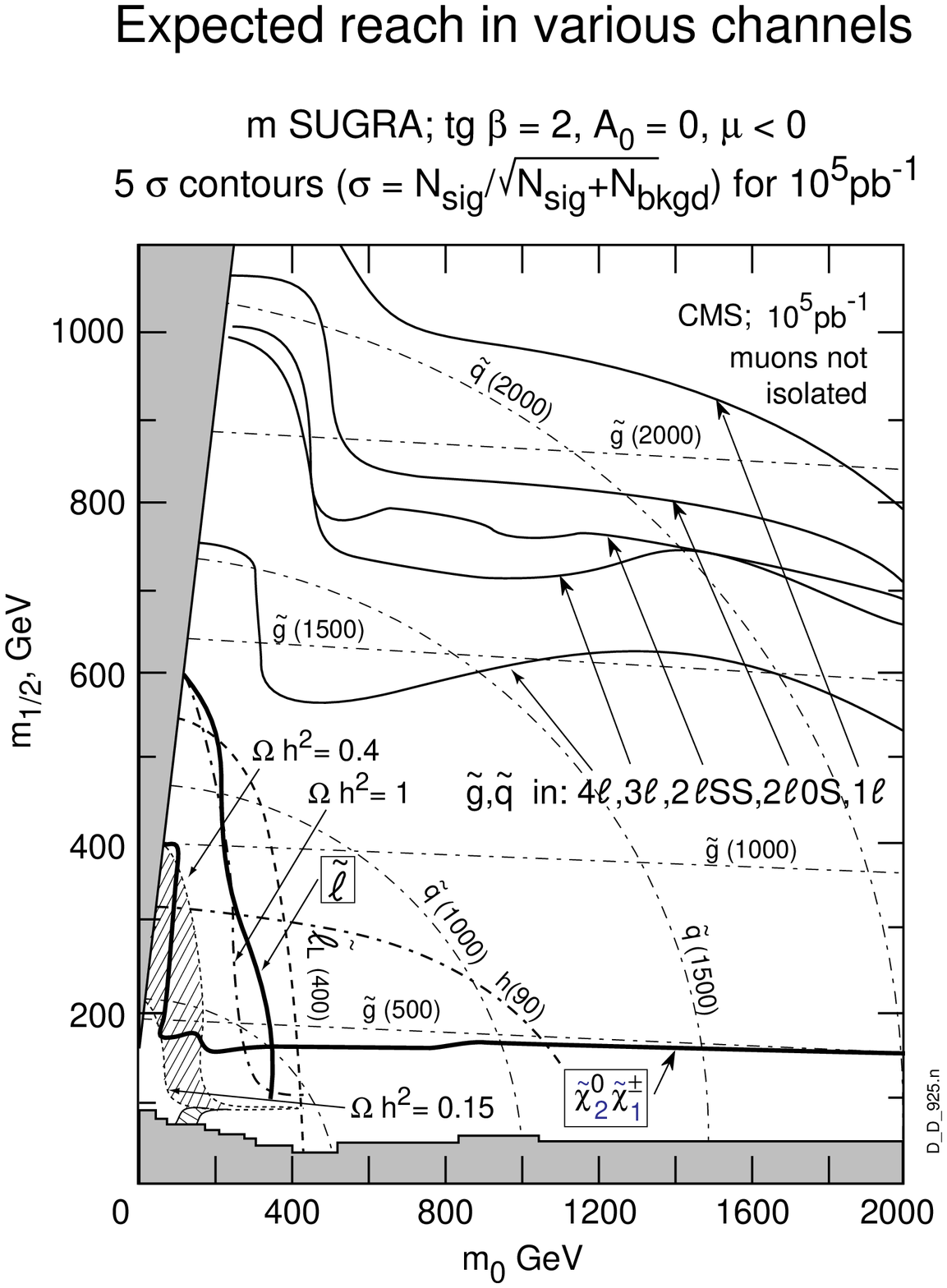}
\caption[fig38]
{Expected ultimate (L=100 fb$^{-1}$) CMS sensitivity 
for squarks and gluinos, sleptons and for 
$\tilde{\chi}^{0}_{2}\tilde{\chi}^{\pm}_{1}$
in the $m_{0}-m_{1/2}$ plane.
The different full lines show the 
expected 5 sigma signal, estimated from S/$\sqrt{S+B_{SM}}$, 
coverage domain for the various signatures
with isolated high $P_{t}$ leptons~\cite{cms98006}. 
The dashed lines indicate the 
corresponding squark and gluino masses.}
\end{center}
\end{figure}

Figure 38 indicates further, that detailed studies of branching ratios 
are possible up to squark or gluino masses of about 1.5 TeV, where 
significant signals can be observed with many different channels.   
Another consequence of the expected large signal cross sections
is the possibility that the ``first day'' LHC luminosity 
$\approx$ 100 pb$^{-1}$ should be sufficient to discover 
squarks and gluinos up to masses of about 600--700 GeV, 
well beyond even the most optimistic 
Tevatron Run III mass range.

Having this exciting discovery potential for squarks and gluinos 
with many different channels,
one might want to know the ``discovery'' or simply the ``best'' channel.
Such a question is unfortunately not easy to answer. 
All potential signals depend strongly on a good understanding  
of various backgrounds and thus the detector systematics. 
Especially the requirements of high efficiency lepton identification 
and a good missing transverse energy measurement demand for a
``perfect'' working and understood detector. 
This requirement of a good understanding 
of complicated ``monster'' like experiments needs thus some time
and is in contradiction with the ``first day'' discovery potential.
We conclude that the best discovery signature is not yet known,
but should be one which is extremely robust and simple and should  
not depend on too sophisticated detector elements and their resolutions.
  
\subsection{SUSY discovered, what can be studied at the LHC?}

Our discussion of the LHC SUSY discovery potential 
has demonstrated the sensitivity of 
the proposed ATLAS and CMS experiments. Being convinced of this 
discovery potential, one certainly wants to know if 
``the discovery'' is consistent with SUPERSYMMETRY and
if some of the many SUSY parameters can be measured.   

To answer the above question one should try find many
SUSY particles and measure their decay patterns as accurately as 
possible. The sensitivity 
of direct exclusive SUSY particle production at the LHC has 
demonstrated the various possibilities and 
cross section limitations for weakly produced SUSY particles.
\begin{figure}[htb]
\begin{center}
\includegraphics*[scale=0.6,bb=-26 -130 373 259 height=1. cm,width=8.cm]
{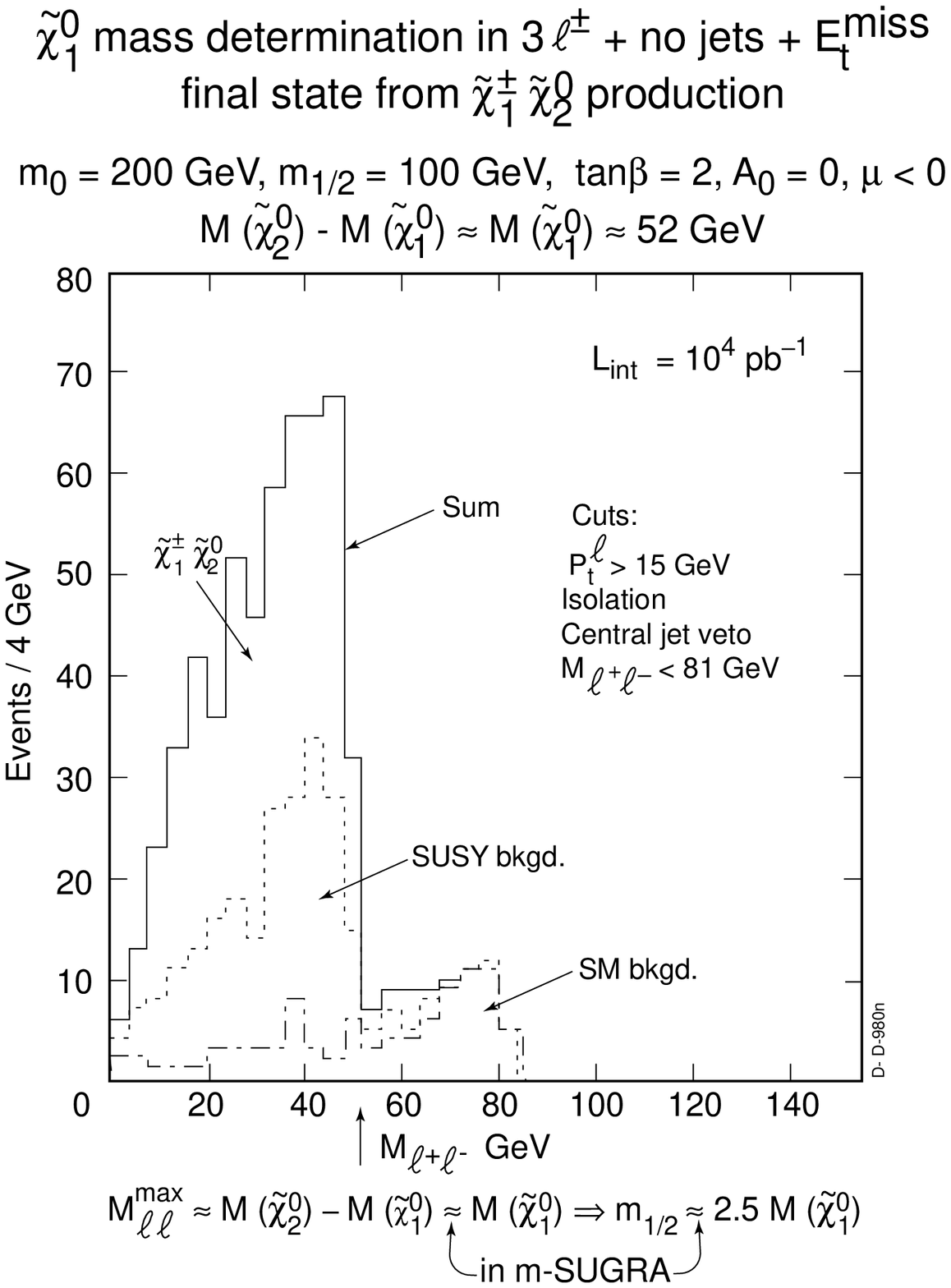}
\caption[fig39]
{Expected CMS dilepton mass distribution with L=10 fb$^{-1}$ and
trilepton events from 
$\tilde{\chi}^{0}_{2}\tilde{\chi}^{\pm}_{1}$~\cite{cms98006}.
The edge in the distribution at about 
50 GeV corresponds to the kinematic 
limit in the decay 
$\tilde{\chi}^{0}_{2} \rightarrow \ell^{+}\ell^{-}\tilde{\chi}^{0}_{1}$
and is thus sensitive to the mass difference 
between $\tilde{\chi}^{0}_{2}$ and $\tilde{\chi}^{0}_{1}$.}
\end{center}
\end{figure}
 
Nevertheless, one finds that the
production and decays of $\tilde{\chi}^{0}_{2}\tilde{\chi}^{\pm}_{1}$
provide good rates for masses below 200 GeV and should allow, 
as indicated in Figure 39, to measure accurately the dilepton mass 
distribution and their relative $p_{t}$ spectra.
The mass distribution and especially the edges 
in the mass distribution are sensitive
to the mass difference between the two neutralinos.
Depending on the used MSUGRA parameters one finds that the 
$\tilde{\chi}^{0}_{2}$ can have two or three body decays. 
The relative $p_{t}$ spectra of the two leptons
can be used to distinguish the two possibilities.

\begin{figure}[htb]
\begin{center}
\includegraphics*[scale=0.45,bb=100 200 480 700]
{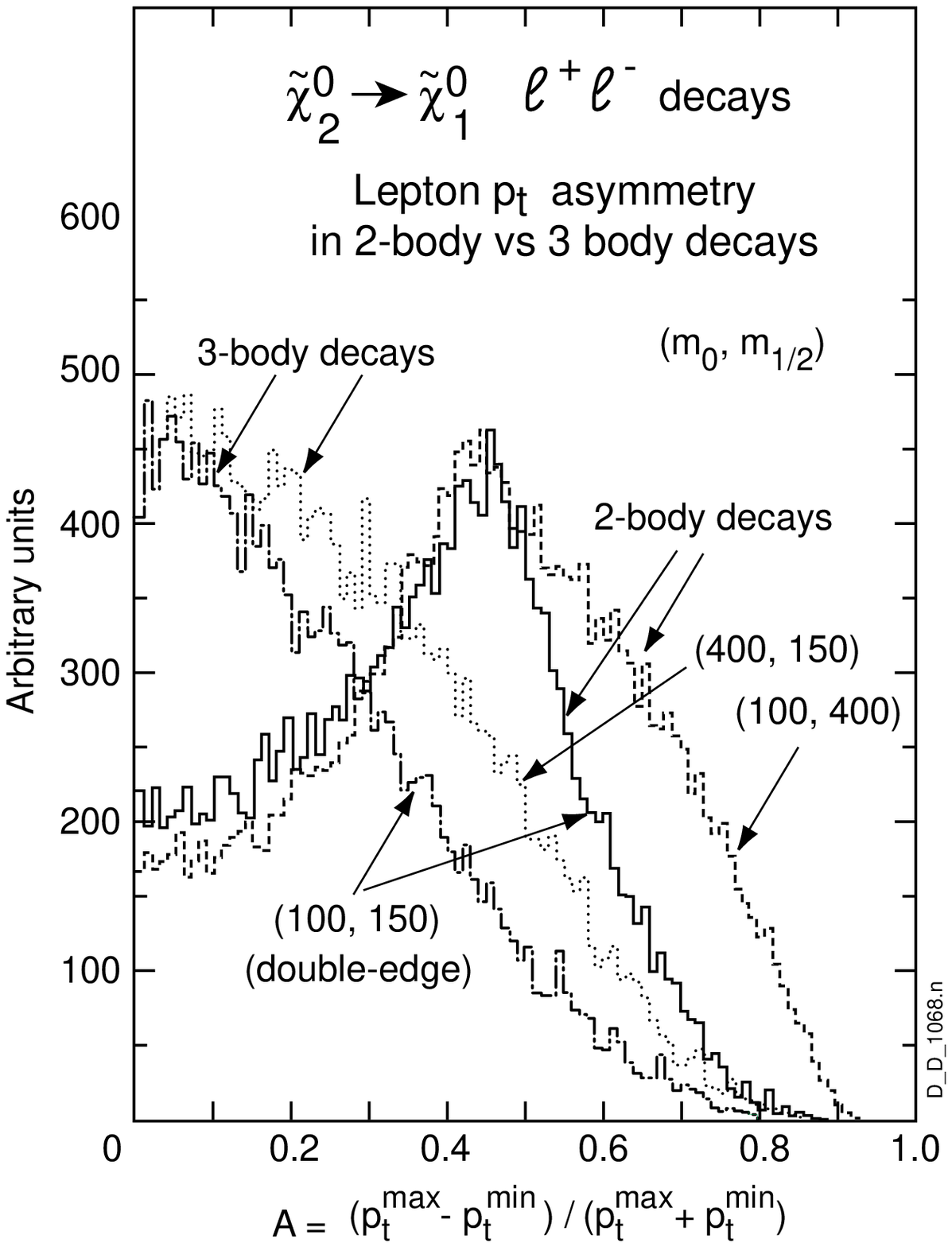}
\caption[fig40]
{Lepton $p_{t}$ asymmetry distributions A in $\tilde{\chi}^{0}_{2}$ decays,
for different choices of $m_{0}$ and $m_{1/2}$ 
and dilepton masses below and above the edge at $\approx$
50 GeV as shown in figure 39~\cite{cms98006}.}
\end{center}
\end{figure}
 
Figure 40 shows the distribution for the variable $A$, defined as 
$A =(p_{t}^{max} -p_{t}^{min}) /(p_{t}^{max} + p_{t}^{min})$ in 
trilepton events and dilepton masses below and above 50 GeV. 
This asymmetry variable originates from 
early investigations of $\tau$ decays~\cite{taudis} where it   
allowed to demonstrate that the leptonic $\tau$ decays 
$\tau \rightarrow \ell \nu \nu$ are three body decays. 
 
In contrast to the rate limitations of weakly produced SUSY 
particles at the LHC, detailed 
studies of the clean squark and gluino events are expected 
to reveal much more information.  
In detail, one finds that the large rate for many 
distinct event channels allows to measure masses and mass 
ratios for several SUSY particles, which are possibly 
being produced in cascade decays of squarks and gluons. 
Many of these ideas have been discussed at a 1996 
CERN Workshop~\cite{cernth96}.
Especially interesting appears to be the idea that 
the $h^{0}$ might be produced and detected in the decay chain 
$\tilde{\chi}^{0}_{2} \rightarrow \tilde{\chi}^{0}_{1} h^{0} $ and 
$h^{0} \rightarrow b \bar{b}$. The simulated mass distribution 
for $b \bar{b}$ jets 
in events with large missing transverse energy is shown 
in Figure 41. Clear Higgs mass peaks above background are found 
for various choices of $\tan \beta$ and $m_{0}, m_{1/2}$.  

\begin{figure}[htb]
\begin{center}
\includegraphics*[scale=1.2,bb=0 300 550 750 height=10. cm,width=14.cm]
{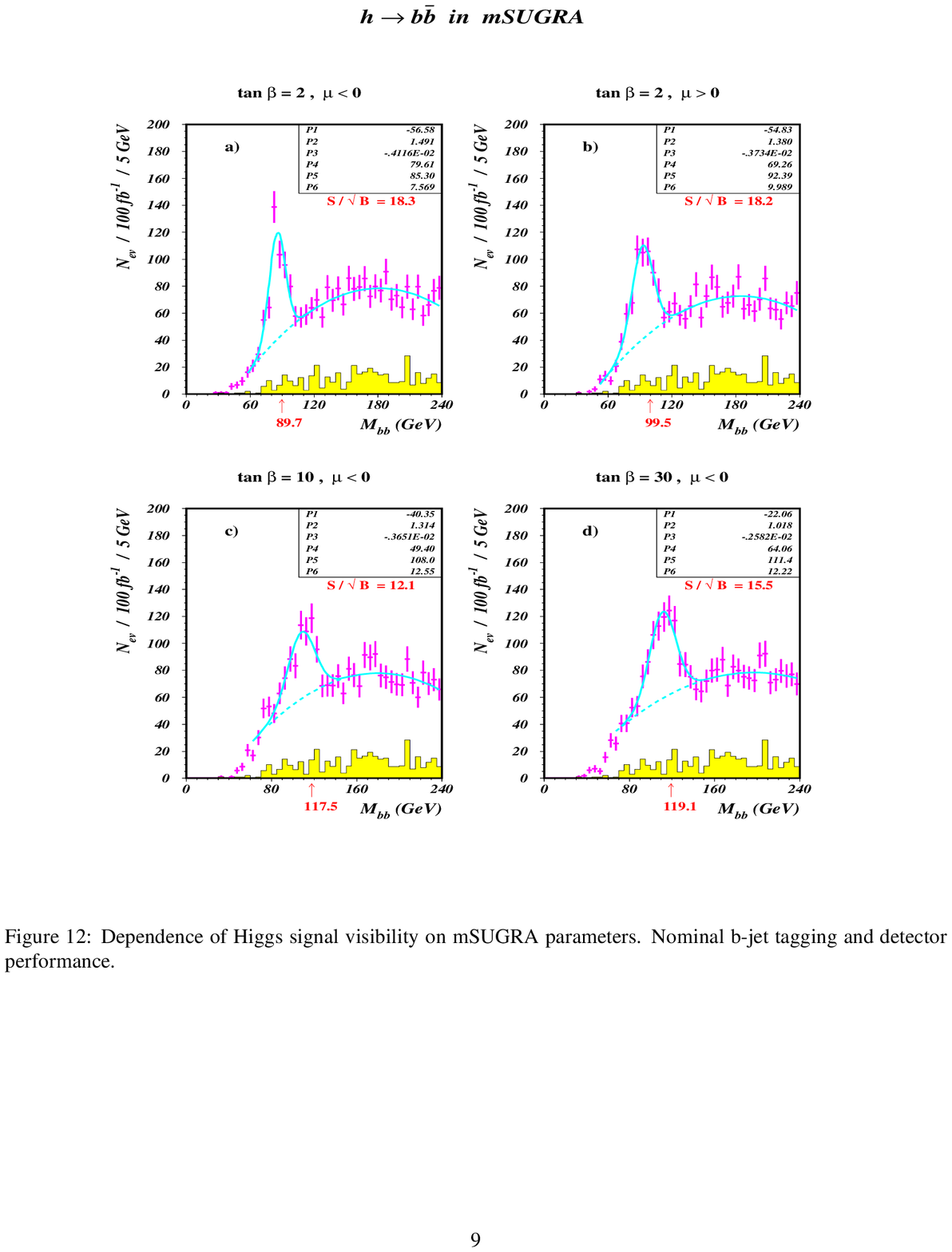}
\caption[fig41]
{Possible inclusive Higgs signals in squark and gluino events,
reconstructed from the invariant mass of $h \rightarrow b \bar{b}$ 
with CMS and a luminosity of  
L=100 fb$^{-1}$~\cite{cms98006}.}
\end{center}
\end{figure}
 
An interesting approach to determine a 
SUSY mass scale has been suggested in a recent ATLAS 
study~\cite{atlas107}.
The idea is to define an effective transverse event mass,  
using the scalar $p_{t}$ sum of the jets with the largest transverse energy
plus the missing transverse energy of the event. 
One finds that this effective mass shows a reasonable linear 
relation to an underlying SUSY mass, defined 
as the minimum of the squark or the gluino mass.
While this idea appears to be very attractive within the MSUGRA
model, the validity of the proposed relation in 
more general SUSY models is not known. 

In addition to the above SUSY studies, 
one would like to to get answers to questions like:
\begin{itemize}
\item
What are the branching ratios of various SUSY particles?
\item
Is the accuracy of the various channels sufficient to determine 
the spin of the new particles?  
\item
Do the data allow to differentiate between specific SUSY models?
\item 
Can one find evidence for CP violation in SUSY decays?
\end{itemize} 
At least some answer might be obtained from future
detailed studies of the various decay chains.
We thus conclude this section with the hope that some 
SUSY enthusiasts will try eventually to answer some of these questions
using the expected performance of the LHC and its planned experiments.

\section {Putting it all together..}

We have discussed the various proposed search/discovery strategies 
for the Higgs, supersymmetry and other exotica at LEP II and  
future hadron colliders with a focus on the LHC. 
Knowing that the LHC experiments will not provide any new 
physics before the year 2005 some time for analysis preparation is left.

The most important aspect for the coming years is the 
confrontation of the assumed detector performance 
with reality. First of all, the   
two huge experiments have to be build according to the 
proposed designs. Any of todays physics case studies has thus to be 
kept realistic and should also reflect the expected experimental 
and theoretical knowledge at day 0. 

For example, LHC studies which show 
a wonderful method on how to discover a SM Higgs boson 
with a mass of 50 GeV can be considered as a waste of time.
A similar judgement could be applied to some ``hard work'' studies, 
which require unrealistic experiments with non existing systematics. 
We do not follow such simple judgement on ``first studies''
as these studies indicate very often the steps towards a realistic 
strategy. 

Realistic and relevant LHC studies should thus in any case   
be aware of possible constraints from near future experiments.
Examples of such possible constraints come from the LEPII Higgs 
search and the new CLEO result on the branching 
ratio for $b \rightarrow s \gamma$, being 
$(3.15 \pm 0.35 (stat.) \pm 0.32 (exp. syst.) \pm 0.26(th)) 
\times 10^{-4}$~\cite{btosgamma}. 
The near future high luminosity b--factory experiments
should allow to decrease the current error by at least a factor of 4.  
The current negative Higgs search results from LEP II
exclude almost the Higgs sector of the 
MSSM model with no mixing and $\tan \beta < 2-3$. 
During the next two years the LEP II sensitivity should increase to  
$\tan \beta$ values of $\leq 4$.

\begin{figure}[htb]
\begin{center}
\includegraphics*[scale=1.,bb=0 100 600 730 height=10. cm,width=12.cm]
{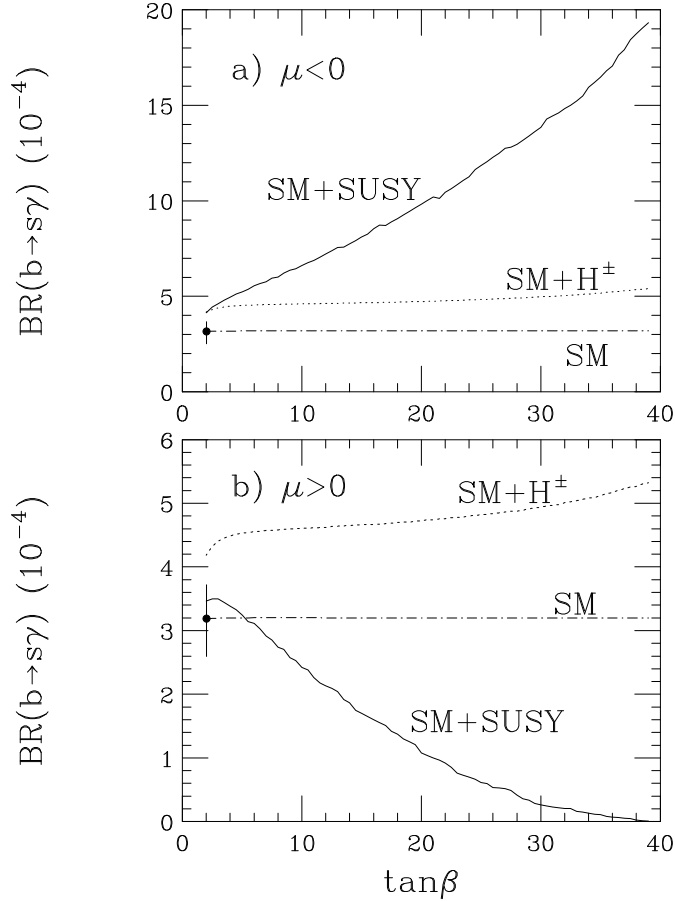}
\caption[fig42]
{Expected branching ratio for $b \rightarrow s \gamma$ 
for the SM and its supersymmetric extension. The  
branching ratio is shown as a function of $\tan \beta$ 
and a negative or positive value of $\mu$~\cite{baer97bsg}.}
\end{center}
\end{figure}
 
Following some theoretical calculations~\cite{baer97bsg}, the new 
CLEO $b \rightarrow s \gamma$ result, as shown in Figure 42
appears to exclude a wide MSUGRA parameter range. In particular, one finds
that the MSUGRA parameter $\mu$ has to be positive and that 
values of $\tan \beta > 10$ are essentially inconsistent 
with the existing data. 

Following strictly the assumed theoretical implications of 
the $b \rightarrow s \gamma$ branching ratios and the LEPII Higgs searches 
one might find that either the MSUGRA model is excluded,
or in case a Higgs is found at LEPII  
that $\mu$ is positive and $\tan \beta$ is 
somewhere between 2 and 4. 

If the Higgs will not be discovered at LEPII and 
the near future $b \rightarrow s \gamma$ branching ratio results 
might give values between $2-3 \times 10^{-4}$, the  
MSUGRA believers should focus on  
$\tan \beta$ values between 4 and 10. Unfortunately, this $\tan \beta$
range appears to be a difficult MSSM Higgs search area 
at the LHC as can be seen from Figure 27.    
In contrast, a branching ratio result 
between $3.5-4 \times 10^{-4}$ could exclude MSUGRA.
We thus conclude this section with the remark that 
one should think twice before a too large effort is put into very  
detailed simulation studies as the possible results might be 
proven irrelevant even before such studies are completed!

\section{Summaries}
Past discoveries of new particles and phenomena have  
demonstrated undoubtfully that searches are  
the most exciting domain of new high energy colliders experiments.
In contrast, the success of the Standard Model of electro--weak 
interactions has put the {\bf Searchers for the New} 
into an esoteric corner group of experimentalists.
Allowed exceptions are however the search for  
the SM Higgs and perhaps the tolerated searches for the 
MSSM Higgs bosons and for SUSY particles within the mSUGRA frame.
Almost unavoidable our guide on ``How to do Searches'' follows todays
theoretical guidance and fashions. 
This restriction is however not too dramatic as the discussed 
methods to isolate the various new signatures are general enough  
to cover even tomorrows fashions.

Assuming that the existing and planned LEP II, Tevatron and LHC 
experiments and colliders behave 
as expected, a large domain of unexploited physics territory 
will be investigated during the coming 10--20 years. 

While the LEPII experiments have reached 
almost the kinematical limit, the future high luminosity 
running of the Tevatron might improve the existing sensitivity for the mass
range of new particles by a factor of about 1.5--2 compared to todays
mass limits. The LHC experiments should increase this mass window 
by another factor of about 6.
 
Starting with the SM Higgs, we find that 
the proposed search methods at LEPII and the LHC
are robust and should lead to the discovery of the SM Higgs. 
We have also studied the question of a potential 
SM Higgs window at RUNIII of the Tevatron (TeV33). Our investigation 
shows that large factors are still missing before one could claim 
that there is a Higgs window 
at the upgraded TeV2000 experiments and a luminosity of 30 fb$^{-1}$. 
We thus disagrees with the optimistic scenarios 
discussed in the literature.

Our discussion of Supersymmetry searches is split into the  
search for the MSSM Higgs sector and for the direct search for SUSY
particles. The proposed and published search methods demonstrate that 
an unambiguous prove of SUPERSYMMETRY can essentially only be obtained 
from the discovery of at least one of the many ``sfermions'' or ``inos''. 
One finds that direct SUSY discovery chances depend mainly on the 
available center of mass energy. 
Only marginal improvements can be expected from the future LEPII
running, which can increase todays chargino sensitivity by perhaps 
another 5 GeV. Nevertheless, even todays limit of $\approx$ 95 GeV
provides strong constraints ($m(\tilde{g} > 270$ GeV) 
on the lowest possible gluino mass. Optimistic studies 
assume that the LEPII charginos range can be improved with 
a few fb$^{-1}$ (the TeV Run II) to masses 
of about 130 GeV and slightly higher for the Tevatron RunIII. 
Thus, negative chargino searches at the RunII exclude 
essentially any mSUGRA possibility to detect 
gluinos with Run III 
(L$>$ 10 fb$^{-1}$) where optimistic studies expect to reach 
a sensitivity to masses between 300--400 GeV. 

In contrast, the squark and gluino searches 
at the LHC are expected to be sensitive up to masses of about 2 TeV.
The LHC experiments should thus be able to 
increase this potential mass window by another factor of about six. 
In addition, the detectable LHC squark and gluino cross sections, even for 
moderate masses well above any possible Tevatron limit, are huge.
Consequently, LHC SUSY discoveries might be possible with a luminosity of 
a few 100 pb$^{-1}$ only, obtainable almost 
immediately at the LHC switch on. Such excellent perspectives
have to be matched however with an almost perfectly working 
full detector and the accurate knowledge of all SM background 
processes. In addition, a well prepared search should consider
a large variety of models and the resulting possible signatures.

\begin{figure}[htb]
\begin{center}
\epsfig{file=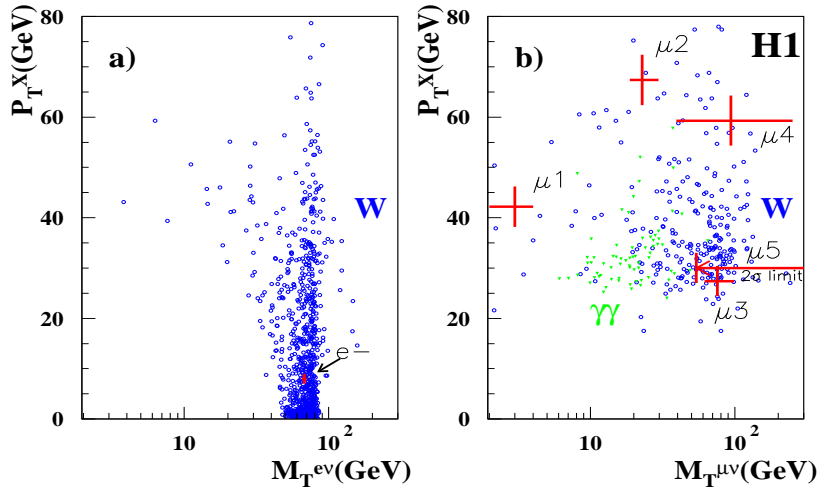,
height=7.5 cm,width=12.cm}
\caption[fig43]
{Observed correlation between the transverse mass of the $\ell \nu$ system 
and the missing transverse momentum for 
the $e-X$ and $\mu-X$ events 
in the data and in the SM Monte Carlo from the 
H1 collaboration~\cite{h1exotics}.}
\end{center}
\end{figure}
\begin{figure}[htb]
\begin{center}
\includegraphics*[scale=0.7,bb=0 100 600 730, height=10. cm,width=10.cm]
{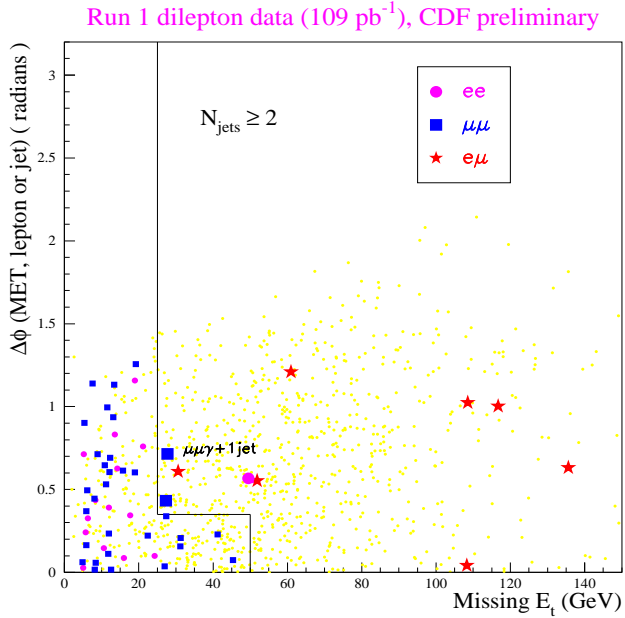}
\caption[fig44]
{Observed correlation between the missing transverse energy and the 
angle between the missing $p_{t}$ vector and the nearest 
lepton or jet for the CDF dilepton $t\bar{t}$ 
candidates in the data and the Monte Carlo~\cite{cdflltop}.}
\end{center}
\end{figure}

Having foccussed on todays fashionable models we would like to finish our 
review with the remark that a succesfull search does not need to please 
a theoretical exotic model.
As an example we would like to mention the existence of a few 
unconnected mysterious events at HERA and at the Tevatron.
The H1 collaboration has recently published the observation of a few 
high $p_{t}$ events which contain isolated muons, jets and large missing 
transverse energy~\cite{h1exotics}. 
The observed 5 $\mu-X$ events are somehow high compared 
to the expected SM rates of about 1 event. In addition, as shown in 
Figure 43, at least three of the five $\mu-X$ events 
show some weird kinematics while the corresponding $e-X$ event 
is in agreement with expectations. 
Another anomaly involving isolated leptons and jets has been 
reported by the CDF collaboration~\cite{cdflltop}. The analysis compares 
events which contain a pair of isolated leptons ($ee$, $\mu\mu$ and $e\mu$)
and at least two jets with expectations from a $t\bar{t}$ Monte Carlo. 
As can be seen from Figure 44, one finds that
the majority of the observed events are well reproduced by the 
Monte Carlo. However, four events have a somehow unexpected large 
missing transverse energy. 

In both cases, an excess of events is found 
in tails of a two-dimensional distribution. The interest in these events 
is enlarged as they are found in an analysis of events 
with isolated leptons plus jets plus missing transverse energy.   
A {\bf trivial} but correct  
statement is that the observed excess does currently 
not allow any discovery claim and that {\bf more data are needed}.
This statement probably satisfies especially the experimentalists 
which are not working at the Tevatron or at HERA.
The {\bf trivial} and correct reply is {\bf ``don't worry''}   
many more data might be available soon.
However, hoping for a real effect, experimentalists 
and theorists should feel encouraged to imagine some new and related 
signatures which might perhaps be tested even with todays 
data.

To finish this ``How to do Searches'' guide we would like to 
quote a few authorities:
\newline
\vspace{0.05cm}

{\it ``What can be measured, results from theory''} Einstein to Heisenberg
\newline
\vspace{0.05cm}

{\it ``Experiments within the next 5--10 years will enable us to decide 
whether supersymmetry, as a solution to the naturalness problem of the 
weak interaction is a myth or reality''} H. P. Nilles 1984~\cite{mssm84}  
\newline
\vspace{0.05cm}

{\it ``One shouldn't give up yet'' .... ``perhaps a correct statement is:
it will always take 5-10 years to discover SUSY''} 
H. P. Nilles 1998~\cite{nilles98}
\newline
\vspace{0.05cm}

{\it ``Superstring, Supersymmetry, Superstition''} Unknown
\newline
\vspace{0.05cm}

{\it ``New truth of science begins as heresy, advances to orthodoxy and ends 
as superstition''} T. H. Huxley (1825--1895).
\newline
\vspace{0.05cm}

{\bf \large Acknowledgements}
{\small I would like to thank the organisers of
the SUMMER school on ``La Physique au Tevatron, \`{e}tape vers le 
LHC'' in Marseille
for the invitation to review ``Search Strategies for the Higgs 
and other new particles'' from LEP to the LHC.
At least I have profited enourmously from many excellent lectures  
about the current results and future prospects of the Tevatron experiments,
which stimulated many interesting discussions about remaining 
problems for the Tevatron and the future LHC experiments.
I am grateful to D. Haztifotiadou, F. Pauss and Z. W\c{a}s 
for their critical suggestions and comments about my lectures and 
this writeup.}
 
\end{document}